\documentclass[letterpaper,twocolumn,10pt]{article}
\usepackage{usenix2020}

\usepackage{caption}
\usepackage{fontenc, inputenc}
\usepackage{amsmath,multirow,multicol,xurl}
\usepackage{enumitem} \setlist[itemize]{leftmargin=*}
\usepackage{graphicx, comment}
\usepackage{blindtext}
\usepackage{tikz,graphics,color,float,epsf,caption,subcaption}
\usepackage[toc,page]{appendix}
\usepackage{color,soul}
\usepackage{authblk}

\begin{document}
%-------------------------------------------------------------------------------

%don't want date printed
\date{}

%\title{\Large \bf Is the Public Cloud Ready for 5G?\\
%  Evaluating Control and User Plane Bottlenecks}

%\title{A First Look at Edge 5G Core Deployments Over the AWS Public Cloud: \\ Evaluating Control and User Plane Bottlenecks}
%\title{A First Look at Edge 5G Core Deployments on Public Cloud: \\ Evaluating Control and User Plane Bottlenecks}
%\title{A First Look at Edge 5G Core Deployments on Public Cloud: \\ Evaluating the Control and User Plane}
\title{A First Look at 5G Core Deployments on Public Cloud: \\ Performance Evaluation of Control and User Planes}

\author[1]{Tolga O. Atalay}
\author[2]{Dragoslav Stojadinovic}
\author[1]{Alireza Famili}
\author[1,2]{Angelos Stavrou}
\author[1]{Haining Wang}
\affil[1]{Department of Electrical and Computer Engineering, Virginia Tech, USA}
\affil[2]{Kryptowire Labs, Arlington, VA, USA}

\maketitle

%-------------------------------------------------------------------------------
\begin{abstract}
The Fifth Generation (5G) mobile core network is designed as a set of Virtual Network Functions (VNFs) hosted on Commercial-Off-the-Shelf (COTS) hardware. This creates a growing demand for general-purpose compute resources as 5G deployments continue to expand. Given their elastic infrastructure, cloud services such as Amazon Web Services (AWS) are attractive platforms to address this need. Therefore, it is crucial to understand the control and user plane Quality of Service (QoS) performance associated with deploying the 5G core on top of a public cloud. To account for both software and communication costs, we build a 5G testbed using open-source components spanning multiple locations within AWS. We present an operational breakdown of the performance overhead for various 5G use cases using different core deployment strategies. Our results indicate that moving specific VNFs into edge regions reduces the latency overhead for key 5G operations. Furthermore, we instantiated multiple user plane connections between availability zones and edge regions with different traffic loads. We observed that the deterioration of connection quality varies depending on traffic loads and is use case specific. Ultimately, our findings provide new insights for Mobile Virtual Network Operators (MVNOs) for optimal placements of their 5G core functions.
\end{abstract}
%-------------------------------------------------------------------------------

\section{Introduction}
The deployment of next-generation mobile networks is gaining momentum. Compared with legacy Long Term Evolution (LTE), the Fifth Generation (5G) networks have been designed to accommodate a wide range of industry verticals with different Quality of Service (QoS) demands. Thus, the delivery of services in 5G takes place over logically isolated segments called ``network slices." %that have been tailored for specific use cases. %In order to systematically classify the requirements of different users, the Third Generation Partnership Project (3GPP) has standardized certain use cases by designating them Slice Service Types (SSTs)~\cite{3gpp23501}. These include, but are not limited to, enhanced Mobile Broadband (eMBB), Ultra Reliable Low Latency Communication (URLLC), massive Internet of Things (mIoT) and Vehicular to Everything (V2X) communications. 
%
%To achieve a Standalone (SA) 5G deployment is made up of a next generation Radio Access Network (RAN) and a decentralized core network. In LTE, the core functions were bound to proprietary hardware as Physical Network Functions (PNFs). %This resulted in a rigid deployment model that was difficult to scale and could not address more fine-grained QoS requirements. 
For the flexible deployment of network slices, 5G leverages Network Functions Virtualization (NFV) as a building block for its core network~\cite{ordonez2017network,yousaf2017nfv}. 
%While Physical Network Functions (PNFs) were used in LTE, 5G mobile network functions are packaged as Virtual Network Functions (VNFs) deployed on top of Commercial off-the-Shelf (COTS) hardware. %This creates a decentralized deployment model that can scale upon demand from the Mobile Virtual Network Operators (MVNOs). 
%Complementary to NFV, SDN enables the near-real time provisioning of these 5G core VNFs to dynamically respond to the QoS requirements of specific use cases. 
%Network slices are constructed by service chaining these VNFs based on instance templates~\cite{3gpp28530} to customize the delivery to the use case.

%This step towards virtualization creates enterprise opportunities for service providers to enter the mobile network ecosystem. 
%Small-to-Medium Enterprises (SME) can build Software-as-a-Service (SaaS), Platform-as-a-Service (PaaS) and Network-Slice-as-a-Service~\cite{gsm2021official,atalay2022network,zhou2016network} business models to create custom 5G offerings for end users.
The adoption of NFV for facilitating 5G deployments creates an increasing demand for computing infrastructure that can scale efficiently. To fill this vacuum, cloud service providers such as Amazon Web Services (AWS), Google Cloud Platform (GCP), and Microsoft Azure have been starting to tailor Infrastructure-as-a-Service (IaaS) offerings for Mobile Virtual Network Operators (MVNOs). Proof-of-concept trials are being conducted by Swisscom and Ericsson to deploy the 5G core on top of AWS~\cite{Swisscom30online}. Furthermore, Deutsche Telekom~\cite{Deutsche96online} and Telefonica~\cite{Telefóni70online} have entered into partnerships with AWS and Azure to explore a cloud-based 5G core network.
%the latter cloud giant has further solidified partnerships with other Telecommunication Service Providers (TSPs). %Currently these include Verizon, KDDI, Vodafone, SK Telekom and Bell~\cite{AWSWavel60online}. 
Given these developments, the objective of this paper is to understand the latency overhead and throughput bottlenecks associated with 5G deployments on the AWS public cloud. We conduct a series of experiments with alternative edge location options in seven countries, across eight AWS regions. %The insight we provide can be leveraged by MVNOs 

\begin{comment}

AWS infrastructure is divided into geographical segments denoted as parent regions. As illustrated in Figure~\ref{fig:awsinfra}, each region consists of multiple Availability Zones (AZs) and, for chosen regions, finer-grained Local Zones (LZs)~\cite{Deployme40online}. While an AZ is a generic deployment zone that exists for reliability through redundancy, LZs exist to enable low-latency applications at the edge of the AWS network. In addition to AZs and LZs, AWS has introduced the ``Wavelength Zone"~\cite{AWSWavel60online} (WZ). The WZs deliver an integrated cloud environment capable of supporting Ultra Reliable Low Latency Communication (URLLC) 5G applications by providing AWS features at the 5G network edge. Compared to LZs, this is achieved by deploying specific applications directly on top of the Telecommunication Service Provider (TSP) infrastructure, in close proximity to the 5G Radio Access Network (RAN). However, due to this displacement, AZ-WZ connections are subject to a higher latency than their AZ-LZ counterparts. %We confirm this through our preliminary benchmarks during evaluation.
\end{comment}

%All these zones offer compute resources in the form of Elastic Compute Cloud (EC2) instances~\cite{AmazonEC25online}. SMEs and MVNOs can leverage these flexible resources to deploy applications close to the network edge using LZs or specifically the 5G edge with WZs.

AWS edge locations denoted as Local Zones (LZs)~\cite{Deployme40online} and Wavelength Zones (WZs)~\cite{AWSWavel60online} are well-suited for hosting Ultra Reliable Low Latency (URLLC) applications to enhance user plane QoS. In addition to user applications, they can be leveraged for deploying 5G core Virtual Network Functions (VNFs) to minimize the control plane latency to the Radio Access Network (RAN). On the one hand, user plane latency and throughput are essential for providing a good Quality of Experience (QoE). On the other hand, control plane latency is crucial in maintaining reliability for Mission Critical Services (MCSs) such as healthcare, energy, and robotics~\cite{borsatti2022mission}. 

\begin{figure}[t]
    \centering
    \begin{subfigure}[t]{0.68\columnwidth}  
        \centering 
        \includegraphics[width=\textwidth,trim={7.9cm 4.6cm 9.5cm 1.5cm},clip]{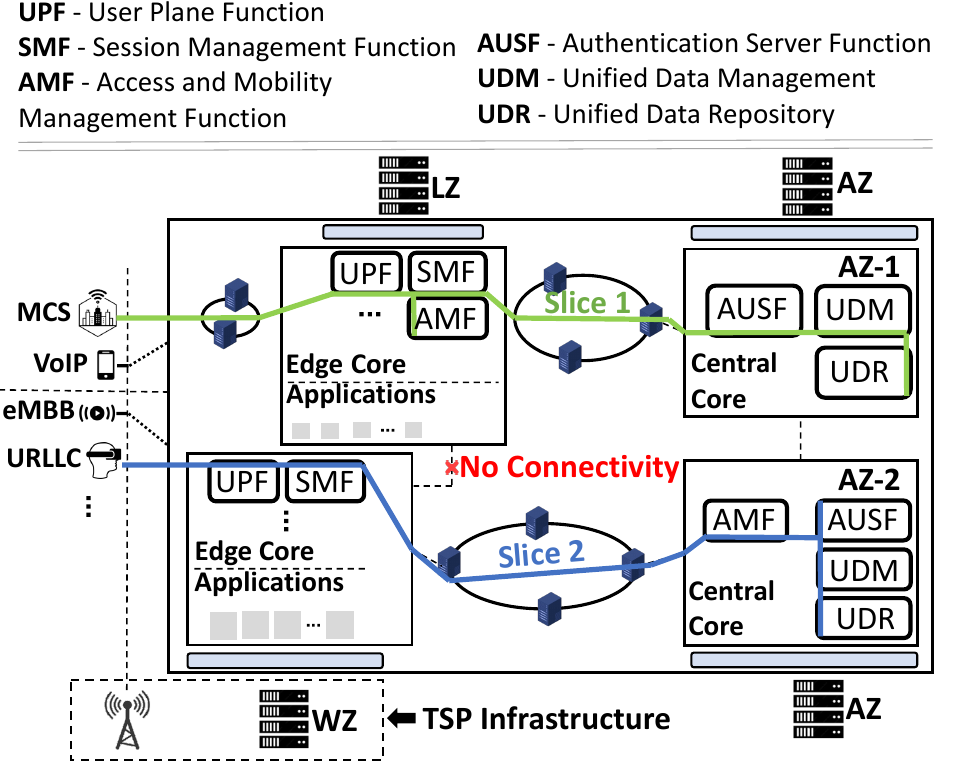}
        \caption[]%
        {{\small }}    
        \label{fig:aws5gdep}
    \end{subfigure}
    \hfill
    \begin{subfigure}[t]{0.24\columnwidth}
        \centering
        \includegraphics[width=\textwidth,trim={14.5cm 6.1cm 13.7cm 3cm},clip]{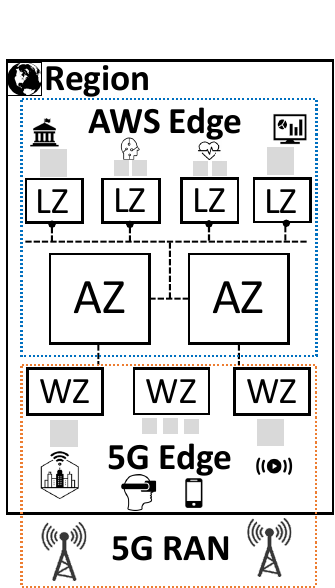}
        \caption[]%
        {{\small }}    
        \label{fig:awsinfra}
    \end{subfigure}
    \caption{(a) Breakdown of an AWS parent region with different infrastructure zones; (b) QoS-sensitive 5G deployment on top of AWS across AZs, LZs, and WZs}
    \label{fig:awsow}
\end{figure}

The significance of prioritizing among these operational planes is illustrated in Figure~\ref{fig:aws5gdep}. Two network slices are depicted over a tentative 5G core deployment in a hybrid Availability Zone (AZ)~\cite{BibEntry2023Sep}, LZ, and WZ layout. For the Mission Critical slice 1, it is more important to manage the network slice setup and transfer times~\cite{3gpp22280} of the clients to maintain high reliability. Therefore, slice 1 is stretched out over an LZ-AZ connection, where the LZ-AZ latency is lower than the WZ-AZ alternative. %This reduces the end-to-end operational time of interactions between the Serving Network (SN) (i.e., UPF, SMF, AMF) and the Home Network (i.e., AUSF, UDM, UDR). 
%
%This minimize the control plane latency between the Home Network VNFs (i.e., AUSF, UDM, UDR) and the Serving Network VNFs (UPF, SMF, AMF). 
On the other hand, for better QoS in latency-sensitive applications such as enhanced Mobile Broadband (eMBB) and URLLC cases, the data session VNFs are deployed closer to the 5G edge in WZs. Therefore, slice 2 is deployed over a WZ-AZ connection.

The advent of moving 5G deployments into a public cloud raises performance concerns at both the control and user planes. Mission critical applications with dense signaling require low latency in control plane handling~\cite{3gpp22280}. On the other hand, applications such as Augmented- and Virtual- Reality (AR/VR) require low latency in the user plane. 

To the best of our knowledge, this work is the first to shed light on large-scale 5G deployments over a public cloud at a global scale. We quantify the control plane latency overhead of critical 5G tasks and user plane bottlenecks in hybrid AZ, LZ, and WZ deployments across multiple locations within AWS. Leveraging our testbed, operators can make informed decisions regarding their 5G deployments to determine favorable placements of control and user plane entities for different use cases.  %We plan to make our 5G core deployment scripts, source code, measurements, and relevant Amazon Machine Images (AMIs) available to the community to render the work reproducible. The guide to replicate this paper is anonymously available as allowed by the operational systems track guidelines~\footnote{https://github.com/anonprvsubs/nsdi2024}. 
Our contributions are summarized below.

%We report the latency overhead for control plane operations in a hybrid central/edge cloud environment. Additionally, our investigation demystifies performance bottlenecks that exist between LZs/WZs and AZs to determine favourable placements for 5G user plane entities (i.e., the UPF and 5G application servers) for different use cases.
%
%Furthermore, the innate flexibilibe ty of 5G deployments as a result of NFV, gives operators the ability to deploy specific core VNFs in foreign countries to facilitate roaming without changing infrastructure domains. For instance, an individual travelling from the United States to Argentina can use edge VNFs located in the Buenos Aires LZ, while using the Verizon central cloud VNFs deployed in Northern Virginia. Therefore it is critical to not only understand the implications of 5G cloud deployments for domestic cases with respect to an operator but also roaming scenarios that include international users. %Depending on the VNF deployment strategy being used, signalling across VNFs 
%
%Ultimately, our goal is to perform a deep dive into the QoS and compute consumption implications of deploying specific network slice VNFs at different cloud hierarchies within AWS. 

\begin{comment}
\begin{figure}[t]
    \centering  
    \includegraphics[width=\columnwidth,trim={7.9cm 4.6cm 9.5cm 2.2cm},clip]{figures/aws5gdepannot.pdf}
    \caption{QoS-sensitive 5G deployment on top of AWS across AZs, LZs, and WZs}
    \label{fig:aws5gdep}
\end{figure}
\end{comment}

\begin{itemize}
    \item For finer-grained control over experimentation, we use the open-source OpenAirInterface~\cite{oaiopena76online} (OAI) 5G core and the gNBSIM~\cite{RohanGnb5online} RAN. This allows us to manage 5G deployments without relying on any proprietary solution. We create multiple self-managed Kubernetes clusters on top of AWS Elastic Compute Cloud (EC2) instances spanning multiple zones. These clusters are used to instantiate mass end-to-end connections between emulated users and Data Network Nodes (DNNs) that we control. To perform analysis on a global scale, we conduct experiments in eight different parent regions for 18 different edge zones.
    
    \item We conduct a measurement campaign to understand the bottlenecks within the AWS edge computing networks. This includes daily benchmarks on latency, throughput, and packet loss. %Firmly establishing these benchmarks within AWS not only elevates our 5G core study, but also provides future researchers that seek to utilize the same infrastructure. 
    Our measurements can be used as a reference point for future studies examining changes in the AWS edge computing architecture.
    
    \item To evaluate the 5G control plane, we build a state-of-the-art instrumentation environment using Jaeger~\cite{Jaegerop88online} and OpenTelemetry~\cite{OpenTele38online}. Integrated into this ecosystem, our custom monitoring Side-car Proxy (SCP) sits adjacent to each VNF, collecting telemetry data. Through this pipeline, we are able to trace the HTTP transactions between individual Representational State Transfer (REST) APIs of the 5G core. %By deploying end-to-end sessions at different AWS AZs, LZs and WZs, the control plane latency overhead of 5G cloud deployments is analyzed. 
    In our evaluations, we derive and experiment with multiple control plane strategies pertaining to real-life use cases (e.g., MCS, URLLC). Each strategy is associated with a placement pattern, where VNFs are shuffled between an edge zone and an AZ. Comparing the overhead of different strategies, we provide insight into what the control plane latency implications are for cloud-based 5G deployments. %This includes experimenting with same and cross region AZs, LZs and WZs in order to investigate roaming scenarios across countries.
    
    \item To understand user plane bottlenecks in different regions, traffic loads of real-life 5G use cases are recorded. Then, end-to-end user connections are created across edge zones and AZs. The captured downlink and uplink traffic is replayed through the 5G user plane to create traffic on top of AWS. Through these experiments, we measure the bandwidth bottlenecks between edge zones and AZs. Our goal is to provide operators with guidance regarding 5G user plane entity placement in the cloud.

\end{itemize}

\section{Background}
%This section provides an overview of our 5G deployment and the tools used in the construction of the monitoring framework. Lastly, the concept of SCPs is explained in the context of micro-service design.

\subsection{5G Core Overview} \label{sec:5gback}
%In order to systematically classify the requirements of different users, the Third Generation Partnership Project (3GPP) has standardized certain use cases by designating them SSTs~\cite{3gpp23501}. These include, but are not limited to, enhanced Mobile Broadband (eMBB), Ultra Reliable Low Latency Communication (URLLC), massive Internet of Things (mIoT) and Vehicular to Everything (V2X) communications. 

%The 5G core is built as a Service-based Architecture (SBA) where multiple VNFs are connected through Service-based Interfaces (SBIs)~\cite{3gpp23501}. %Each VNF has multiple microservices where through deterministic control plane operations they manage different procedures of the User Equipment (UE) lifecycle. 
The 5G core VNFs communicate with one another using standardized Representational State Transfer (REST) APIs named according to the Common API Framework (CAPIF)~\cite{3gpp23222} described by the 3rd Generation Partnership Project (3GPP). %Regardless of how the internal implementation of the microservice is customized, the API endpoint names of the 5G core components must conform to CAPIF in order to remain vendor-agnostic. 
%
%To deploy a fully functional end-to-end 5G core network, we utilize the open-source OAI 5G core~\cite{oaiopena76online} along with the gNBSIM~\cite{RohanGnb5online} RAN entity. This approach allows us to control and scale the deployment for finer-grained experimentation. 

\begin{figure}[t]
    \centering
    \includegraphics[width=0.8\columnwidth,trim={14cm 8.75cm 12.6cm 7.15cm},clip]{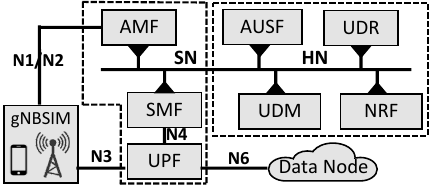}
    \caption{OAI 5G core deployment}
    \label{fig:oailogi}
    \vspace{-0.1in}
\end{figure}

The logical architecture of our deployment is presented in Figure~\ref{fig:oailogi}. The VNFs are divided into the Home Network (HN) and Serving Network (SN) components. As a critical entity of the 5G core, the Access and Mobility Management Function (AMF) typically resides in the SN, acting as the communication highway among the remaining VNFs, the RAN, and the User Equipment (UE). % It oversees various registration procedures and handles the authentication of the user in the SN. 
AMF receives authentication credentials from the Authentication Server Function (AUSF) in the HN. Along with the Unified Data Management (UDM) and Unified Data Repository (UDR), AUSF takes part in the 5G Authentication and Key Agreement (AKA)~\cite{3gpp33501}. In the SN, the Session Management Function (SMF) and the User Plane Function (UPF) are respectively the control and user plane anchors for the UE data session. The UPF tunnels the user traffic toward a Data Network Node (DNN) container, which represents the application servers in real life. The initial discovery and communication between these VNFs are facilitated by the Network Functions Repository Function (NRF) acting as a metadata database. Last but not least, connecting to this network is the gNBSIM entity, which sends both gNB and UE signals to the core from the same sandbox encapsulation. %A detailed overview of the 5G core VNF interactions in our deployment is illustrated in Appendix~\ref{app:5gmsgflow}.

\subsection{AWS Local and Wavelength Zones}

\begin{comment}
\begin{figure}[t]
    \centering  
    \includegraphics[width=0.9\columnwidth,trim={11cm 8.5cm 10.7cm 6.3cm},clip]{figures/awsinfrahor.pdf}
    \caption{Breakdown of an AWS parent region with different infrastructure zones}
    \label{fig:awsinfra}
\end{figure}
\end{comment}

AWS infrastructure is divided into geographical segments denoted as parent regions. As illustrated in Figure~\ref{fig:awsinfra}, each region consists of multiple AZs and, for chosen regions, finer-grained LZs~\cite{Deployme40online}. 
While an AZ is a generic deployment zone that exists for reliability through redundancy, LZs exist to enable low-latency applications at the edge of the AWS network. In addition to AZs and LZs, AWS has introduced the WZ~\cite{AWSWavel60online}. The WZs deliver an integrated cloud environment capable of supporting URLLC 5G applications by providing AWS features at the 5G network edge. Compared to LZs, this is achieved by deploying specific applications directly on top of the Telecommunication Service Provider (TSP) infrastructure, in close proximity to the 5G RAN. However, due to this displacement, AZ-WZ connections are subject to a higher latency than their AZ-LZ counterparts.

\section{Measurement Methodology}
%In the upcoming evaluation sections, we carry out 5G network slice deployments in multiple regions around the world. Our goal is to evaluate the control and user plane performance of the core network using different deployment strategies where VNFs and DNNs are deployed across AZs, LZs and WZs. 
\begin{figure}[t]
    \centering
    \includegraphics[width=\columnwidth,trim={0cm 1cm 4.22cm 3cm},clip]{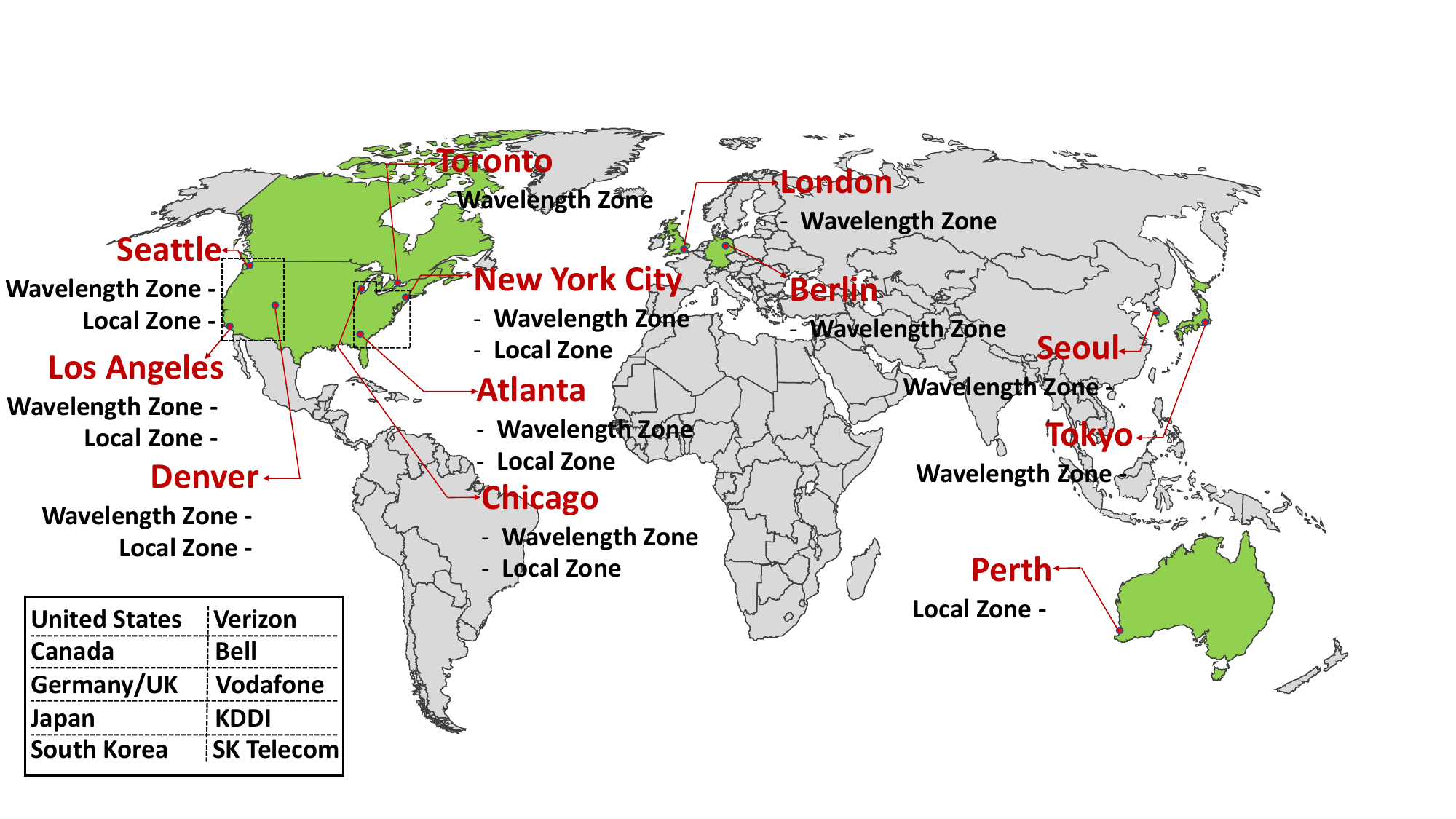}
     \caption{The Local and Wavelength zones used across the world for experimentation}
     \label{fig:locationmap}
\end{figure}

In this section, we first present the global locations selected for experimentation. Next, the instrumentation pipeline for HTTP transaction tracing is introduced. Subsequently, we delve into the VNF placement strategies related to actual Slice Service Types (SSTs)~\cite{3gpp23501} employed in control plane assessments. Following that, the user plane measurement methodology is described using the selected traffic patterns. Finally, we present our experimentation environment within AWS.

\subsection{Global Locations}
To represent a wide range of regions, we have selected edge zones corresponding to different locations within the AWS infrastructure. The specific cities and their WZ/LZ support are illustrated in Figure~\ref{fig:locationmap}.

To compare LZ and WZ edge locations within a single city, we selected six cities from the United States: three from the US-East (Northern Virginia) region and three from the US-West (Oregon) region. The remaining locations offer only one of the two edge zone variants, making it difficult to conduct a comparison. Nevertheless, we still evaluate the individual edge performance of LZ/WZ deployments across different locations in Perth, Tokyo, Seoul, London, and Berlin. %The AWS parent regions for the edge zones in Figure~\ref{fig:locationmap} are listed in Figure~\ref{tbl:parents}.

\subsection{Monitoring Framework}

\begin{figure}[t]
    \centering
    \includegraphics[width=0.9\columnwidth,trim={11.1cm 5.4cm 10.5cm 3.4cm},clip]{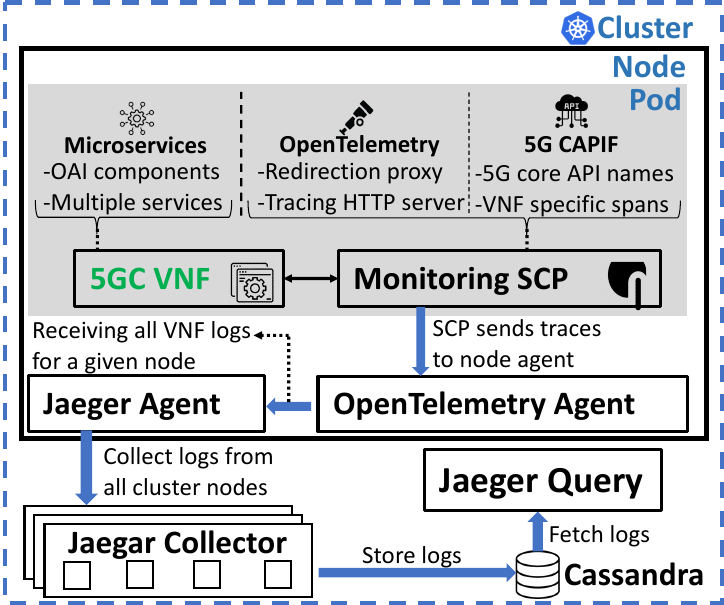}
     \caption{Overview of the monitoring framework for 5G core transaction tracing}
     \label{fig:frameov}
\end{figure}

An instrumentation pipeline is constructed using Jaeger~\cite{Jaegerop88online} and OpenTelemetry~\cite{OpenTele38online} as illustrated in Figure~\ref{fig:frameov}. Each building block is shown within its respective deployment hierarchy in the Kubernetes experimentation environment, which can occur at a pod-, node- or cluster-level. 

%Deployed in individual pods, 5G core VNFs are the applications of interest that are traced by the monitoring framework. Each VNF is comprised of multiple microservices dedicated to a specific core network operation. 
To avoid modifications to the source code of the 5G core VNFs for gathering telemetry information, we design a custom Side Car Proxy (SCP)~\cite{WhatIsaS8online} using OpenTelemetry for intercepting and redirecting the HTTP messages between the 5G pods. This is denoted as indirect communication and has been standardized by 3GPP in Release 16 as a viable VNF-to-VNF interaction method~\cite{3gpp23501}.

%Sitting adjacent to each 5G core VNF in the pod is the monitoring SCP designed using OpenTelemetry. %We generate the necessary logs for our control plane measurements by ine. 
The SCP functions as a two-way redirection proxy that forwards messages to and from the 5G application residing in the same pod. When a message enters the pod network sandbox, it is processed at the OpenTelemetry HTTP server within the SCP. As the HTTP request is forwarded to its destination, a span is created for the given transaction based on the 5G CAPIF path in the URL. This allows us to distinguish between individual 5G core control plane messages and identify sender/receiver VNFs. Once the spans are created, the SCP sends them to the node-level OpenTelemetry agent (see Appendix~\ref{app:msginterflow} for details on message interception).

\begin{comment}
\begin{figure}[t]
    \centering
    \includegraphics[width=\columnwidth,trim={3.3cm 9.7cm 3.8cm 3.2cm},clip]{figures/monovshor.pdf}
     \caption{Overview of the monitoring framework for 5G core transaction tracing}
     \label{fig:frameov}
\end{figure}
\end{comment}

The OpenTelemetry agent functions as the initial consolidation point for the gathered logs. After receiving telemetry from individual SCPs, it forwards the information to the respective Jaeger agent deployed alongside it. A central Jaeger collector receives logs from all the Jaeger agents in the cluster and stores them in a Cassandra database as JSON files. Finally, the Jaeger query is used to fetch them from the database. Both OpenTelemetry and Jaeger agents are deployed as Daemonsets within the Kubernetes cluster, and communications with them take place over Kubernetes DNS.

\subsection{VNF Deployment Strategies} \label{sec:vnfstrat}

\begin{figure}[t]
    \centering
    \includegraphics[width=0.9\columnwidth,trim={3.4cm 5.1cm 6.8cm 4.6cm},clip]{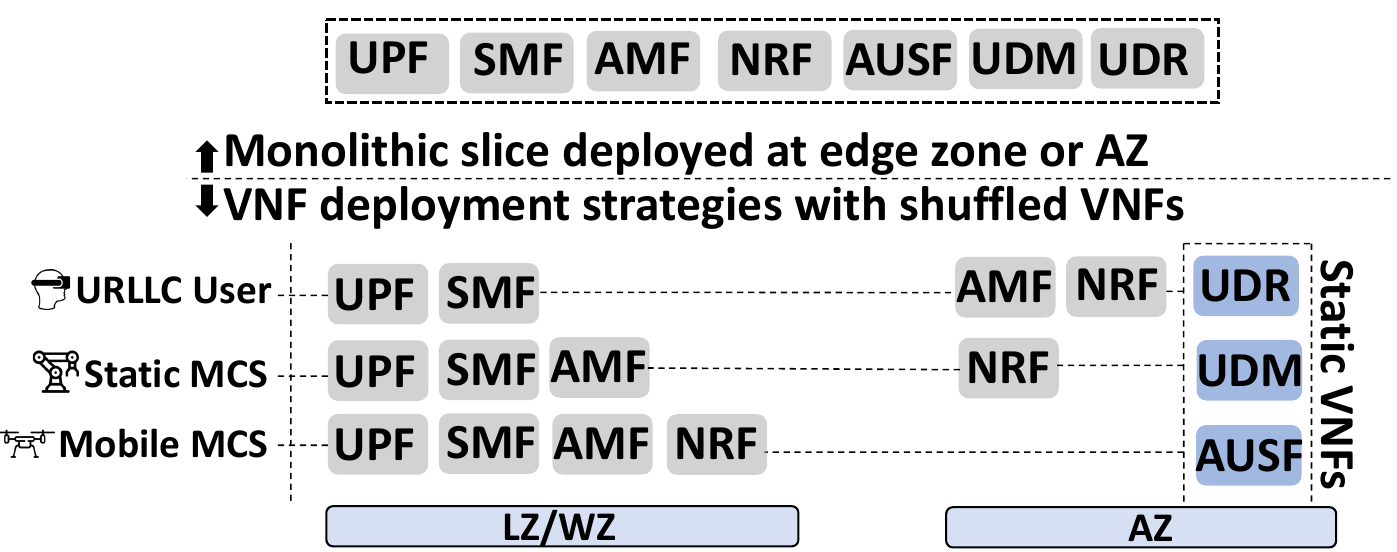}
     \caption{5G core VNF deployment strategies on AWS infrastructure for different use cases}
     \label{fig:depstrats}
     %\vspace{-0.1in}
\end{figure}

We employ three hybrid placement strategies, as illustrated in Figure~\ref{fig:depstrats}, wherein the VNFs are positioned either at the network edge or an AZ. Additionally, a monolithic network slice is used as a benchmark, where all the VNFs are grouped together in the same physical location. Since computational resources are scarcer and more expensive in edge zones, control plane VNFs are placed there only when use cases require low latency between specific control plane interactions. %Each strategy we have chosen is associated with the control plane requirements of a real-life SST. The HN VNFs (i.e., UDR, UDM, and AUSF) are statically placed in the AZ, instead of being distributed to edge zones. 

For the first strategy, we consider the generic \textbf{URLLC} SST for the user plane, where only the data session VNFs (i.e., the UPF and the SMF) have been placed in edge zones. The primary goal for this deployment is to maintain high QoS in the user plane for use cases such as AR/VR and gaming. Since any added control plane latency will not deteriorate the QoE of users, the remaining VNFs are instantiated in AZs rather than edge zones.

The second strategy is where the AMF is moved into the edge alongside the UPF and SMF. This network slice configuration is suited for \textbf{static MCSs} where the clients are not moving, but there is high connection density. Specific use cases include mIoT in smart warehouses, hospitals, and government buildings. By moving the AMF to the edge zone, the session setup interaction with the SMF is minimized. 

The final strategy is for \textbf{mobile MCSs} such as drones, vehicular to everything (V2X) communication users, and other high mobility clients. For these devices, it is reasonable to expect a high volume of network slice transfer traffic as the target moves across different physical locations. To accommodate this behavior, new network slices might need to be instantiated to host these incoming users. Alternatively, VNFs from existing slices may need to be discovered~\cite{3gpp29510} to establish the connection of the user to the new slice. To facilitate these operations with minimal overhead, the NRF is moved into the edge along with the UPF, SMF, and AMF.

\subsection{User Traffic Generation}

\begin{table}[t]
\vspace{12pt}
\centering
\small\selectfont
\caption{Traffic loads in user plane experiments} 
\label{tbl:usecases}
  \begin{tabular}{p{0.12\textwidth}|p{0.14\textwidth}|p{0.13\textwidth}}
    \multirow{1}{*}{\textbf{SST}} & \multirow{1}{*}{\textbf{Use Case}} & \textbf{Application} \\ \hline
    \multirow{2}{*}{eMBB} &  Streaming & Netflix \\ \cline{2-3}
                          &  Video browse & Tiktok \\ \hline
    \multirow{1}{*}{URLLC} & Gaming & Fortnite \\ \hline
    \multirow{1}{*}{eMBB + URLLC} & VR & Horizon Venues \\\hline 
    \multirow{1}{*}{VoIP} & Video call & Zoom \\
  \end{tabular}
\end{table}

For user data, we use the pre-recorded traffic of different use cases listed in Table~\ref{tbl:usecases}. Netflix and Tiktok are considered for the eMBB SST. For re-creating a gaming session, we record the traffic during Fortnite gameplay to represent URLLC. As a more hybrid SST example, with both eMBB and URLLC requirements, VR traffic pattern from an Oculus Quest 2 is captured during a Horizon Venues session. Finally, for Voice over IP (VoIP), a Zoom session with video is recorded. The traffic patterns of each use case are shown in Appendix~\ref{app:traffics}.

\subsection{Experimentation Environment}

 \begin{figure}[b]
    \centering
    \includegraphics[width=0.8\columnwidth,trim={12.8cm 7.2cm 12.2cm 7.3cm},clip]{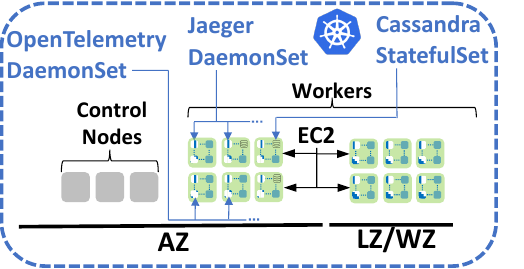}
     \caption{Complete experimentation Kubernetes cluster deployed on top of AWS with all entities illustrated}
     \label{fig:expsetup}
\end{figure}

For a High Availability (HA) Kubernetes cluster, the full experimental setup is depicted in Figure~\ref{fig:expsetup}, spanning multiple EC2 instances in a hybrid cloud deployment. Three control nodes are deployed in the AZ along with six workers. Another six workers are placed in edge zones. Jaeger and OpenTelemetry agents are running on every worker, deployed as DaemonSets. Finally, Cassandra is deployed as a StatefulSet for synchronization and scalability. %More details can be found in Appendix~\ref{app:clusterdet}. 
In Appendix~\ref{app:clusterdet}, we detail the deployment process and outline key issues for anybody seeking to construct such a testbed.

%We use the Ranchers Kubernetes Engine (RKE)~\cite{RKE72online} to set up a self-managed production-grade Kubernetes cluster on top of 15 EC2 instances. This allows us to have greater visibility into the container management infrastructure. We are able to manipulate worker node labels during setup to manage the location-specific VNF deployment during experimentation. 

%In our flavor selection, we are limited to the t3.medium, t3.xlarge, and r5.2xlarge general-purpose compute nodes at WZs~\cite{5GEdgeCo55online}. With only 2 vCPUs, t3.medium is not ideal for worker nodes and can lead to instability in the Kubernetes control plane. Therefore, we choose to use the more conservative t3.xlarge with 4 vCPUs, instead of the r5.2xlarge with 8 vCPUs. This enables us to spawn more EC2 instances within our regional vCPU quotas. The same flavor is used in AZs and LZs as well to preserve equality across the experiments.

\section{Control Plane Results} \label{sec:eval1}
Notable research studies have focused on increasing the reliability of cellular control plane operations~\cite{ahmad2022enabling, ahmad2020low}. Thus, to assess the eligibility of 5G deployments in the cloud, our first set of evaluations revolves around the operational latency induced in the control plane. 
In this section, we first perform extensive latency benchmarks across AZs for each edge zone. Next, we present the results that highlight the operational overhead for the chosen deployment strategies in Section~\ref{sec:vnfstrat}.

\subsection{AWS Latency Benchmarks} \label{sec:latbench}

\begin{figure}[t]
    \centering
    \begin{subfigure}[t]{0.442\columnwidth}
        \centering
        \includegraphics[width=\textwidth,trim={8.8cm 1.4cm 8.2cm 1cm},clip]{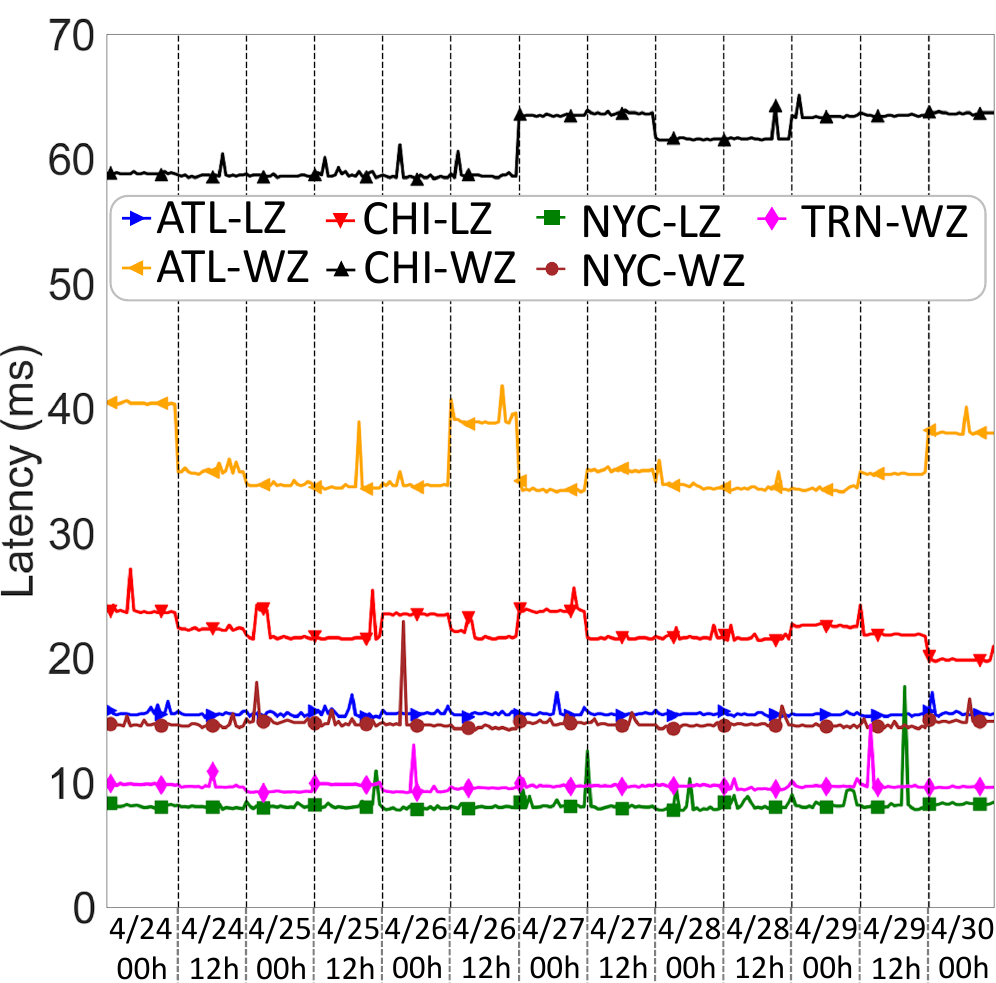} 
        \caption{North America - East}
        \label{fig:benchalllatnaesq}
    \end{subfigure}%
    \begin{subfigure}[t]{0.446\columnwidth}
        \centering
        \includegraphics[width=\textwidth,trim={7.3cm 2.2cm 9.6cm 0.6cm},clip]{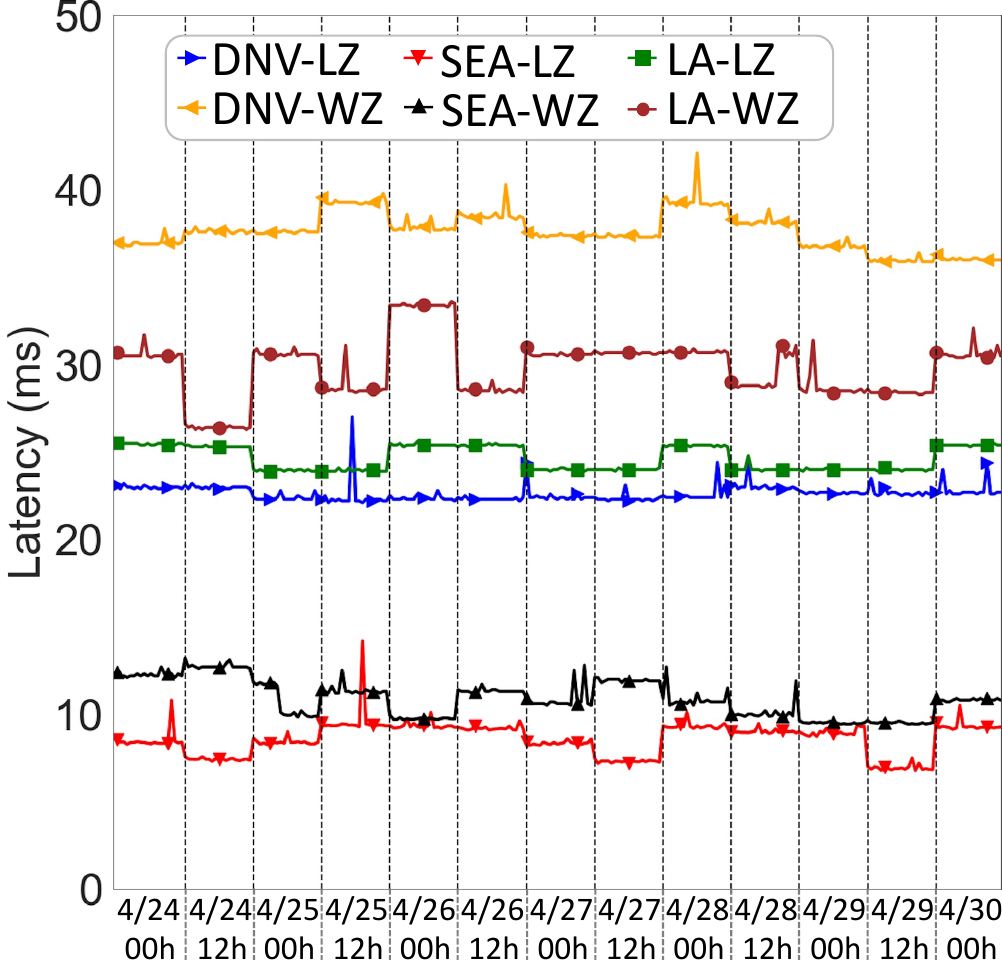}        
        \caption{North America - West}
        \label{fig:benchalllatnawsq}
    \end{subfigure}
    \begin{subfigure}[t]{0.445\columnwidth}
        \centering
        \includegraphics[width=\textwidth,trim={6.5cm 0.6cm 9.8cm 1.3cm},clip]{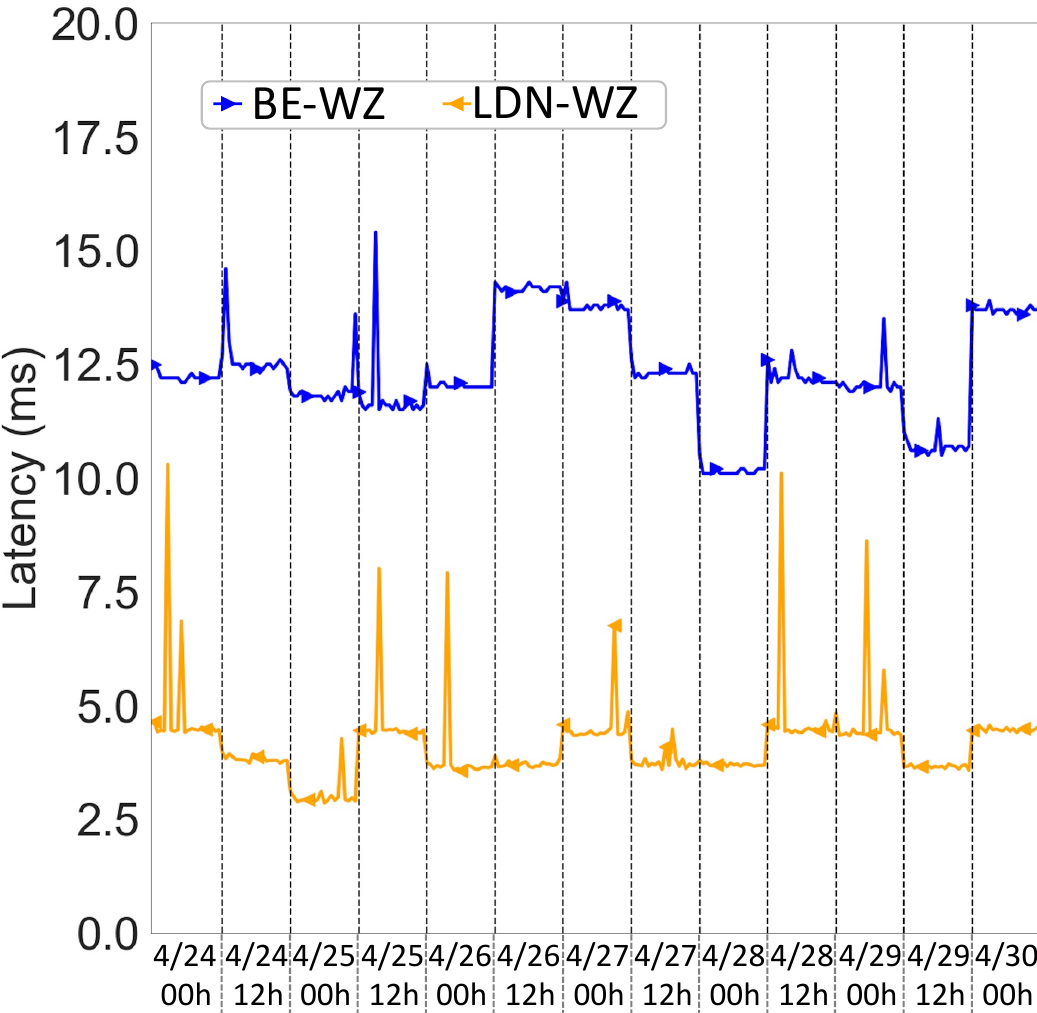}        
        \caption{Europe}
        \label{fig:benchalllatneusq}
    \end{subfigure}
    \begin{subfigure}[t]{0.445\columnwidth}
        \centering
        \includegraphics[width=\textwidth,trim={8.8cm 1.4cm 8.2cm 1cm},clip]{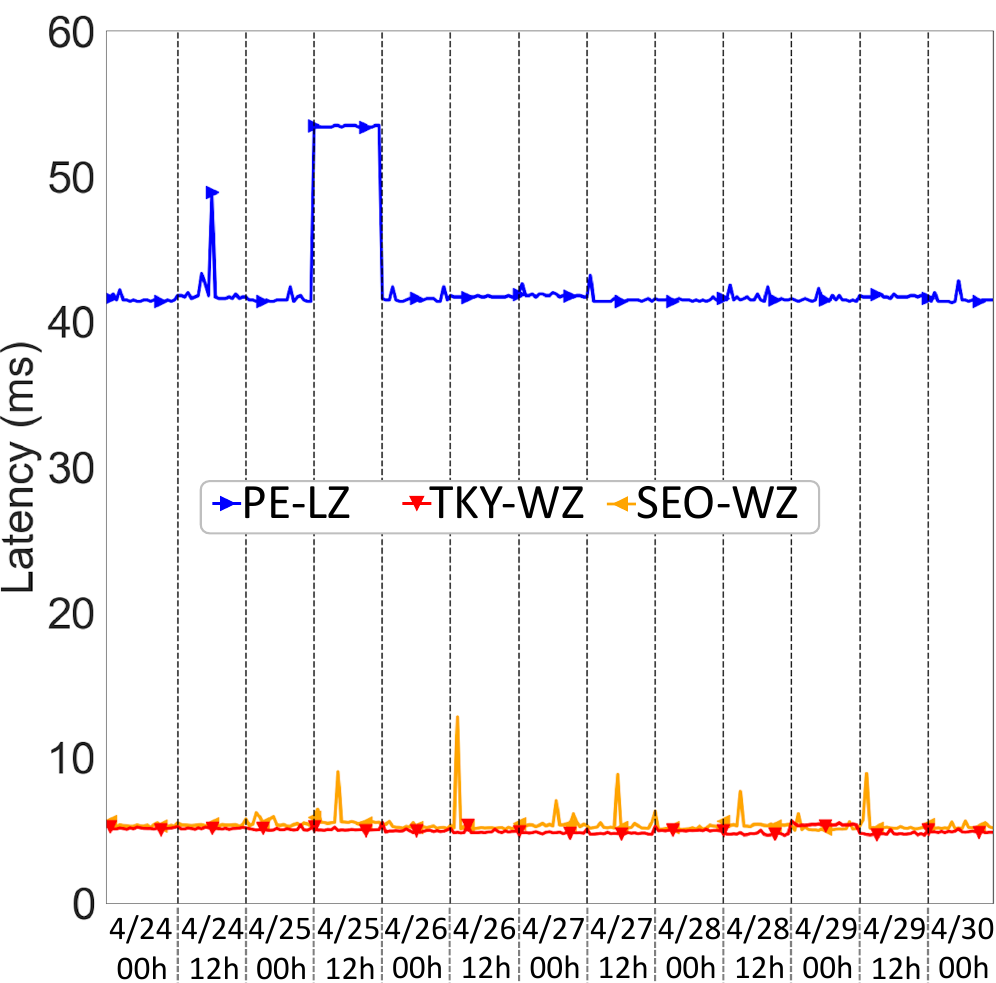}        
        \caption{Asia-Pacific}
        \label{fig:benchalllatapsq}
    \end{subfigure}
    \caption{Latency measured over a week with 12-hour intervals between edge zones and the relevant AZ. Measurement recordings are started at the reported times according to Eastern Standard Time. (April 24-30th 2023)}
    \label{fig:benchlatallsq}
\end{figure}

In order to achieve fidelity in our control plane experiments, we start by performing latency benchmarks between AZs and edge zones. To make a fair comparison across locations, we determine if selecting one AZ over another will significantly affect the connection. Our measurements (see Appendix~\ref{app:azbench}) show that the difference in latency to different AZs from a given edge zone is less than 1~ms. Since such latency is negligible, we use the first AZ (i.e., AZ-a) in each region during our experiments throughout the rest of the measurements.

%Our next benchmark is to verify that there is a consistent behavioral pattern over time. Since  
While connections between edge zones and AZs do display consistent behaviors across different AZs, measurements in Internet-based studies are still prone to fluctuations. Thus, we attempt to ensure that our control plane evaluations are not significantly affected by any time-related inconsistencies. To that end, we measure the latency from each edge location to its respective AZ over a week with 12-hour intervals. Thirty iterations of ``ping" with 100 samples each are averaged for the reported value. The fluctuations during the week of April 24th to April 30th, 2023 are shown in Figure~\ref{fig:benchlatallsq}, while the P50, P90, and P99 latency values are given in Table~\ref{tbl:pvalues}.
Based on the p-values, 
the highest volatility is observed for the London-WZ while the remainder of the zones have relatively stable latency readings. Thus, during our 5G core HTTP transaction trace in Section~\ref{sec:5gctrlmeas}, we gather ten measurements per trace which are averaged (see Appendix~\ref{app:httptran}). The latency fluctuations for each HTTP transaction are less than 5ms.

%To better demonstrate the fluctuations in the measurements over the entire week, Figure~\ref{fig:benchlatcdf} shows the Cumulative Distribution Functions (CDFs) for the four umbrella regions. Certain edge zones such as the Perth-LZ, Toronto-WZ and NYC-LZ have very stable latency measurements. On the other hand, several edge zones in western North America show higher fluctuations.  

\begin{comment}
\begin{figure}[t]
    \centering
    \includegraphics[width=1\columnwidth,trim={0cm 0cm 24cm 0cm},clip]{figures/benchlat_vert.eps}
     \caption{Bi-daily latency average between AZs and edge zones for the week of April 24th-30th 2023}
     \label{fig:latovertimebench}
     %\vspace{-0.1in}
\end{figure}
\end{comment}

\begin{table}[t]
\vspace{12pt}
\centering
\small\selectfont
\caption{P50, P90, and P99 values for the weekly latency (ms) measurements conducted between AZs and edge zones} 
\label{tbl:pvalues}
  \begin{tabular}{c|c|c|c|c}
    \textbf{City} & \textbf{Zone} & \textbf{P50} & \textbf{P90} & \textbf{P99} \\
    \hline \hline
    \multirow{2}{*}{Atlanta}
                         & LZ & 15.5 & 15.7 & 16.7 \\
                         & WZ & 34.1 & 39.1 & 40.6 \\ \hline 
    \multirow{2}{*}{New York City} 
                         & LZ & 8.05 & 8.36 & 10.5 \\
                         & WZ & 14.6 & 14.9 & 16.3 \\ \hline
    \multirow{2}{*}{Chicago} 
                         & LZ & 21.8 & 23.7 & 24.69  \\
                         & WZ & 61.6 & 63.7 & 63.9  \\ \hline
    \multirow{2}{*}{Denver} 
                         & LZ & 22.5 & 23.0 & 24.4  \\
                         & WZ & 37.6 & 39.2 & 39.6  \\ \hline
    \multirow{2}{*}{Seattle} 
                         & LZ & 8.92 & 9.35 & 10.2  \\
                         & WZ & 10.8 & 12.3 & 12.8 \\ \hline
    \multirow{2}{*}{Los Angeles}  
                         & LZ & 24.1 & 25.4 & 25.5  \\
                         & WZ & 30.5 & 31.1 & 33.5  \\ \hline
    \multirow{1}{*}{Toronto} 
                         & WZ & 9.67 & 9.84 & 10.5   \\ \hline
    \multirow{1}{*}{London} 
                         & WZ & 3.87 & 4.49 & 8.25  \\ \hline
    \multirow{1}{*}{Berlin}  
                         & WZ & 12.2 & 13.81 & 14.3   \\ \hline
    \multirow{1}{*}{Tokyo} 
                         & WZ & 5.29 & 5.57 & 8.87   \\ \hline
    \multirow{1}{*}{Seoul} 
                         & WZ & 4.97 & 5.27 & 5.48   \\ \hline
    \multirow{1}{*}{Perth}  
                         & LZ & 41.6 & 42.5 & 53.5    \\ 
  \end{tabular}
\end{table}

For four zones, the difference between the minimum and maximum latency throughout the week exceeds 5ms. In the US, only the Atlanta, Chicago, and Los Angeles WZs display such behavior. On the other hand, the LZ alternatives in the same cities have stable latency readings with a sub-2ms difference for the entire week. This goes to show that the connection between the AWS-TSP infrastructure is more prone to fluctuations than the AZ-LZ link. Another discrepancy can be seen at the noon measurement of April 25th for the Perth LZ. This is primarily due to an anomaly as this zone has a sub~1ms difference across the remainder of the intervals.

\subsection{5G Measurement Results} \label{sec:5gctrlmeas}

\begin{figure}[t]
\vspace{-0.1in}
    \centering
    \includegraphics[width=\columnwidth,trim={9.5cm 4cm 9.8cm 4cm},clip]{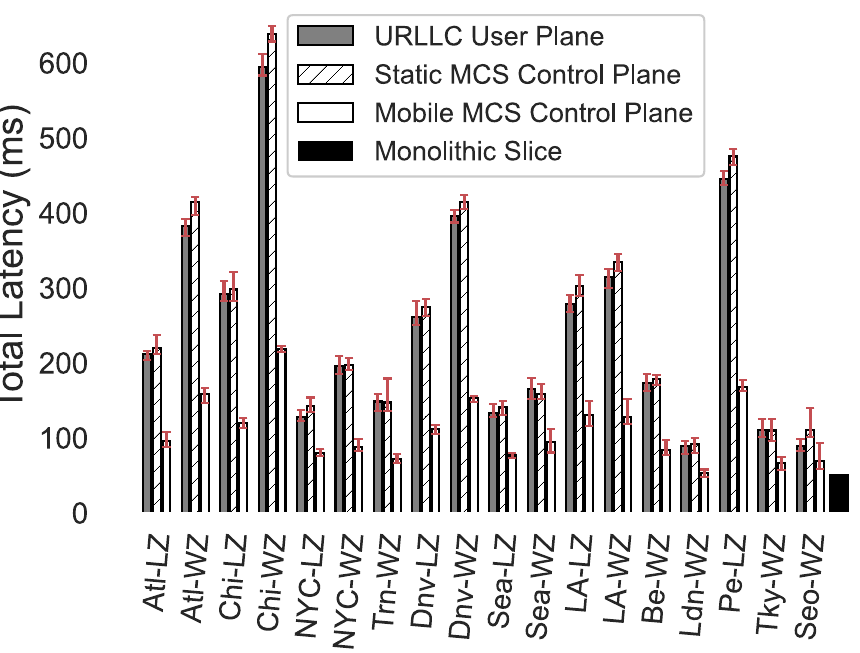}
     \caption{Total latency comparison for the different strategies 
     %shown in Figure~\ref{fig:depstrats} 
     across edge locations}
     \label{fig:totlat}
     %\vspace{-0.1in}
\end{figure}

For the first set of evaluations given in Figure~\ref{fig:totlat}, we analyze the total latency overhead of different 5G core deployment strategies illustrated in Figure~\ref{fig:depstrats}. Total latency is the summation of the processing times seen at the CAPIF endpoints of individual VNFs. To simplify the comparison, we do not include the latency introduced as a result of the communication between the static VNFs in Figure~\ref{fig:depstrats}. These are the 5G-AKA message exchanges taking place between the AUSF, UDM, and UDR. Since these will be the same across all the strategies, their contribution to the total latency is omitted. However, we still take into account the HTTP transactions between the AMF and the AUSF during 5G-AKA. This allows us to capture the resultant effect on the authentication process when the AMF is moved to the edge zones for the Static and Mobile MCS control plane strategies. %the 5G-AKA procedure is affected as a result of the AMF-AUSF communication delay. 

Omitting the 5G-AKA interactions between the HN VNFs, total latency in Figure~\ref{fig:totlat} is obtained by adding up the delay in all the remaining message exchanges. Each message has been categorized into one of three groups in Table~\ref{tbl:apicallscateg} depending on which operation they are tied to. The operations include: 5G-AKA messages taking place between the AMF and AUSF; session setup messages among the NRF, AMF, and SMF; NRF registration and update messages (detailed message flows can be found in Appendix~\ref{app:5gmsgflow}). %Three key observations are made based on the results in Figure~\ref{fig:totlat}.

\begin{table}[t]
\vspace{12pt}
\centering
\small\selectfont
\caption{Categorization of 5G core HTTP messages into groups based on operational significance} 
\label{tbl:apicallscateg}
  \begin{tabular}{p{0.22\columnwidth}|p{0.7\columnwidth}} 
    \textbf{Category} & \textbf{Interaction} \\ \hline
    \multirow{4}{*}{5G-AKA} & AMF --> AUSF - UE authentication \\ %\cline{2-2}
                        & AMF --> UE - mutual authentication \\ %\cline{2-2}
                        & AMF --> AUSF - auth. confirmation \\ \hline    
    \multirow{6}{*}{Session Setup} & AMF --> NRF - SMF discovery request \\ %\cline{2-2}
                        & AMF --> SMF - Context creation request \\ %\cline{2-2}
                        & SMF --> AMF - N1-N2 context creation \\ %\cline{2-2}
                        & AMF --> UE - Session resource setup \\ %\cline{2-2}
                        & AMF --> SMF - Context update \\ \hline
    \multirow{2}{*}{NRF Register} & AMF, SMF, UPF --> NRF - registration %and patch update to profiles 
                        \\ %\cline{2-2}
                        & NRF --> AMF, SMF, UPF - update \\
  \end{tabular}
\end{table}

%\textbf{A. Proximity between AZ and edge zones.} 
%We can see that the WZs located outside the United States have overall lower latency, which is expected based on the latency benchmarks conducted in Section~\ref{sec:latbench}. Especially, Tokyo, Seoul, and London WZs have relatively lower total operational latency than other edge locations. The obvious reason is due to the AZ and edge zone proximity. Specifically for Tokyo, Seoul, and London, the AZs are located in the same city as the WZs. This creates a significant advantage for these locations when deploying the 5G core in a hybrid public cloud. %For the other Europe (EU) WZ in Berlin, the AZ is located in Frankfurt. While the total delay is less than some of the other edge locations, it is still noticeably larger than those of Tokyo, Seoul, and London. %Nevertheless, the EU-based WZs will be able to host 5G MCS network slices with better reliability guarantees on top of the AWS public cloud.

\textbf{Comparing LZs and WZs in the same US cities.} To analyze the performance of the North America edge locations, it is important to consider the relative geographical distances between AZs and edge zones. In Tokyo, Seoul, and London, the AZ is located within the same city as the edge zone. For the majority of US cities, this is not the case. In terms of the performance of the Atlanta and Chicago WZs, which are tied to the Northern Virginia AZ, the total overhead is relatively higher as expected. Still, for MCSs, it is unnecessary to host the 5G core VNFs on WZs because the user plane latency to the RAN will not be a primary concern. Instead, operators can opt to use the LZs in the same cities to significantly reduce the latency of critical control plane operations. Especially in the US-East edge locations, choosing to use an LZ rather than a WZ for MCS network slices can achieve comparable performance to EU-based edge zones. The same pattern, however, cannot be observed in the US-West edge locations. For Seattle and Los Angeles, we observe that both LZ and WZ options yield similar results. This grants operators greater flexibility in deploying network slices using the public cloud.

\begin{figure*}[t]
\vspace{-0.1in}
    \centering
    \includegraphics[width=\textwidth,trim={0.5cm 7.5cm 0.5cm 0.2cm},clip]{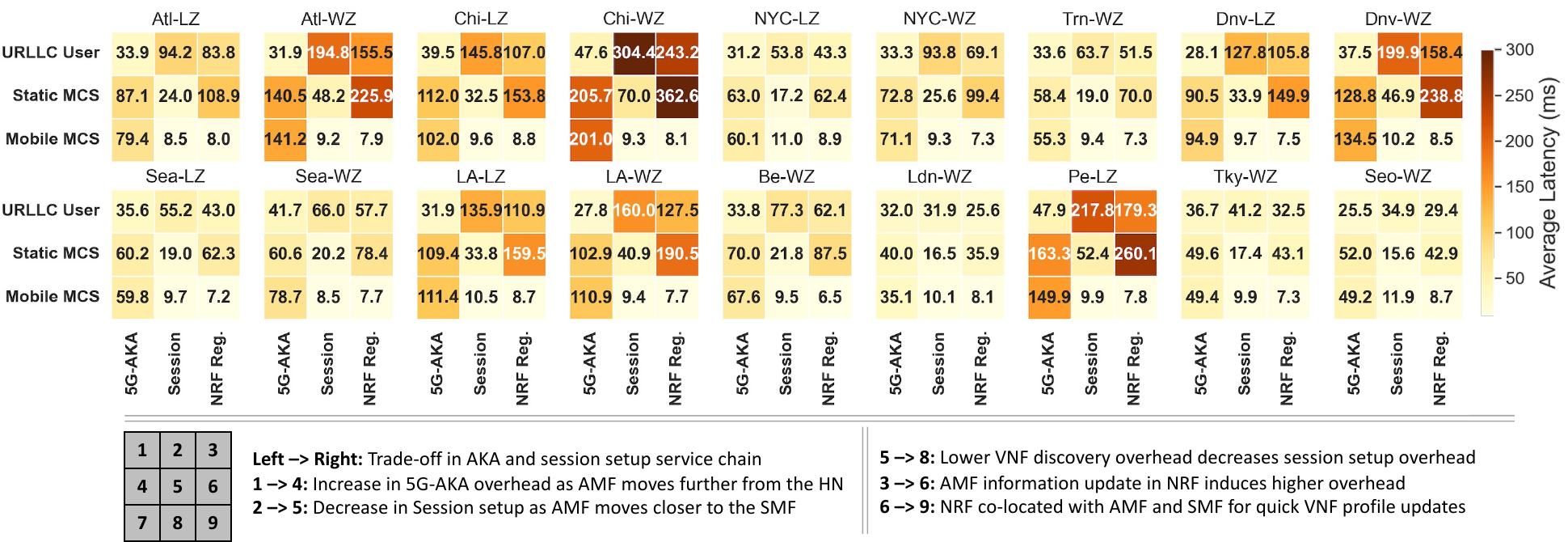}
     \caption{Operational overhead breakdown of 5G core interactions for different VNF placement strategies for all evaluated edge locations. Complete HTTP transaction measurements are given in Appendix~\ref{app:httptran}
     }
     \label{fig:opbreak}
\end{figure*}

\textbf{Deceptive Increase in Total Latency Moving from URLLC to Static MCS Strategy}. As the AMF is moved into an edge zone, Figure~\ref{fig:totlat} indicates that the total end-to-end session setup latency of a user slightly increases. This paints an incomplete picture because it does not specify the source of the latency for each strategy. While examining the total latency overhead provides a direct comparison across edge locations, it cannot reveal the source of the latency for each strategy. To better understand how different strategies in Figure~\ref{fig:depstrats} could impact operational latency, we present the operational breakdown in Figure~\ref{fig:opbreak}. These results dissect the total latency of each strategy for edge zones according to the categorization in Table~\ref{tbl:apicallscateg}. This makes it possible to demonstrate how specific strategies are able to lower the latency of certain operations while increasing it for others. There are three arguments to be drawn from Figure~\ref{fig:opbreak}. 

%\textbf{(IV-1) Increased delay in 5G AKA and its significance.} The 5G AKA service chain experiences an increase in delay when switching from the URLLC user plane slice to the MCS slice. When the AMF is moved to an edge zone, the SN - HN messages between the AMF and the AUSF are subjected to higher latency. However, this does not pose a significant issue for MCS users because 5G-AKA is a procedure that takes place during the initial registration and connection management procedure~\cite{3gpp33501}. While there could be other secondary authentication operations (e.g., network slice specific authentication and authorization~\cite{3gpp29526}), these can be conducted using secondary authentication servers independent from the HN VNFs. Therefore, any latency induced into 5G-AKA as a result of moving the AMF into the edge for the Static MCS network slice, will not impact the reliability of the control plane. Furthermore, there is no difference in 5G-AKA service chain moving from the Static to Mobile MCS strategy. Any slight difference observed in Figure~\ref{fig:opbreak} for 5G-AKA between the Static and Mobile MCS is the result of outlier processing overhead rather than connection latency. %Similarly the value of 

\textbf{(I) Increased delay in 5G AKA and reduced delay in session setup.} The 5G AKA service chain experiences an increase in delay when switching from the URLLC user plane slice to the MCS slice. When the AMF is moved to an edge zone, the SN - HN messages between the AMF and the AUSF are subjected to higher latency. However, this does not pose a significant issue for MCS users, because 5G-AKA is a procedure that takes place during the initial registration and connection management procedure~\cite{3gpp33501}. While there could be other secondary authentication operations (e.g., network slice specific authentication and authorization~\cite{3gpp29526}), these can be conducted using secondary authentication servers independent from the HN VNFs. On the other hand, moving the AMF next to the SMF into the edge zones also leads to a considerable drop in session setup latency. During network slice construction, both the AMF and SMF need to send a discovery request to the NRF to discover a suitable SMF and UPF~\cite{3gpp29510}, respectively. This leads to two HTTP transactions between the edge zone and the AZ for the Static MCS slice. However, for the users of this slice, such a discovery procedure only causes a one-time delay. A potential consumer of this type of slice can be a mIoT client. Such a client may require additional packet data unit (PDU) sessions on multiple slices to accommodate the high data volume. To that end, it can establish multiple PDU sessions on single or multiple network slices. The VNF discovery will only be prompted initially during the construction of these slices. Afterward, for an additional session setup, only the AMF-SMF interaction will take place, which has lower latency. Therefore, the static MCS use case can be well accommodated through the designated VNF placement as shown in Figure~\ref{fig:depstrats}.

%\textbf{(II) Reduced delay in session setup for static MCS slices.} Switching from a URLLC user slice to a Static MCS slice, a considerable drop in session setup latency is observed. As the AMF is moved into the edge next to the SMF, the only source of significant latency for the session setup is the VNF discovery procedure for both the AMF and SMF. During network slice construction, both the AMF and SMF need to send a discovery request to the NRF to discover a suitable SMF and UPF~\cite{3gpp29510}, respectively. This leads to two HTTP transactions between the edge zone and the AZ for the Static MCS slice. However, for the users of this slice, such a discovery procedure only causes a one-time delay. A potential consumer of this type of slice can be a mIoT client. Such a client may require additional packet data unit (PDU) sessions on multiple slices to accommodate the high data volume. To that end, it can establish multiple PDU sessions on single or multiple network slices. The VNF discovery will only be prompted initially during the construction of these slices. Afterwards, for an additional session setup, only the AMF-SMF interaction will take place, which has lower latency. Therefore, the static MCS use case can be well accommodated through the designated VNF placement shown in Figure~\ref{fig:depstrats}.

\textbf{(II) Convenience of the Mobile MCS network slice.} Moving the NRF into the edge zone lowers the total delay of both the session setup and registration. For clients with high mobility, this is an important advantage, as a network slice transfer can be required to properly accommodate them in case of mission criticality~\cite{orsino2017effects,petrov2018achieving}. When network slice transfer takes place, the components of the existing network slice need to communicate with the VNFs of the target slice, or a new slice needs to be constructed. Either way, the VNF discovery operation as well as fetching up-to-date metadata from the NRF becomes a time-sensitive operation. Given the sub-15ms session setup and sub-10ms NRF interaction overhead, these tasks can be achieved with reduced latency through the Mobile MCS strategy.

\textbf{(III) How to make a choice between LZ and WZ?} The objective of an MVNO is to deliver the optimal QoS while managing operational overhead. Given that computational costs are higher in edge zones, MVNOs will aim to minimize the number of VNFs they relocate to the edge (see Appendix~\ref{app:costtot}). The operational breakdown provides a thorough comparison between WZs and LZs for cities that have both. As expected, being hosted on TSP infrastructures may have lower latency to the RAN, but it has a negative impact on the backhaul. This is especially apparent in Atlanta, Chicago, and Denver, where the latency overhead of LZs is significantly lower than that of WZs. If operators seek to prioritize the control plane latency for MCSs, they should opt for LZs rather than WZs. For Seattle, Los Angeles, and New York City edge zones, the difference between WZs and LZs remains comparatively less significant. This provides operators with higher flexibility when deploying network slices at these locations.

%Up until now, we have analyzed the trends present across all the heatmaps in Figure~\ref{fig:opbreak}. 

\section{User Plane Results}

\subsection{AWS Throughput Benchmarks} \label{sec:thrbench}

\begin{figure}[t]
%\vspace{-0.1in}
    \centering
    \begin{subfigure}[t]{0.45\columnwidth}
        \centering
        \includegraphics[width=\textwidth,trim={10.5cm 6cm 10cm 3.2cm},clip]{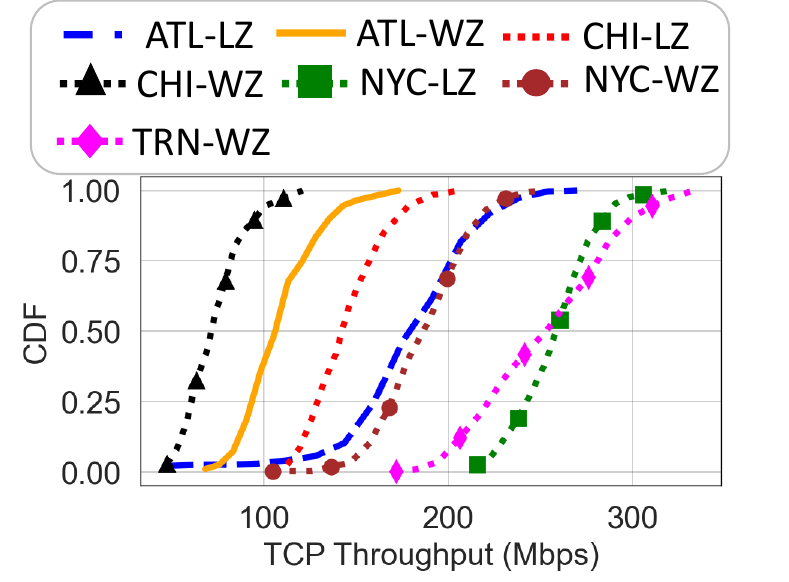} 
        \caption{North America - East}
        \label{fig:benchcdfthrnae}
    \end{subfigure}%
    \begin{subfigure}[t]{0.45\columnwidth}
        \centering
        \includegraphics[width=\textwidth,trim={10.5cm 6cm 10cm 3.2cm},clip]{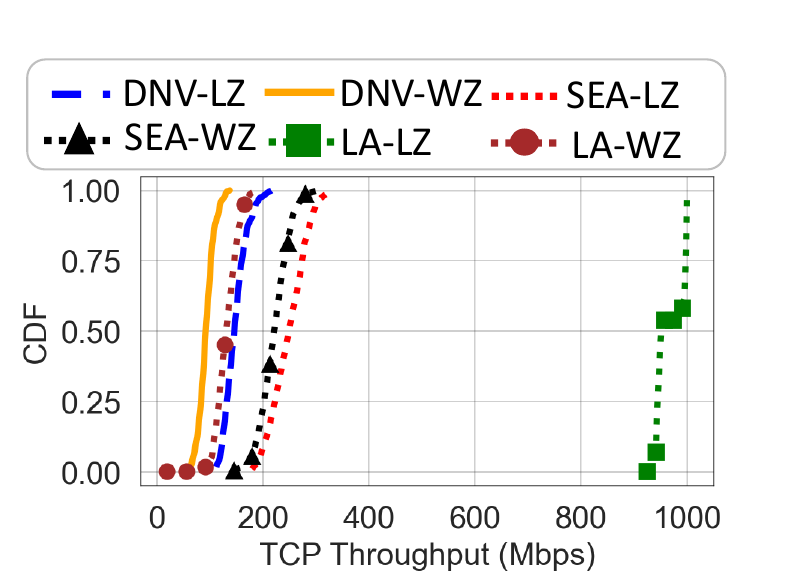}        
        \caption{North America - West}
        \label{fig:benchcdfthrnaw}
    \end{subfigure}
    \begin{subfigure}[t]{0.45\columnwidth}
        \centering
        \includegraphics[width=\textwidth,trim={10.5cm 6cm 10cm 3.2cm},clip]{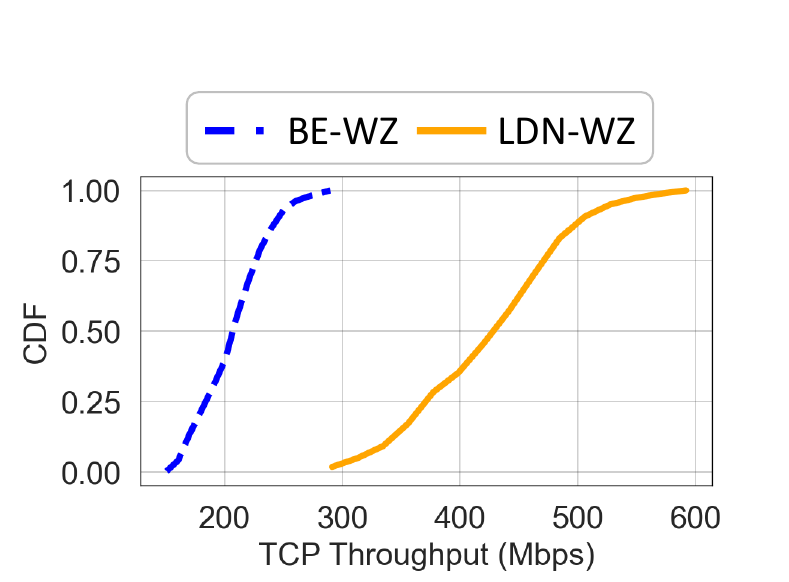}        
        \caption{Europe}
        \label{fig:benchcdfthrneu}
    \end{subfigure}
    \begin{subfigure}[t]{0.45\columnwidth}
        \centering
        \includegraphics[width=\textwidth,trim={10.5cm 6cm 10cm 3.2cm},clip]{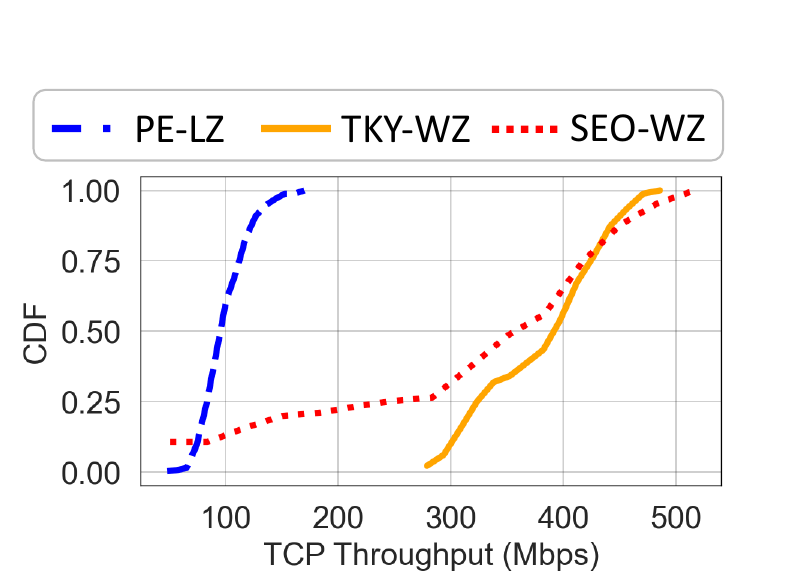}        
        \caption{Asia-Pacific}
        \label{fig:benchcdfthrap}
    \end{subfigure}
    \caption{CDF of weekly TCP throughput measurements between AZs and edge zones for April 24-30th 2023. (Raw measurements are given in Appendix~\ref{app:rawbench})}
    \label{fig:benchthrcdf}
\end{figure}

\begin{figure}[]
%\vspace{-0.1in}
    \centering
    \begin{subfigure}[t]{0.45\columnwidth}
        \centering
        \includegraphics[width=\textwidth,trim={10.5cm 6cm 10cm 3.2cm},clip]{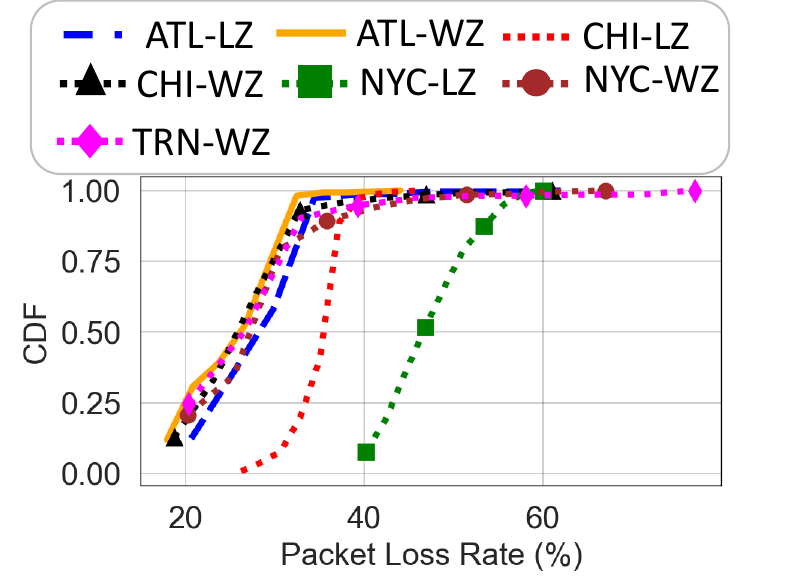} 
        \caption{North America - East}
        \label{fig:benchcdfplnae}
    \end{subfigure}%
    \begin{subfigure}[t]{0.45\columnwidth}
        \centering
        \includegraphics[width=\textwidth,trim={10.5cm 6cm 10cm 3.2cm},clip]{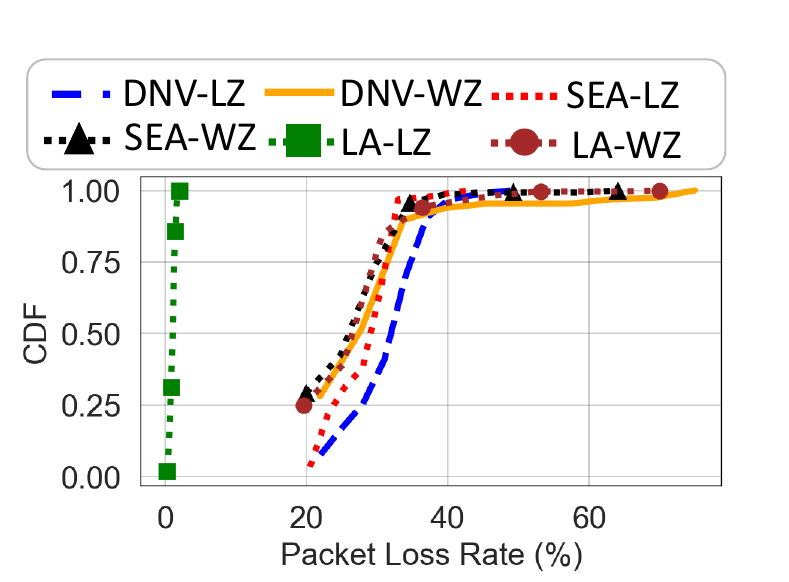}        
        \caption{North America - West}
        \label{fig:benchcdfplnaw}
    \end{subfigure}
    \begin{subfigure}[t]{0.45\columnwidth}
        \centering
        \includegraphics[width=\textwidth,trim={10.5cm 6cm 10cm 3.2cm},clip]{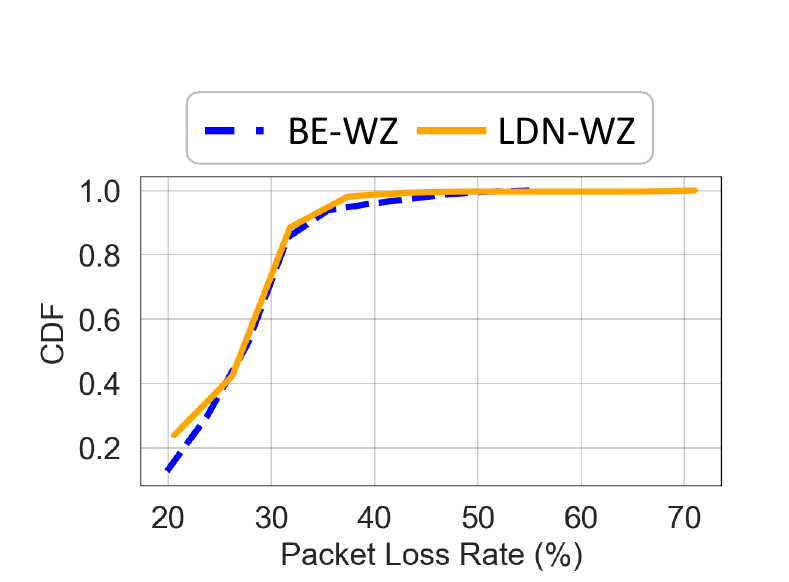}        
        \caption{Europe}
        \label{fig:benchcdfplneu}
    \end{subfigure}
    \begin{subfigure}[t]{0.45\columnwidth}
        \centering
        \includegraphics[width=\textwidth,trim={10.5cm 6cm 10cm 3.2cm},clip]{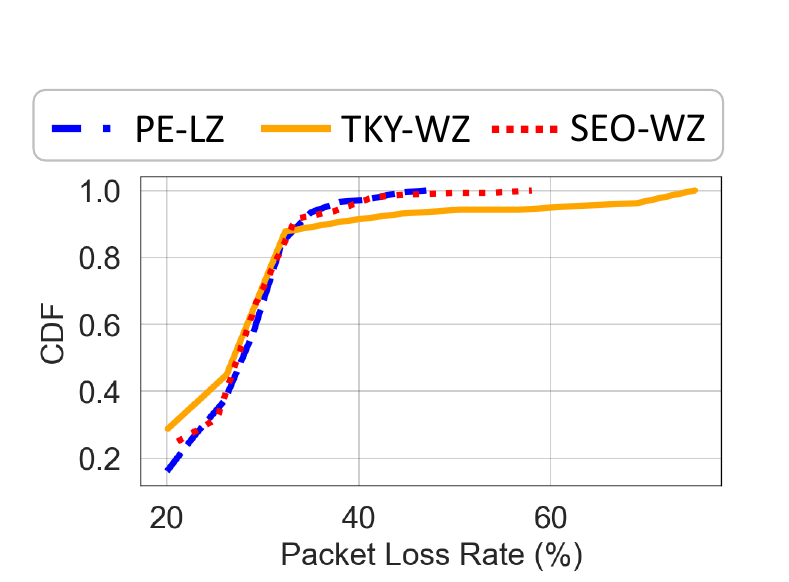}        
        \caption{Asia-Pacific}
        \label{fig:benchcdfplap}
    \end{subfigure}
    \caption{CDF of weekly UDP Packet Loss Rate measurements between AZs and edge zones for April 24-30th 2023. (Raw measurements are given in Appendix~\ref{app:rawbench})}
    \label{fig:benchplcdf}
\end{figure}

Benchmarks are conducted over a one-week period with 12-hour intervals. We use iperf3~\cite{iPerfDow98online} with a single stream to evaluate TCP throughput and measure packet loss over a 1~Gbps UDP connection. The iperf3 client runs for 100 samples with 20 iterations, which are then averaged for each zone. The results are shown in Figures~\ref{fig:benchthrcdf},~\ref{fig:benchplcdf}.

\begin{comment}
\begin{figure}[t]
    \centering
    \begin{subfigure}[t]{0.445\columnwidth}
        \centering
        \includegraphics[width=\textwidth,trim={6.3cm 1.2cm 6cm 1.2cm},clip]{figures/benchcdfthrNAe.pdf} 
        \caption{North America - East}
        \label{fig:benchcdfthrnae}
    \end{subfigure}%
    \begin{subfigure}[t]{0.445\columnwidth}
        \centering
        \includegraphics[width=\textwidth,trim={1.1cm 1cm 11.5cm 1.5cm},clip]{figures/benchcdfthrNAw.pdf}        
        \caption{North America - West}
        \label{fig:benchcdfthrnaw}
    \end{subfigure}
    \begin{subfigure}[t]{0.445\columnwidth}
        \centering
        \includegraphics[width=\textwidth,trim={5.1cm 3.3cm 6.8cm 3cm},clip]{figures/benchcdfthrEU.pdf}        
        \caption{Europe}
        \label{fig:benchcdfthrneu}
    \end{subfigure}
    \begin{subfigure}[t]{0.445\columnwidth}
        \centering
        \includegraphics[width=\textwidth,trim={5.5cm 3.3cm 6.4cm 3cm},clip]{figures/benchcdfthrAP.pdf}        
        \caption{Asia-Pacific}
        \label{fig:benchcdfthrap}
    \end{subfigure}
    \caption{CDF of weekly TCP throughput measurements between AZs and edge zones for April 24-30th 2023. (Raw measurements are given in Appendix~\ref{app:rawbench})}
    \label{fig:benchthrcdf}
\end{figure}
\end{comment}

\begin{figure}[]
\vspace{-0.3in}
    \centering
    \begin{subfigure}[t]{0.95\columnwidth}
        \centering
        \includegraphics[width=\columnwidth,trim={0cm 7cm 23cm 0cm},clip]{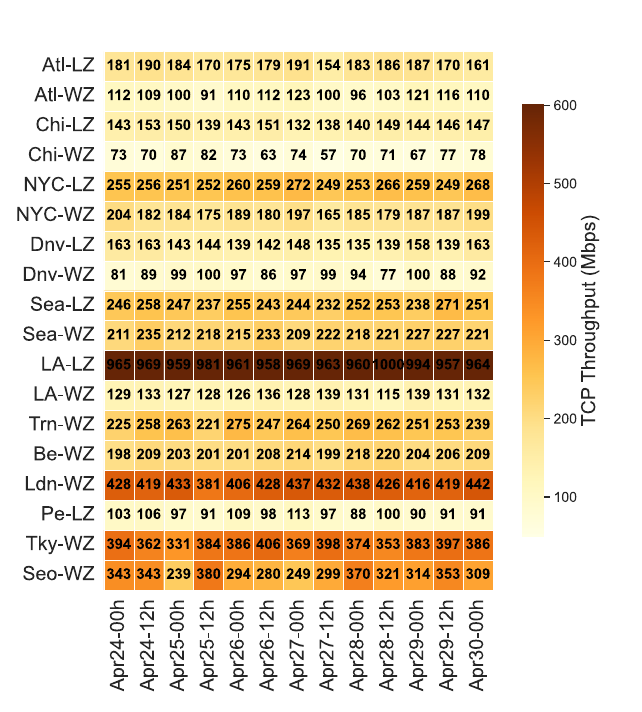} 
        \caption{TCP throughput}
        \label{fig:thrbench}
    \end{subfigure}%
    \hfill
    \vskip\baselineskip
    \begin{subfigure}[t]{0.95\columnwidth}
        \centering
        \includegraphics[width=\columnwidth,trim={0cm 7cm 23cm 0cm},clip]{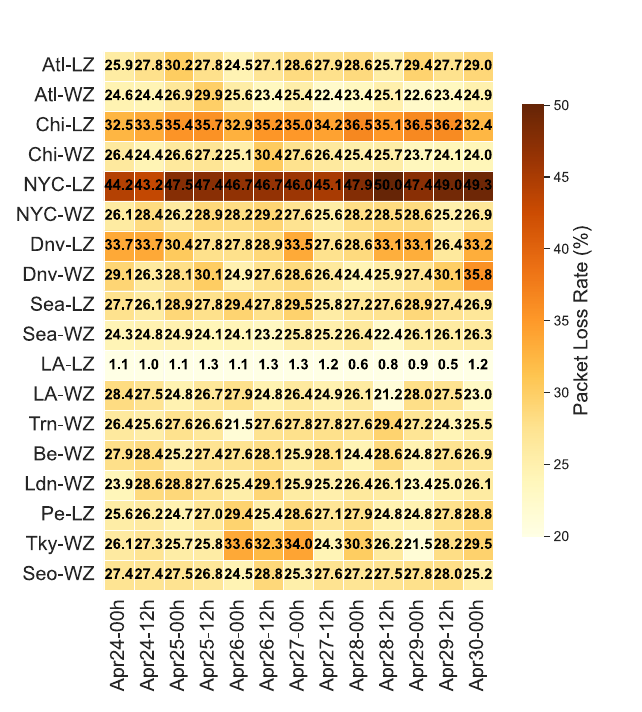}        
        \caption{UDP packet loss over 1 GBps bandwidth}
        \label{fig:plbench}
    \end{subfigure}
    \caption{Throughput measured over a week with 12-hour intervals between edge zones and AZs. Measurement recordings are started at the reported times according to Eastern Standard Time. (This figure represents the bi-daily averages. Raw throughput measurements are given in Appendix~\ref{app:rawbench}.)}
    \label{fig:qosbench}
\end{figure}

%The TCP benchmarks in 
%Figure~\ref{fig:thrbench} shows the consistency of TCP throughput over a weekly average across edge zones over time.
%With the exception of the Los Angeles LZ, 
%
%One very clear observation from Figure~\ref{fig:thrbench} is that the EU-based WZs have much higher throughput than the zones in North America, except the Los Angeles LZ. This significant difference between the Los Angeles LZ and other edge zones clearly indicates the superiority of the infrastructure between the US-West (Oregon) AZs and the Los Angeles LZ. Compared with the rest of the zones, the Seoul-WZ experiences the highest fluctuations.

The bi-daily averages are given in Figure~\ref{fig:qosbench}. While with sufficient measurements the daily averages indeed reflect consistent patterns both for TCP throughputs and UDP packet losses, the variations are more apparently observed from the Cumulative Distribution Functions (CDFs) in Figures~\ref{fig:benchthrcdf} and~\ref{fig:benchplcdf}. Since these fluctuations are non-negligible, our measurement approach for 5G user plane evaluation, as described in Figure~\ref{fig:measmetuser}, spans over an 8-hour time period for each use case. This allows us to account for the time-related fluctuations in both TCP and UDP connections and obtain a reliable average.

\textbf{WZ-LZ throughput-packet loss discrepancy.} Examining the results of the packet loss benchmarks shown in Figure~\ref{fig:plbench}, we notice a discrepancy in comparison with Figure~\ref{fig:thrbench}. The measurements in Atlanta, Chicago, and New York City have conflicting TCP throughput and UDP packet loss rate results. While throughput over TCP is higher in LZs than in WZs of these cities, packet loss rates at WZs are lower than those at LZs. TCP has built-in congestion control with retransmission to recover from packet losses and throttle the throughput. By contrast, UDP is a connectionless protocol without loss recovery mechanisms. %making it more susceptible to packet losses. %Nevertheless, it is expected that higher packet loss would ultimately lead to a more throttled TCP connection. 
Thus, this behavior is a clear irregularity since given the same queue, a higher packet loss rate should ultimately lead to a more throttled TCP throughput as a result of congestion control. 

We corresponded with AWS to understand the underlying infrastructure and explain this behavior. The higher TCP throughput in the connection between an AZ and an LZ is due to the fact that AWS uses higher bandwidth links between AZs and LZs compared to WZs. However, this does not necessarily result in lower latency or packet loss rate. On the other hand, WZs are specifically optimized for low-latency connections to the 5G access networks, which require high reliability. To achieve this, AWS uses specialized hardware and software in WZs to prioritize UDP traffic and reduce packet losses. Therefore, if reliable connections are critical for an application, connecting to a WZ is the better option. However, if users require high bandwidth and can tolerate packet loss, connecting to a LZ provides higher TCP throughput.

%Furthermore, the packet loss issue could be due to a variety of factors related to network architecture, physical distance, and the specific use case and workload being run. 
MVNOs should consider these factors when selecting an edge zone for a particular application or workload. Since the details of network paths or hardware are not public, it is not possible to make a definitive claim regarding this issue. Additional infrastructure customizations are itemized below that may lead to the resultant behaviors.

\begin{itemize}
    \item \textbf{Different network paths}. A variety of techniques, such as multiple redundant paths, load balancing, and intelligent routing are employed in AWS network paths. Depending on the network conditions and load on the infrastructure, different routes may be used for TCP and UDP packets to optimize network performance. 

    \item \textbf{QoS policies}. The network path to LZs prioritizes TCP traffic over UDP, while the opposite is true for WZs. This explains the higher TCP throughput in LZs, while also justifying a higher relative packet loss rate in LZs than WZs. 
    \item \textbf{Edge hardware}. The TSP hardware in the WZs handles UDP traffic more efficiently than AWS LZ hardware. For example, the hardware in the New York City LZ may have lower processing power, smaller buffers, or slower interfaces than the hardware in the New York City WZ. These limitations could result in dropped UDP packets, even if the traffic rate is below link capacity. 
\end{itemize}

\subsection{5G Measurement Results} \label{sec:thrplresults}

\begin{figure}[t]
\vspace{-0.1in}
    \centering
    \includegraphics[width=0.7\columnwidth,trim={8.8cm 4.6cm 12.3cm 5.8cm},clip]{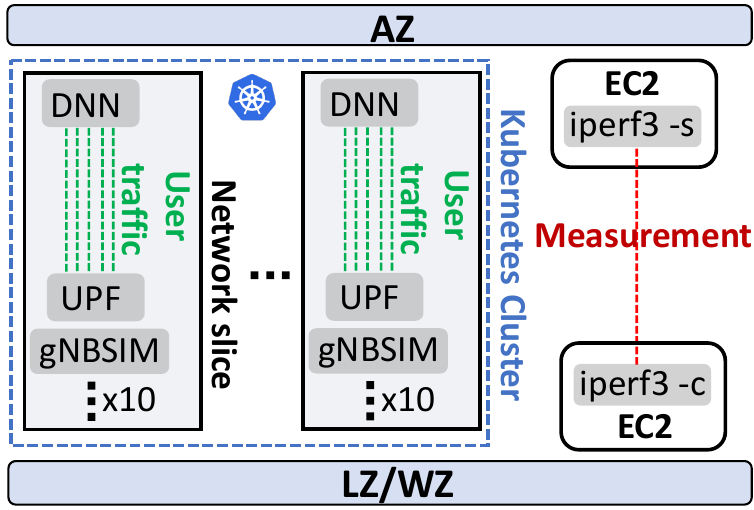}
     \caption{Measurement methodology for user plane experiments across edge zones and AZs}
     \label{fig:measmetuser}
     \vspace{-0.1in}
\end{figure} 

%The traffic is replayed in the 5G user plane using the setup described in Figure~\ref{fig:logdep}. 
For stress testing the AZ-LZ/WZ connection with 5G traffic, the UPF is placed at the network edge while the DNN container is located in the AZ. The UPF routes the user traffic from the DNN to the gNBSIM in the edge zone. Our measurement methodology is illustrated in Figure~\ref{fig:measmetuser}. We instantiate up to 8 network slices with each slice running 10 pairs of gNBSIM-DNN connections. Each pair represents a single-user session. This requires instantiating 10 pairs of gNBSIMs and DNNs per slice with a single UPF. We gather results for 40, 80 users, with 4, 8 network slices. 

\begin{figure*}[t]
    \centering
    %\vspace{-0.1in}
    %\includegraphics[width=\textwidth,trim={0cm 0cm 0cm 0cm},clip]{figures/thrAll_2_9.pdf}
    \includegraphics[width=\textwidth,trim={0cm 0.68cm 0cm 0.3cm},clip]{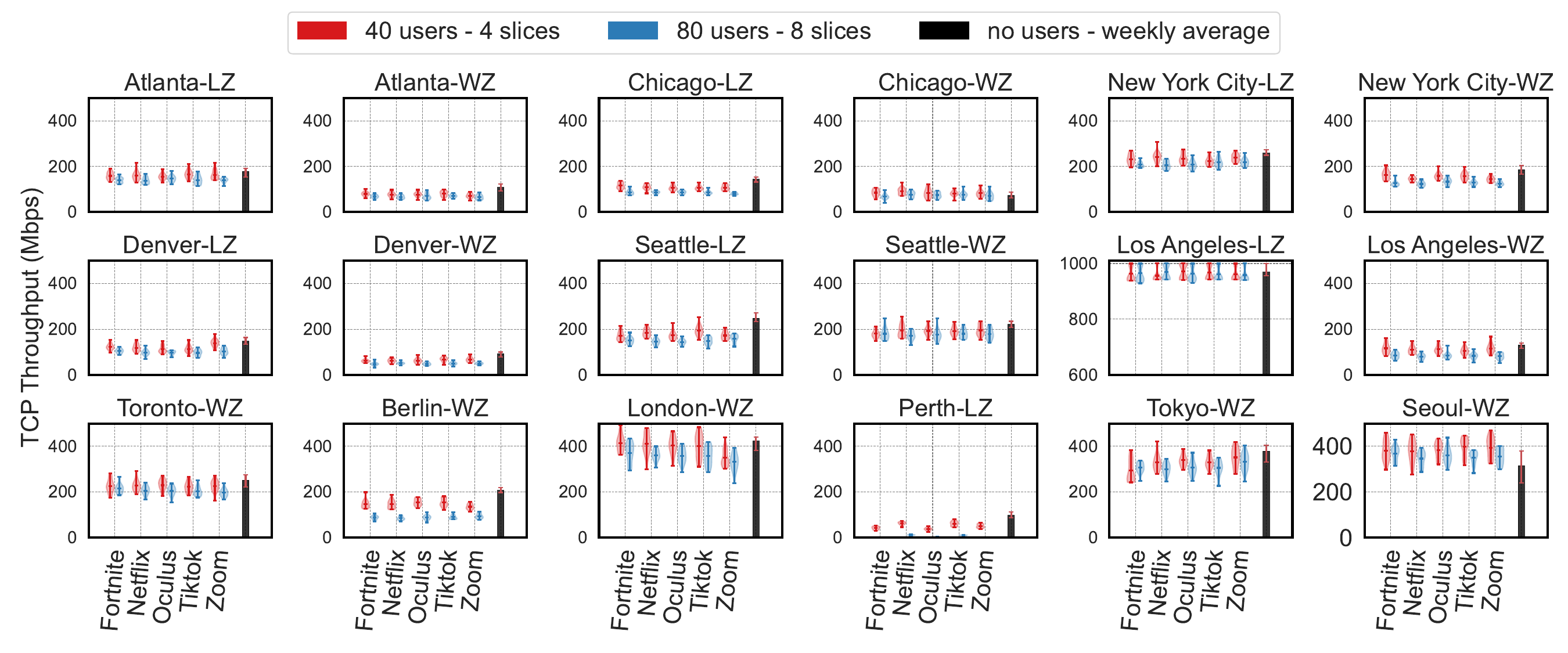}
     \caption{TCP Throughput measured between an edge zone and AZ host with 5G user plane traffic.}
     \label{fig:thrall}
     \vspace{-0.in}
\end{figure*}

\begin{figure*}[]
    \centering
    \vspace{-0.1in}
    \includegraphics[width=\textwidth,trim={0cm 0.68cm 0cm 0.3cm},clip]{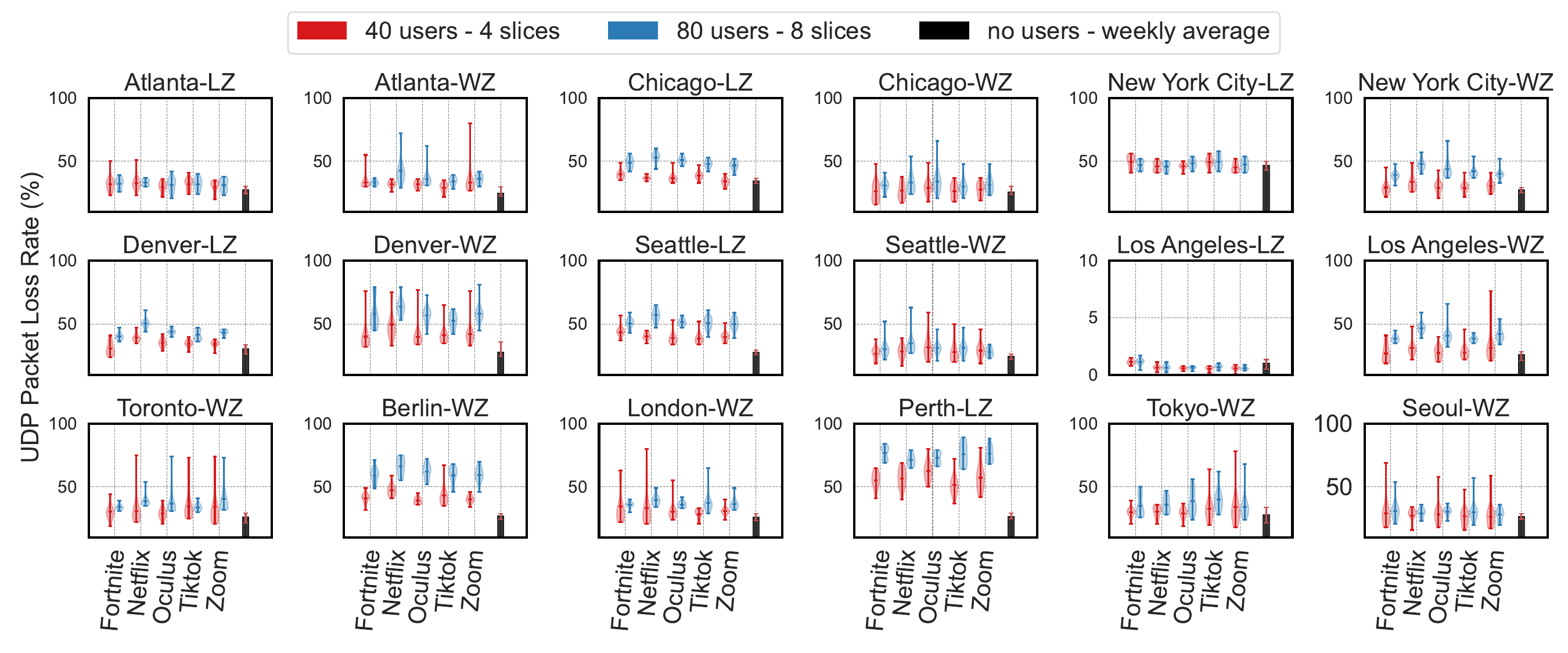}
     \caption{UDP Packet loss rate over a 1~Gbps connection between LZs/WZs and AZs with 5G user plane traffic.}
     \label{fig:plall}
     %\vspace{-0.in}
\end{figure*}

To understand how each traffic pattern affects the connection, we run the same experiment separately for each use case in Table~\ref{tbl:usecases}.
While scaling the user plane in this manner, we avoid overloading a single UPF. Thus, not limited by processing bottlenecks that exist in software, we can test the AZ-LZ/WZ link capacity. As a result, the focus is on measuring the deterioration of the 5G user plane connection due to the consumption of the AZ-LZ/WZ bandwidth.

To understand the impact of user plane traffic, a pair of reference EC2 instances are booted in each zone, independent from the Kubernetes cluster nodes. We start an iperf3 client between these reference nodes and run the client for 100 samples with 20 iterations. Then, we observe the behavior of the AZ-LZ/WZ connection. 

\vspace{-0.06in}
\subsubsection{TCP Throughput Measurements.}

The first set of results in Figure~\ref{fig:thrall} demonstrate how different traffic patterns affect the TCP throughput across zones. 
%The scale for each plot is the same except for the Los Angeles LZ, which has been adjusted to capture the relevant interval. 
For the edge zones in US-East and US-West regions, the deterioration experienced by LZs and WZs is consistent across all the use cases. Compared with the weekly average bandwidth measurements, the instantiation of the first four network slices has a negligible impact on the edge zone connection for the LZs. Going from four to eight slices, however, the decrease in throughput becomes more apparent for the WZ connections.
For the majority of the EU and Asia-Pacific regions, the throughput fluctuations are very high. Especially in London, Tokyo, and Seoul, it is harder to observe a specific pattern because the native throughput capacity is higher. 
%Furthermore, for Seoul, we observe that the weekly average is in fact lower than measurements gathered while traffic patterns were active. This indicates that there were times during the weekly average when the native activity in the region was higher compared to when experiments were conducted. 

To understand how each traffic pattern affects the connection, we analyze the results for cities where the deterioration is most noticeable. These are primarily the US zones and the Berlin WZ. The lowest TCP throughputs are reported when the Netflix and Zoom traffic patterns are active, followed by the Oculus VR traffic. For URLLC slices, such as Oculus and Fortnite traffic, the UPF is better hosted on WZs. However, hosting application servers in edge zones is more expensive, due to the scarcity of computing resources. Thus, for chosen zones that do not suffer significant latency penalties (e.g., Seoul, Tokyo, and London), it is acceptable for the application servers of these use cases to be placed in the AZ. Furthermore, for use cases such as Netflix and Tiktok that are not ultra-latency sensitive, the servers can similarly be placed in AZs. It is crucial that both service providers and MVNOs are aware of the traffic load on AZ-LZ/WZ connections when committing to a placement model.

\vspace{-0.1in}
\subsubsection{UDP Packet Loss Rate Measurements.}

In certain use cases, TCP is favored for its reliable data delivery, while in other scenarios, user data is transmitted over UDP. 
Through our traffic analysis, we can see that the data in Zoom and VR gaming use UDP. The results in Figure~\ref{fig:plall} illustrate how UDP packet loss rate over a 1 Gbps connection changes as more 5G users are instantiated. Currently, only the Los Angeles LZ has packet loss rate below 2\% over a 1Gbps connection. To keep the comparison fair, we use the same network bandwidth while evaluating other regions.

In contrast to the TCP measurements, there are more outliers in the UDP evaluations. Nevertheless, we still observe noticeable patterns. 
%
%We saw in Figure~\ref{fig:thrall} that the TCP throughput deterioration caused by Zoom users was on average higher compared to VR. However, in Figure~\ref{fig:thrall}, VR load has a more noticeable negative impact on packet loss rate compared to Zoom.
%
In the US region LZs, the difference between the weekly benchmark results and the case with four active slices is lower for LZs than WZs. This is most apparent by checking the results of New York City and Denver. %This indicates that while base packet loss in WZs is lower, the deterioration of the packet loss in WZs is higher compared to LZs with increasing number of users. 
Even though the average packet loss rate remains lower in WZs than LZs, the deterioration in the WZs is higher than LZs with an increasing number of data sessions. 

In our benchmarks (Section~\ref{sec:thrbench}), we mentioned the correspondence with AWS to confirm that WZs (with the notable exception of the Los Angeles LZ) are specialized for handling UDP traffic. When stress testing with traffic, we observe that for Denver and Atlanta, the WZs in fact do not outperform the LZs in the same city. This indicates that, while WZs are designed for handling UDP traffic, the AZ-WZ connection traversing AWS-TSP infrastructure is not yet fully stable when subjected to high traffic loads. Thus, for selective edge regions, the user plane can be shared between the LZs and WZs for better performance.

The takeaway from the measurements in Figure~\ref{fig:plall} is the consistent increase in packet loss rate for the US regions compared to the relative stability of London, Tokyo, and Seoul. This goes to show that the bottlenecks are more apparent in the US due to the physical distance between AZs and edge zones. With the ever-expanding 5G deployments, this will eventually create a backhaul bottleneck that needs to be addressed if the AWS public cloud is to become a valid candidate for hosting the 5G user plane in US regions. 
Overall, our measurements shown in Figures~\ref{fig:thrall} and~\ref{fig:plall} provide valuable insights into the network performance of multiple edge locations when subjected to different traffic loads.

%\section{Evaluation: Cost Analysis}
%\input{eval3.tex}

%\section{Deductions on AWS Infrastructure}
%\input{geog.tex}

\section{Discussion}
This section discusses the limitations of our study as well as the recommendations for improving the AWS infrastructure. 

%\subsection{AWS vs GCP vs Azure}
%In this study, our experimentation has been exclusively conducted on top of AWS.  
\vspace{-0.06in}
\subsection{Inter-WZ Handover} 

\begin{figure}[t]
    \centering
    \includegraphics[width=0.85\columnwidth,trim={8.5cm 7.6cm 15cm 5.1cm},clip]{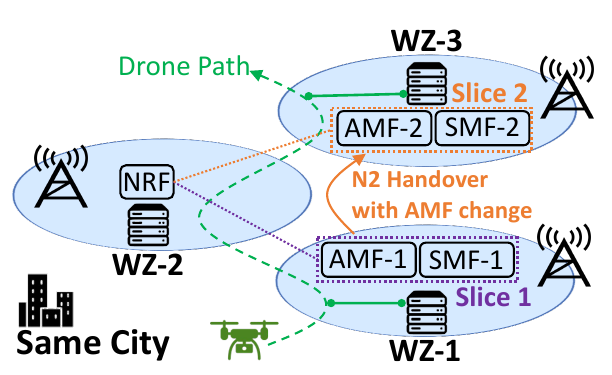}
     \caption{N2 handover scenario with AMF change illustrated in the case of multiple WZs existing in a single city}
     \label{fig:handovervisual}
     \vspace{-0.03in}
\end{figure}

Our analysis regarding Mobile MCS clients assumes that the VNFs of the current and target slice are situated in the same edge zone. This is necessary for evaluation purposes, but ultimately it is a counter-intuitive assumption. 
%For a mobile UE, it is expected that the device is moving away from the micro-datacenter hosting the VNFs of its current slice. 
A case study is depicted in Figure~\ref{fig:handovervisual} that presents a scenario with multiple WZs and a drone traversing through their coverage jurisdiction. In such a scenario, the VNFs of the target slice are located in WZ-3 along the movement path of the drone. Therefore, the N2 handover process with AMF change takes place between slice 1 in WZ-1 and slice 2 in WZ-3, where the VNFs of both slices are registered with the NRF in WZ-2.

The existing infrastructure provisions, such as the distribution and density of edge zones, are insufficient to assess this scenario because the required setups are not in place. In the future, every Central Unit (CU) of a 5G gNB in a heterogeneous network might be paired with COTS hardware~\cite{taleb2017multi}, which can host the VNFs of Mobile MCS slices. Under this assumption, what becomes significant is the delay between these individual locations. While we currently cannot understand the implications of this setup, our results in Figure~\ref{fig:opbreak} serve to demonstrate the feasibility of using WZs and LZs with AZs to host Mobile MCS slices by moving the NRF and AMF into the edge. A discussion point is provided in Section~\ref{sec:interwzlz} that further elaborates on how establishing inter-LZ-WZ connectivity can improve QoS for hybrid use cases.

\subsection{LZ-WZ Inter-Connectivity} \label{sec:interwzlz}
We have confirmed through experimental attempts and correspondence with the AWS technical team that currently no direct connectivity exists between LZs and WZs. Thus, for any deployment, only one of the edge zone variants can be used at the same time. 
AWS should consider establishing this connectivity to unlock different deployment options for operators. For instance, a hybrid deployment can be constructed with edge, distributed, and central cloud domains, represented by WZ, LZ, and AZ, respectively. In such an ecosystem, both VNFs and application servers can be distributed to inter-connected WZs and LZs, depending on operational and user-plane requirements. Ultimately, the deployments will have higher flexibility in addressing different use cases.

\begin{figure*}[]
    \centering
    %\vspace{-0.1in}
    %\includegraphics[width=\textwidth,trim={0cm 0cm 0cm 0cm},clip]{figures/plAll_2_9.pdf}
    \includegraphics[width=0.85\textwidth,trim={2cm 7.5cm 5.6cm 1.7cm},clip]{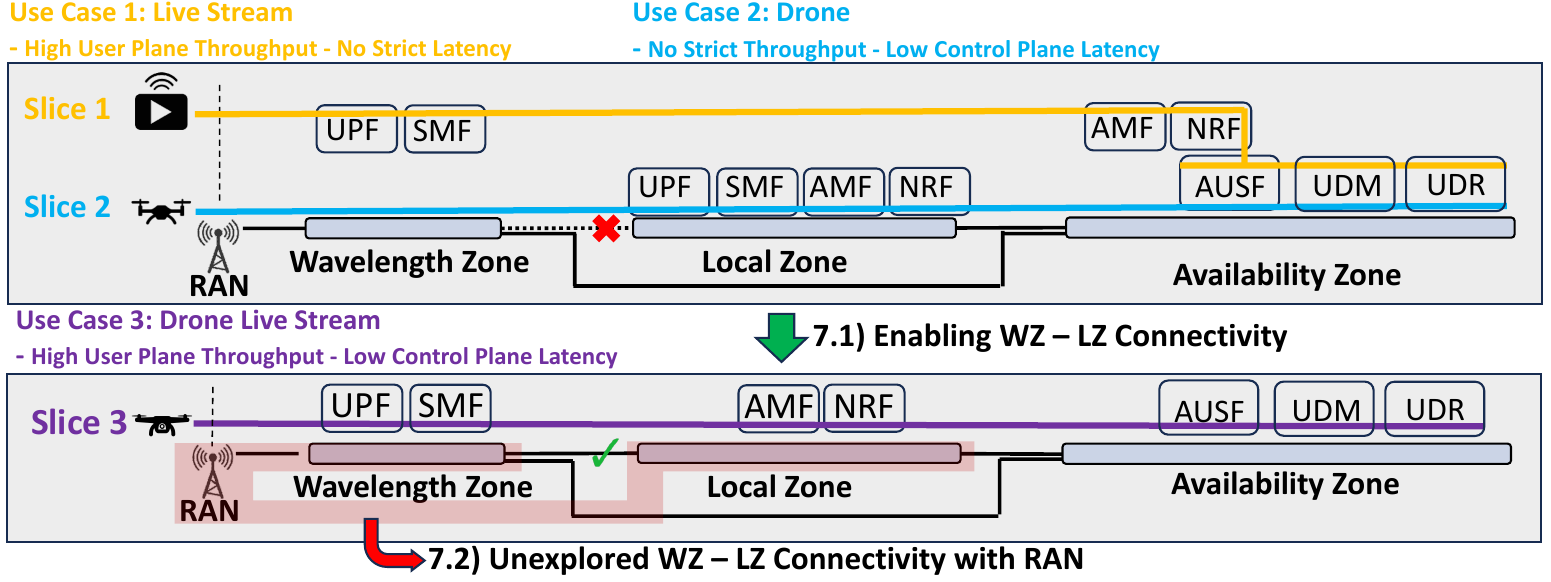}
     \caption{Summarizing the discussions and limitations. \textbf{Suggestion:} Enabling Interconnectivity between WZ and LZ. \textbf{Limitation:} Unexplored measurement domain between RANs and edge zones for respective operators in each country.}
     \label{fig:discsummary}
\end{figure*}

A specific case study is illustrated in Figure~\ref{fig:discsummary}, where for a live stream and drone communications, standalone WZ and LZ connectivity is sufficient. However, for enhancing the support to a hybrid drone live stream use case, establishing connectivity between WZs and LZs will promote high user plane throughput by placing UPF and SMF within WZs while also maintaining low control plane latency with AMF and NRF placed in LZs. Ultimately, the user plane (i.e., UPF) remains in close proximity to the 5G RAN and the edge control plane VNFs (i.e., AMF and NRF) have lower latency to the 5G core HN (i.e., AUSF, UDM, and UDR) than the case where they would have been placed in WZs.

\vspace{-0.04in}
\subsection{WZ-LZ Connectivity to the RAN}
To fully compare WZs and LZs, their connections to the available RANs need to be evaluated. While this is beyond the scope of our study, it is desirable to assess how WZs are superior to LZs for latency-sensitive user plane traffic.  
Without understanding the end-to-end latency between a real-life user and the application server, it will be challenging to fully unveil the limitations of 5G deployments over AWS. %Nevertheless, understanding how different traffic patterns affect the AZ-LZ/WZ connection is an important first step in deploying mass 5G connections that generate terabytes of data transfer across regions. 

\section{Related Work}
%The deployment of 5G SA networks is an ongoing process in many countries. %To the best of our knowledge, there exist no widely accessible 5G core deployments that reside in the public cloud. 
%Nevertheless, private testbeds do exist at either enterprise or smaller levels that enable large scale testing with various 5G and beyond features. In this section, we first summarize the relevant academic studies. Next we provide an overview of the existing work in industry for realizing 5G core deployments in the public cloud.
%Both the academia and industry have investigated deployment with various 5G and beyond features. 

%In this section, we first summarize the relevant academic studies. Next an overview is provided of the existing work in industry for cloud-based 5G core deployments.

%\subsection{Research Studies}

In~\cite{atalay2022network}, the authors proposed a network slice-as-a- service~\cite{gsm2021official} delivery framework for different use cases. They evaluated network slice deployments in a local testbed and calculated cost projections for deployment in a cloud based on the computing consumption of 5G components. %However, their evaluation takes place over a single physical node with multiple VMs, isolated from the internet. %Hence it is not a realistic evaluation.
A similar scaling study is conducted in~\cite{atalay2022scaling}, where slices are deployed in different topologies pertaining to different use cases. The computing resource consumption of individual VNFs is stress-tested in response to the number of UEs being serviced. 
However, both evaluations take place over a single physical node with multiple VMs, isolated from the Internet. %Both works suffer from being conducted in isolated environment compared to over a public cloud network. %isolated from the network characteristics of a public cloud environment.

%Another work~\cite{ungureanu2021collaborative} explores the utilization of large scale Management and Orchestration (MANO) units for virtualized 5G core eployments in the cloud. Authors propose a novel orchestration framework for edge 5G VNFs where they compare the orchestration overhead of their proposed solution with existing frameworks. However, there is no evaluation regarding the performance of the actual 5G core control or user plane both in regards to QoS and compute consumption.

%Other research~\cite{slamnik2022realistic,slamnik2022building,esmaeily2020cloud} has focused on creating proof-of-concepts for 5G deployments by carrying out network slicing in isolated environments. However, similar to the works mentioned before, they only carry out testing in lab environments, therefore deteriorating the fidelity of the deployment. 
%

A commercial measurement study is conducted in~\cite{narayanan2020lumos5g}, where a machine learning model is proposed to predict the performance of millimeter-wave (mmWave) 5G deployments. However, at the time of the study, 5G deployments were reliant on the LTE core. While the study provides insight into the workings of a mmWave commercial 5G deployment, its focus is on RAN measurements rather than core.

With the focus on the 5G core, ~\cite{ahmad2022enabling} seeks to improve the control plane reliability. An edge-based 5G core deployment design is presented with fault tolerance. While the study identifies key issues related to mobility management, the experiments take place over co-located hardware, isolated from the Internet, and thus they are not exposed to real-life conditions.

A recent study focuses on high mobility use cases amongst 5G and LTE~\cite{pan2022first}. An in-depth analysis provides coverage of how high mobility affects various 5G procedures. While the authors provide very important findings on how 5G deals with mobility, they do not dig into the intricacies of cloud-based 5G core networks. Rather, their study is focused on measuring the performance of existing 5G networks instead of analyzing what cloud-based 5G networks could be like.

Other research studies have delved into the analysis of more specific user ecosystems. In~\cite{lutu2020things}, the roaming of IoT devices connected to mobile networks is investigated. A data-driven approach for troubleshooting the user-plane performance degradation is presented in~\cite{shi2022towards}. Surveys have been conducted over the past years for qualitatively investigating the challenges of deploying 5G networks in clouds~\cite{esmaeily2021small,gupta2015survey}. However, to the best of our knowledge, this paper is the first to present a global analysis of 5G core control and user plane bottlenecks using real public cloud deployments. 

Last but not least, Microsoft has launched the Azure Private 5G core~\cite{AzurePri91online}, their advent towards enabling enterprises to build and operate private 5G networks. Azure provides their own set of 5G core VNFs to be used by MVNOs for creating on-premises or cloud-based deployments. For businesses without in-house 5G experts, this solution provides easy access to a 5G stack.

\begin{comment}
\vspace{-0.05in}
\subsection{Industry Efforts}

AWS and Azure have both made moves as mobile network deployments started transitioning into an NFV environment. AWS has introduced WZs~\cite{AWSWavel60online}, which have played a central role in our work, to facilitate the delivery of low-latency 5G applications. %by allowing them to be hosted on the TSP infrastructure. 
This has encouraged developers to start testing low-latency requirement applications on EC2 instances 
%capable of supporting 
with a large array of AWS features.

Independently, Microsoft has launched the Azure Private 5G core~\cite{AzurePri91online}, which has been their advent towards enabling enterprises to build and operate private 5G networks. Azure provides their own set of 5G core VNFs to be used by MVNOs for creating on-premises or cloud-based deployments. For businesses that do not have in-house 5G experts, 
%capable of deploying an end-to-end mobile network, 
this solution provides easy access to a full-fledged 5G stack.

%Last but not least, Swisscom and Ericsson have announced partnerships with AWS~\cite{Swisscom30online} to pursue the deployment of the 5G core in the cloud. While the project is still at the proof-of-concept stage, this enterprise displays commitment from operators in exploring various core deployment options. 

\end{comment}

%\vspace{-0.14in}

\section{Conclusion}
The 5G network deployments have been expanding at a rapid pace. With their extensive computing infrastructure, AWS has partnered with chosen operators to build an integrated cloud environment for 5G. To explore this ecosystem, we built a large scale 5G testbed spanning multiple edge LZs and WZs across different AWS regions. In our campaign, we have demystified the operational control plane latency implications of hosting different 5G core VNFs in alternative edge locations. Furthermore, we instantiated 5G user plane traffic loads across zones to stress test the AZ-LZ/WZ connections. This helped us identify bottlenecks for US-based regions while showing the superiority of the connection for the chosen EU and Asia-Pacific zones. Leveraging our findings, operators around the world can glance into the existing limitations of using edge zones for specific use cases.

%\newpage

\bibliographystyle{plain}
\bibliography{reference}

\appendix
%\section{Statement of Ethics}
%\input{ethics.tex}

\section{Message Interception Flow} \label{app:msginterflow}

\begin{figure}[t]
    \centering
    \includegraphics[width=0.9\columnwidth,trim={8.8cm 10.3cm 11.5cm 6.5cm},clip]{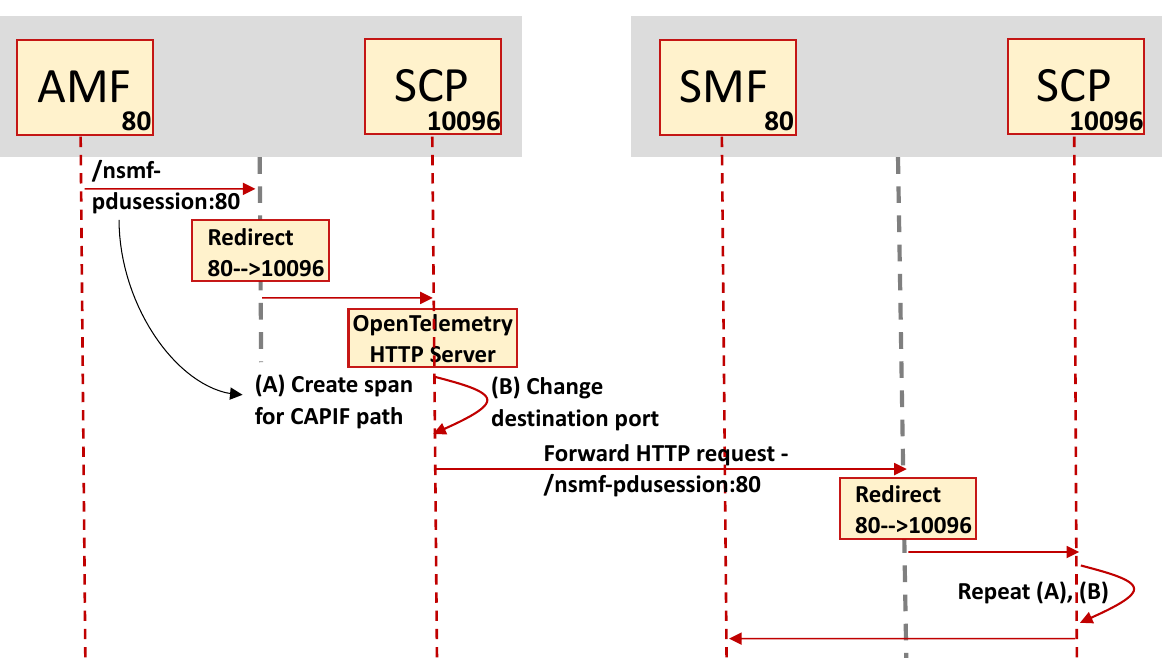}
     \caption{Sample API call from AMF to SMF illustrating monitoring SCP interception}
     \label{fig:framemsg}
\end{figure}

To clarify the message interception process in our monitoring pipeline, Figure~\ref{fig:framemsg} depicts a sample message flow for a single HTTP transaction taking place between AMF and SMF. The gray outline corresponds to individual pod deployments within the Kubernetes cluster. The IPTABLES of all the pods have been modified to perform port forwarding on all messages originating from the user identifier (UID) of the 5G core VNFs. Therefore, every outgoing and incoming message to the VNFs is redirected through the monitoring SCP.

In this example, AMF is making an API call to SMF with the CAPIF path $/nsmf- pdusession$ on port 80. The pod network sandbox redirects this HTTP request to the localhost on port 10096, where the monitoring SCP is running an OpenTelemetry HTTP server. Upon receiving the request, SCP extracts the CAPIF path from the URL and creates a span corresponding to the specific source and destination VNF (i.e., AMF and SMF, respectively). Finally, it changes the destination port on the outgoing message and forwards it to the pod of the target VNF. The same process is executed within the SMF pod upon the arrival of the HTTP message. Combining this interception and redirection with the instrumentation pipeline in Figure~\ref{fig:frameov} allows us to perform distributed tracing on the HTTP transactions in the 5G core. Ultimately, we are able to monitor the end-to-end execution time of individual operations and group them together in concise logs.

\section{AZ - Edge Zone Measurements} \label{app:azbench}

For experimentation, we have multiple AZs in each parent region to choose from. This is mainly for redundancy purposes, where applications with high-reliability requirements can be deployed in multiple AZs. However, we are primarily interested in the connection characteristics between AZs and a given edge zone. 

\begin{table}[t]
\vspace{12pt}
\centering
\small\selectfont
\caption{Average latency between edge zones and all AZs in a given parent region. In some regions, the AZ designation skips a letter (e.g., Tokyo - a, c, d)} 
\label{tbl:latbench}
  \begin{tabular}{p{0.063\columnwidth}|p{0.077\columnwidth}|p{0.08\columnwidth}|p{0.08\columnwidth}|p{0.08\columnwidth}|p{0.08\columnwidth}|p{0.08\columnwidth}|p{0.08\columnwidth}}
    \multirow{2}{*}{\textbf{City}} &
    \multirow{2}{*}{\textbf{Zone}} & \multicolumn{6}{|c}{\textbf{Avg. AZ Latency (ms)}} \\\cline{3-8}
    & & a & b & c & d & e & f \\
    \hline \hline
    \multirow{2}{*}{NYC} & LZ & 8.36 & 8.23 & 7.94 & 8.07 & 8.60 & 7.98 \\ %\cline{2-8}
                        & WZ & 14.9 & 14.1 & 14.2 & 14.6 & 14.8 & 14.6 \\ \hline                         
                        
    \multirow{2}{*}{CHI} & LZ & 24.6 & 23.2 & 23.5 & 23.8 & 23.0 & 24.3 \\ %\cline{2-8}
                        & WZ & 58.6 & 59.0 & 58.7 & 58.6 & 59.5 & 58.4 \\ \hline

    \multirow{2}{*}{ATL} & LZ & 15.8 & 16.2 & 16.1 & 15.4 & 16.5 & 15.8 \\ %\cline{2-8}
                        & WZ & 34.1 & 33.7 & 33.5 & 33.8 & 34.2 & 47.5 \\ \hline
    %\multirow{1}{*}{BA} & LZ & 140.0 & 139.7 & 140.1 & 140.1 & 142.1 & \textbf{139.3} \\ \hline
    %\multirow{1}{*}{LI} & LZ & 103.3 & 103.1 & 103.1 & \textbf{102.5} & 103.8 & 102.7 \\ \hline
    \multirow{2}{*}{SEA} & LZ & 8.9 & 8.1 & 8.5 & 8.2 & - & - \\ %\cline{2-8}
                        & WZ & 9.3 & 9.6 & 9.4 & 11.8 & - & - \\ \hline
    \multirow{2}{*}{LA} & LZ & 25.4 & 26.3 & 26.5 & 25.4 & - & - \\ %\cline{2-8}
                        & WZ & 26.5 & 27.8 & 28.0 & 26.2 & - & - \\ \hline
    \multirow{2}{*}{DNV} & LZ & 23.0 & 23.1 & 22.3 & 22.1 & - & - \\ %\cline{2-8}
                        & WZ & 36.0 & 36.8 & 35.6 & 37.5 & - & - \\ \hline
   \multirow{1}{*}{BE} & \multirow{1}{*}{WZ} & \multirow{1}{*}{12.0} & \multirow{1}{*}{12.4} & \multirow{1}{*}{11.2} & \multirow{1}{*}{-} & \multirow{1}{*}{-} & \multirow{1}{*}{-} \\
    \hline
   \multirow{1}{*}{LDN} & \multirow{1}{*}{WZ} & \multirow{1}{*}{4.6} & \multirow{1}{*}{4.3} & \multirow{1}{*}{4.9} & \multirow{1}{*}{-} & \multirow{1}{*}{-} & \multirow{1}{*}{-} \\
    \hline
    \multirow{1}{*}{TRN} & \multirow{1}{*}{WZ} & \multirow{1}{*}{9.7} & \multirow{1}{*}{10.2} & \multirow{1}{*}{-} & \multirow{1}{*}{10.8} & \multirow{1}{*}{-} & \multirow{1}{*}{-} \\
    \hline
    \multirow{1}{*}{SEO} & \multirow{1}{*}{WZ} & \multirow{1}{*}{5.3} & \multirow{1}{*}{4.4} & \multirow{1}{*}{5.9} & \multirow{1}{*}{5.3} & \multirow{1}{*}{-} & \multirow{1}{*}{-} \\
    \hline
    \multirow{1}{*}{TKY} & \multirow{1}{*}{WZ} & \multirow{1}{*}{5.3} & \multirow{1}{*}{-} & \multirow{1}{*}{6.8} & \multirow{1}{*}{6.3} & \multirow{1}{*}{-} & \multirow{1}{*}{-} \\
    \hline
    \multirow{1}{*}{PE} & \multirow{1}{*}{LZ} & \multirow{1}{*}{41.5} & \multirow{1}{*}{42.3} & \multirow{1}{*}{42.3} & \multirow{1}{*}{-} & \multirow{1}{*}{-} & \multirow{1}{*}{-} \\
  \end{tabular}
\end{table}

Since their exact location has not been disclosed by AWS, there is no way to select the optimal AZ for a given edge zone based on geographical proximity. However, according to AWS architecture guidelines~\cite{Architec10online}, all these AZs are within 100 kilometer distance from each other. To fully understand the difference between opting for one AZ or another in the same region, we benchmark the latency between all AZs and edge zones in Figure~\ref{tbl:latbench}. These benchmarks show that the difference in latency to different AZs from a given edge zone is less than 1ms. Since such a latency is negligible, we use the first AZ (i.e., AZ-a) in each region for experimentation. %Using this initial assessment, we determine the optimal AZ to serve as the central core network for a given edge zone. After determining the best AZ pairings it becomes possible to compare how each city can accommodate a 5G deployment with respect to others under optimal conditions. The chosen AZs are highlighted bold in Table.~\ref{tbl:latbench}.

\section{OAI 5G Core Message Flow} \label{app:5gmsgflow}
\begin{figure*}[h]
  \centering
  \includegraphics[width=\textwidth,trim={6.2cm 8.5cm 8cm 5.3cm},clip]{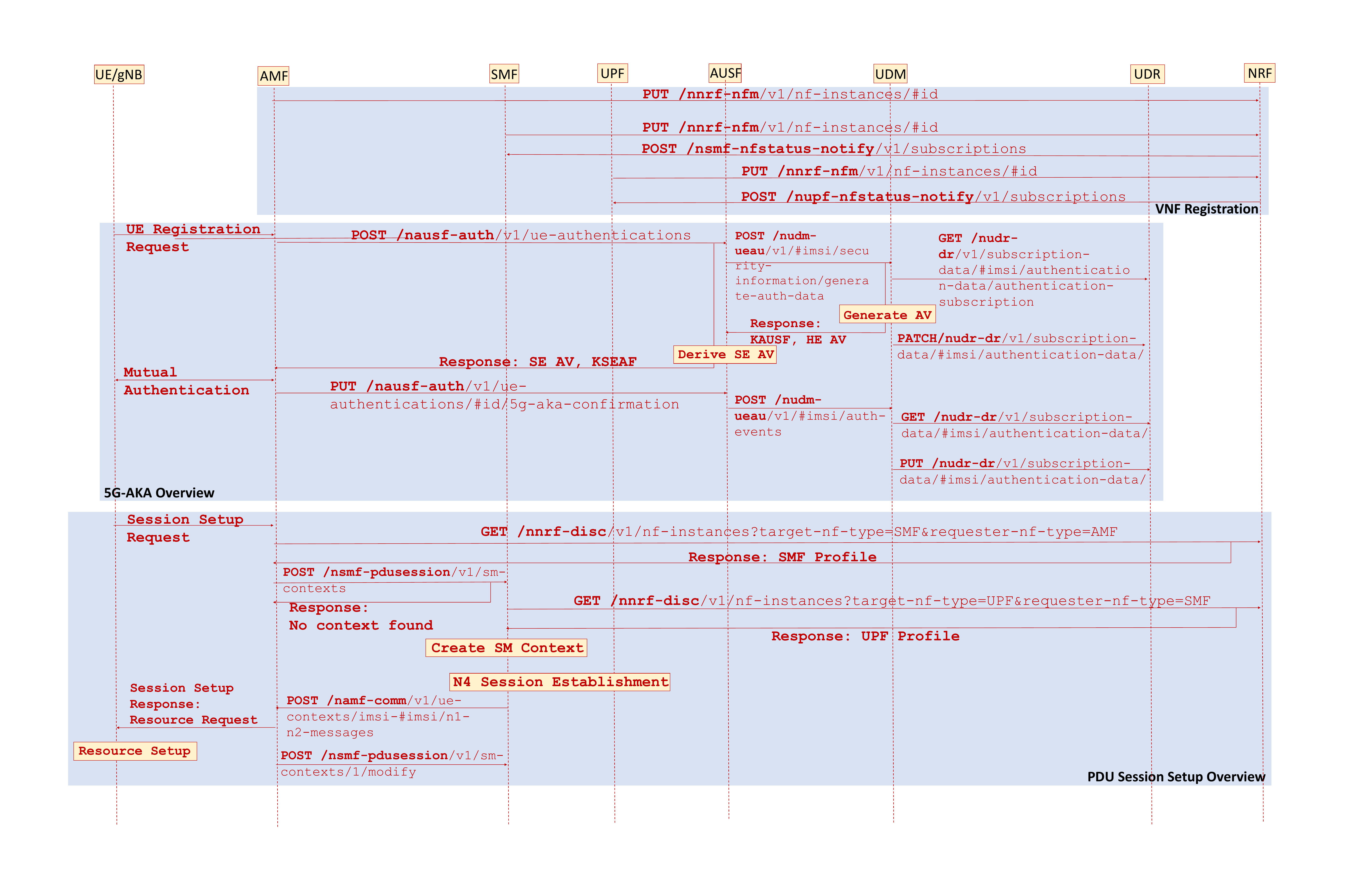}
  \caption{5G core VNF message exchange for end-to-end session setup}
  \label{fig:oaimsg}
\end{figure*}

The message flow in Figure~\ref{fig:oaimsg} has been divided into three sub-blocks. These are the VNF registration block, 5G-AKA overview, and the Packet Data Unit (PDU) session setup. 

In the VNF registration, the AMF, SMF, and the UPF send their metadata profiles to the NRF and receive a confirmation after the latter registers them in a local database. Afterwards, heartbeat updates are sent from the AMF, SMF, and UPF to the NRF to report any metadata alterations during run-time. In case of a modification, the NRF will respond with a status notification to each VNF. 

In the 5G-AKA service chain, the AMF is responsible for handling the SN credentials, while the AUSF, UDM, and UDR are parts of the HN service chain. We have abstracted out some of the extreme details of this process and 
%the interested reader is directed to
more detailed information can be found at~\cite{3gpp33501}.

Finally, in the PDU session setup, AMF, SMF, and UPF establish a user plane data connection for the UE. In this process,  AMF and SMF need to respectively discover an SMF and UPF for network slice construction~\cite{3gpp29510}. The discovery is facilitated by the NRF. After discovery, AMF and SMF set up a data connection through a series of HTTP transactions as shown in the \textit{PDU Session Setup Overview} block in Figure~\ref{fig:oaimsg}.

\section{User Traffic Patterns} \label{app:traffics}
To capture the traffic patterns for the user plane experimentation, we start by connecting our devices that generate traffic to a  Linksys WRT3200ACM programmable OpenWRT router. When a device connected to this router generates traffic, we capture the traffic pattern to be replayed later during the user plane experiments. For the Zoom video call and Netflix streaming use case, we use an IPhone 13 Pro. For the virtual reality case, we use an Oculus Quest 2.% The first 60-100 seconds of the downlink and uplink traffic patterns are presented in Figure~\ref{fig:5gtrafficpatterns}. %The traffic is replayed in the 5G user plane using the setup described in Figure~\ref{fig:logdep}.

\begin{figure*}[h]
    \centering
    \begin{subfigure}[b]{0.19\textwidth}
        \centering
        \includegraphics[width=1\textwidth,trim={7.5cm 0cm 7.3cm 0cm},clip]{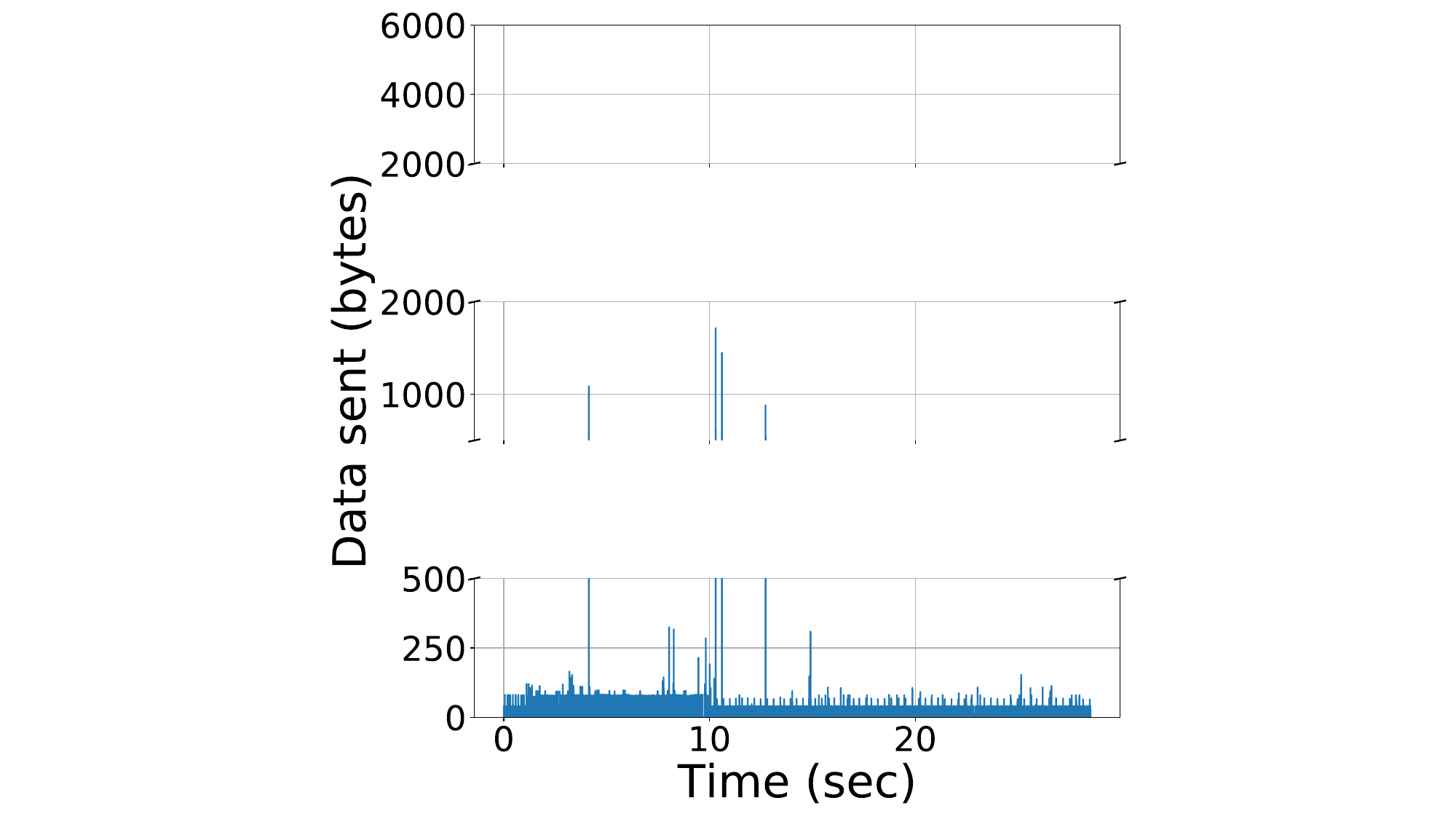}
        \label{fig:ps5gamein}
        \centering 
        \includegraphics[width=1\textwidth,trim={7.5cm 0cm 7.3cm 0cm},clip]{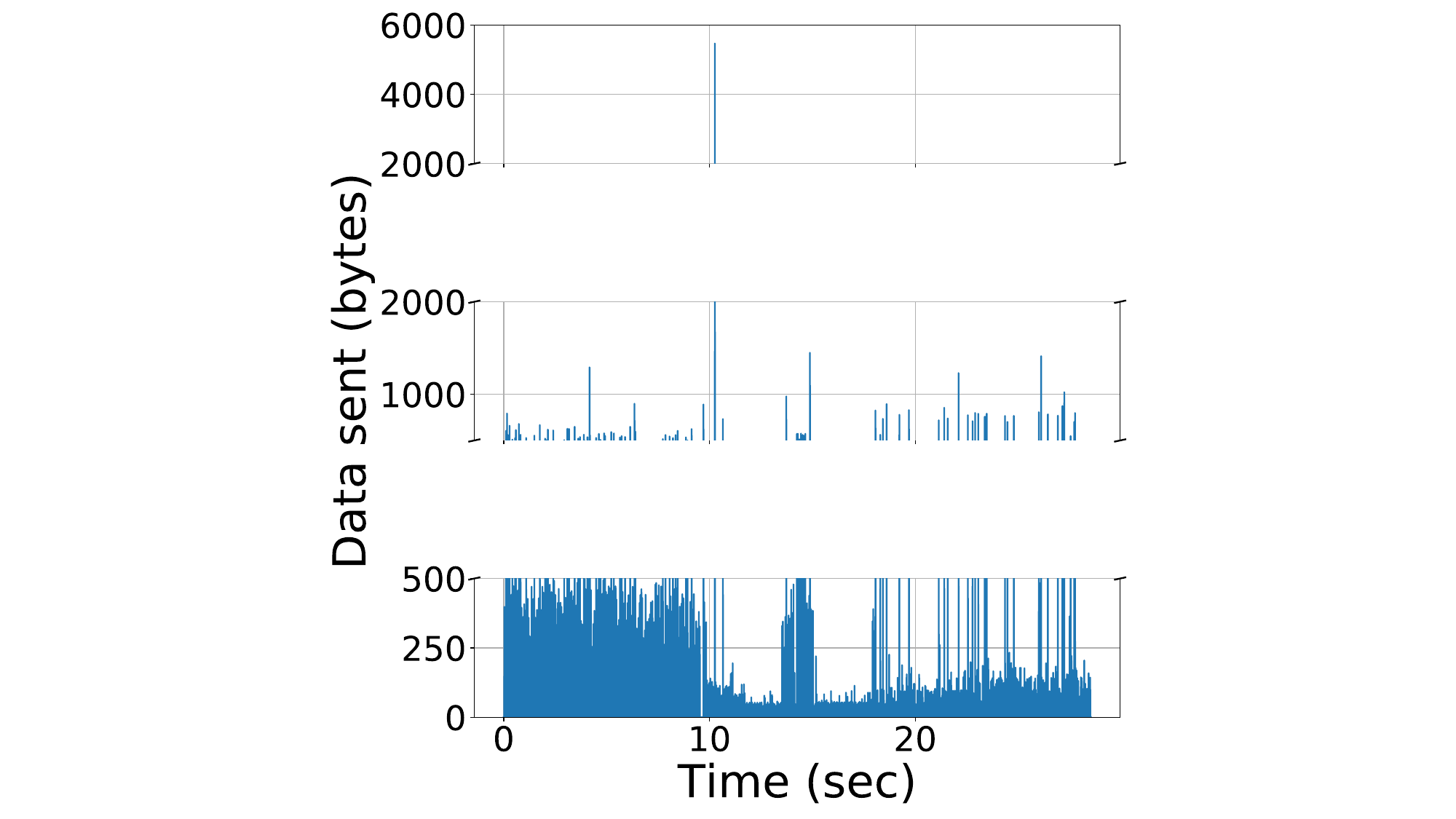}
        \label{fig:ps5gameout}
        \caption{ Fortnite}
    \end{subfigure}
    %\vskip\baselineskip
    \begin{subfigure}[b]{0.19\textwidth}   
        \centering 
        \includegraphics[width=1\textwidth,trim={7.5cm 0cm 7.3cm 0cm},clip]{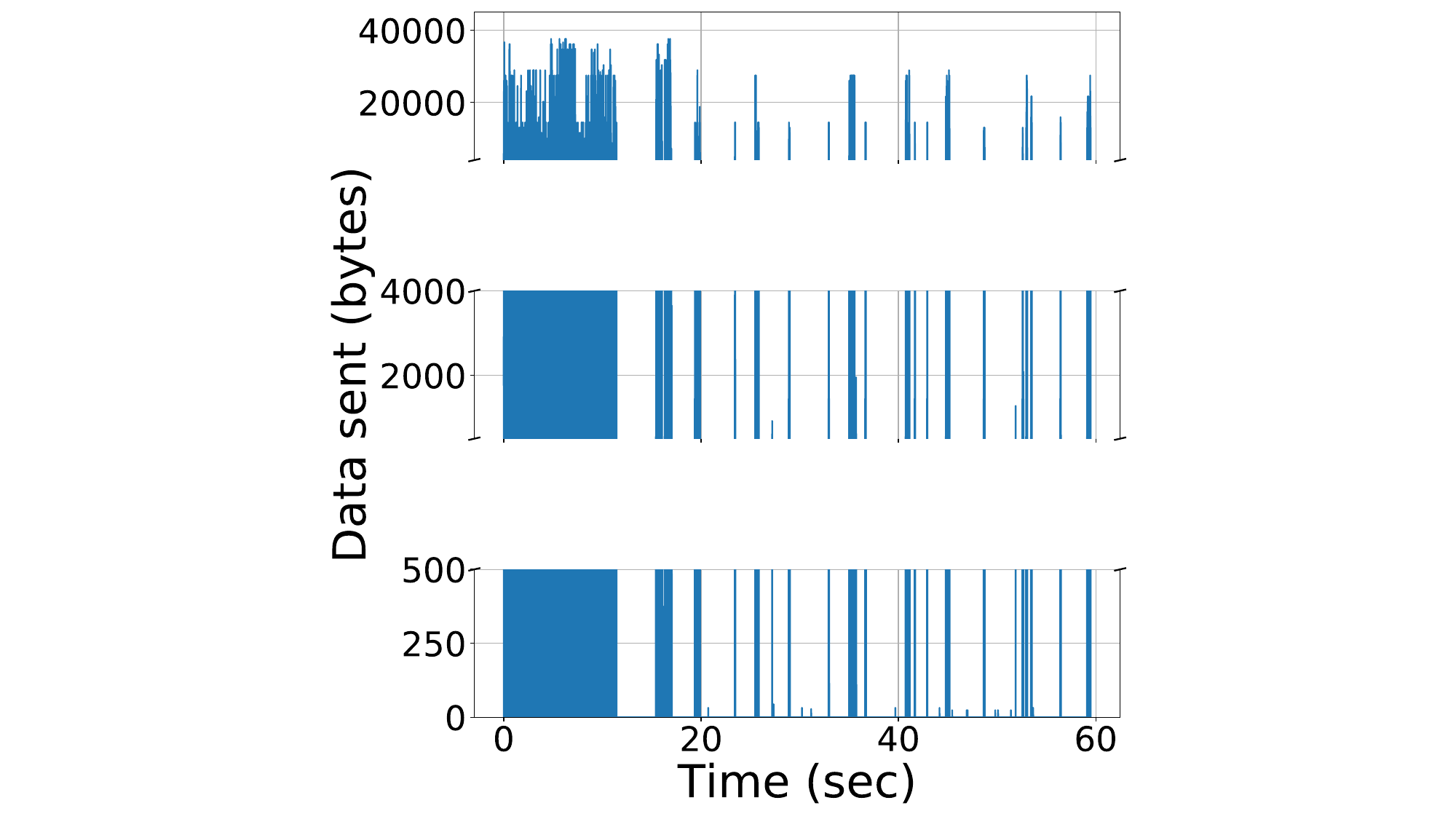}  
        \label{fig:ps5netout}  
        \centering 
        \includegraphics[width=1\textwidth,trim={7.5cm 0cm 7.3cm 0cm},clip]{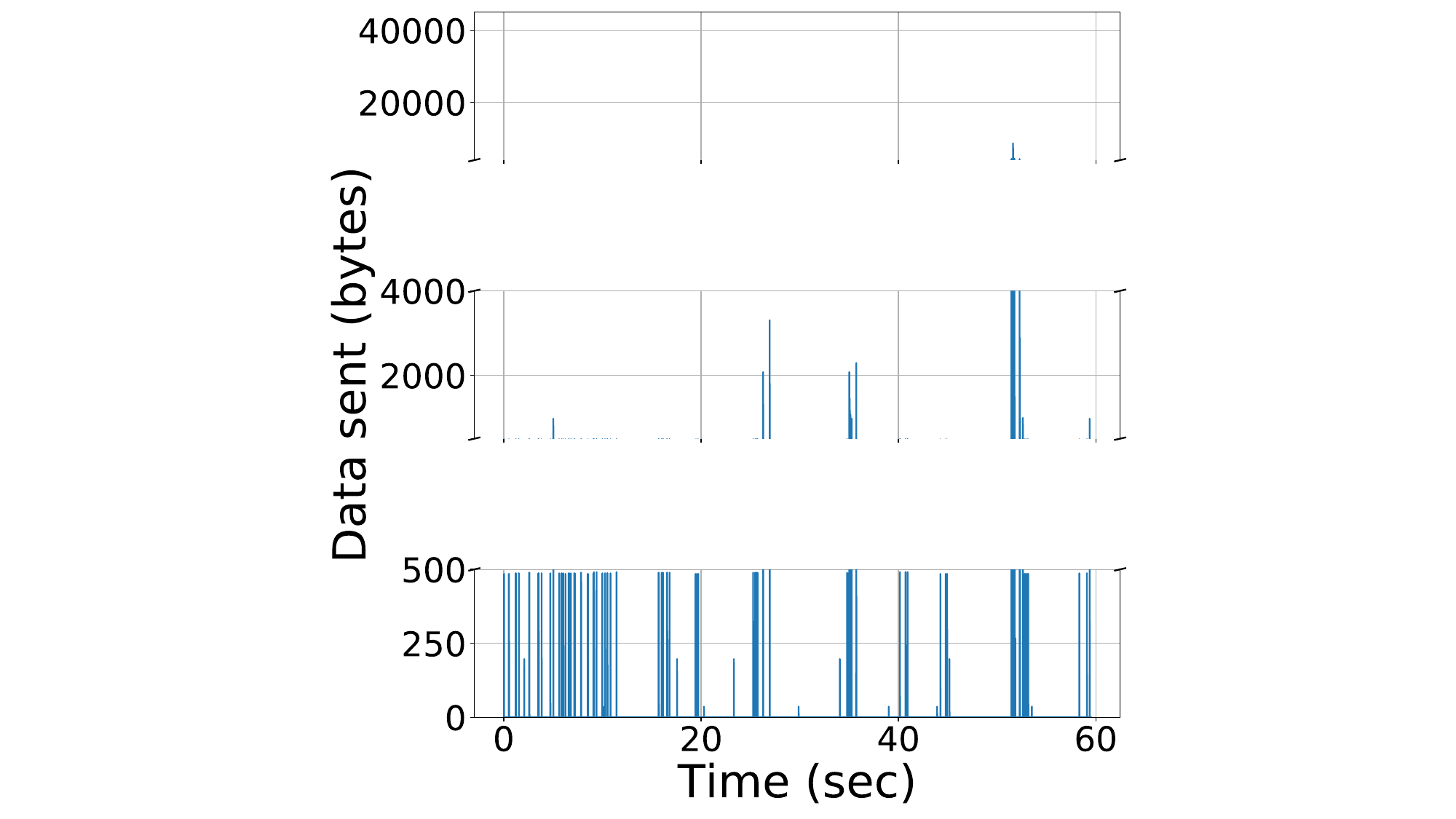}  
        \label{fig:ps5netin}
        \caption{ Netflix}
    \end{subfigure}
    %\vskip\baselineskip
    \begin{subfigure}[b]{0.19\textwidth}   
        \centering 
        \includegraphics[width=1\textwidth,trim={7.5cm 0cm 7.3cm 0cm},clip]{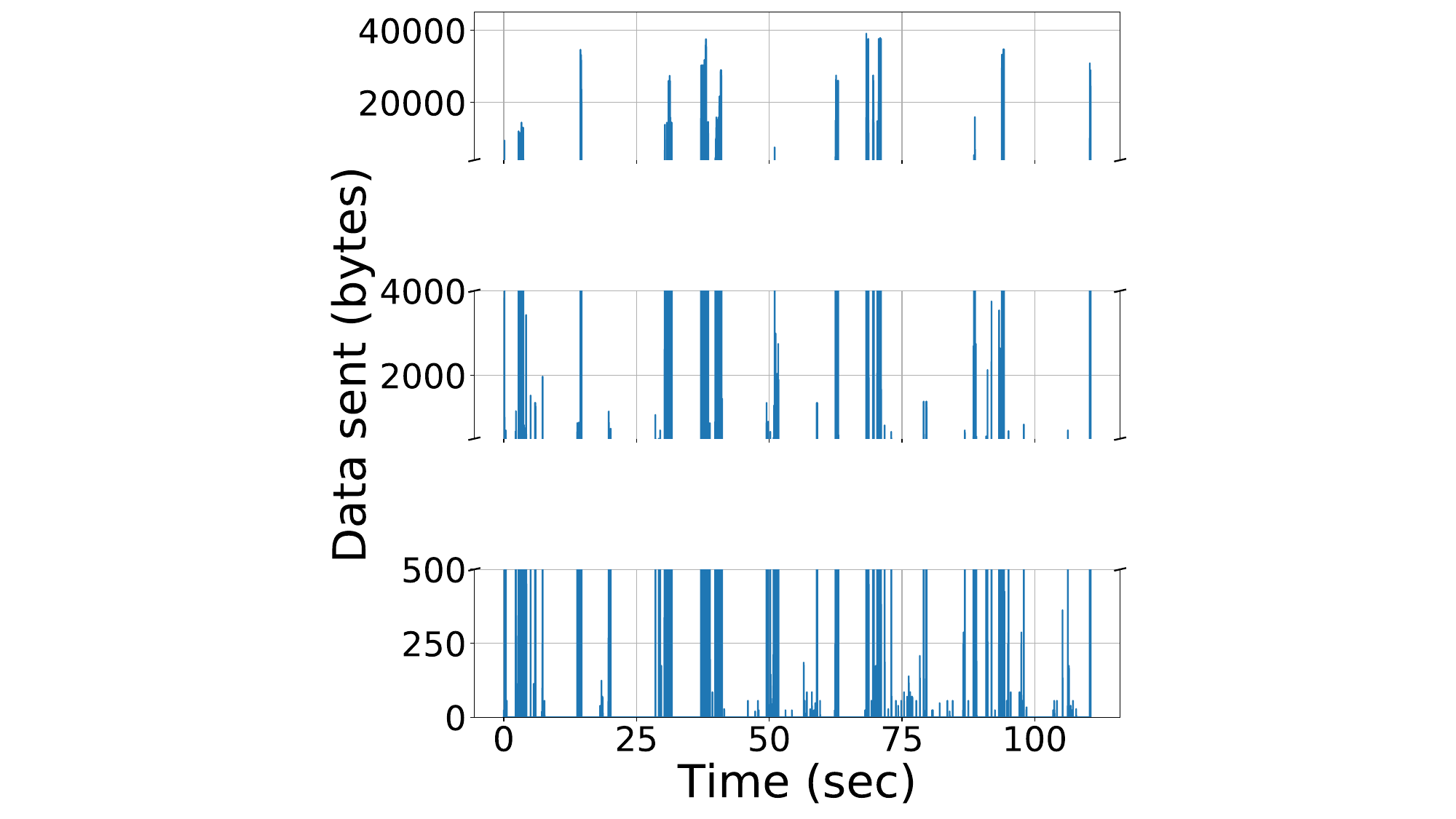}  
        \label{fig:tiktokout}  
        \centering 
        \includegraphics[width=1\textwidth,trim={7.5cm 0cm 7.3cm 0cm},clip]{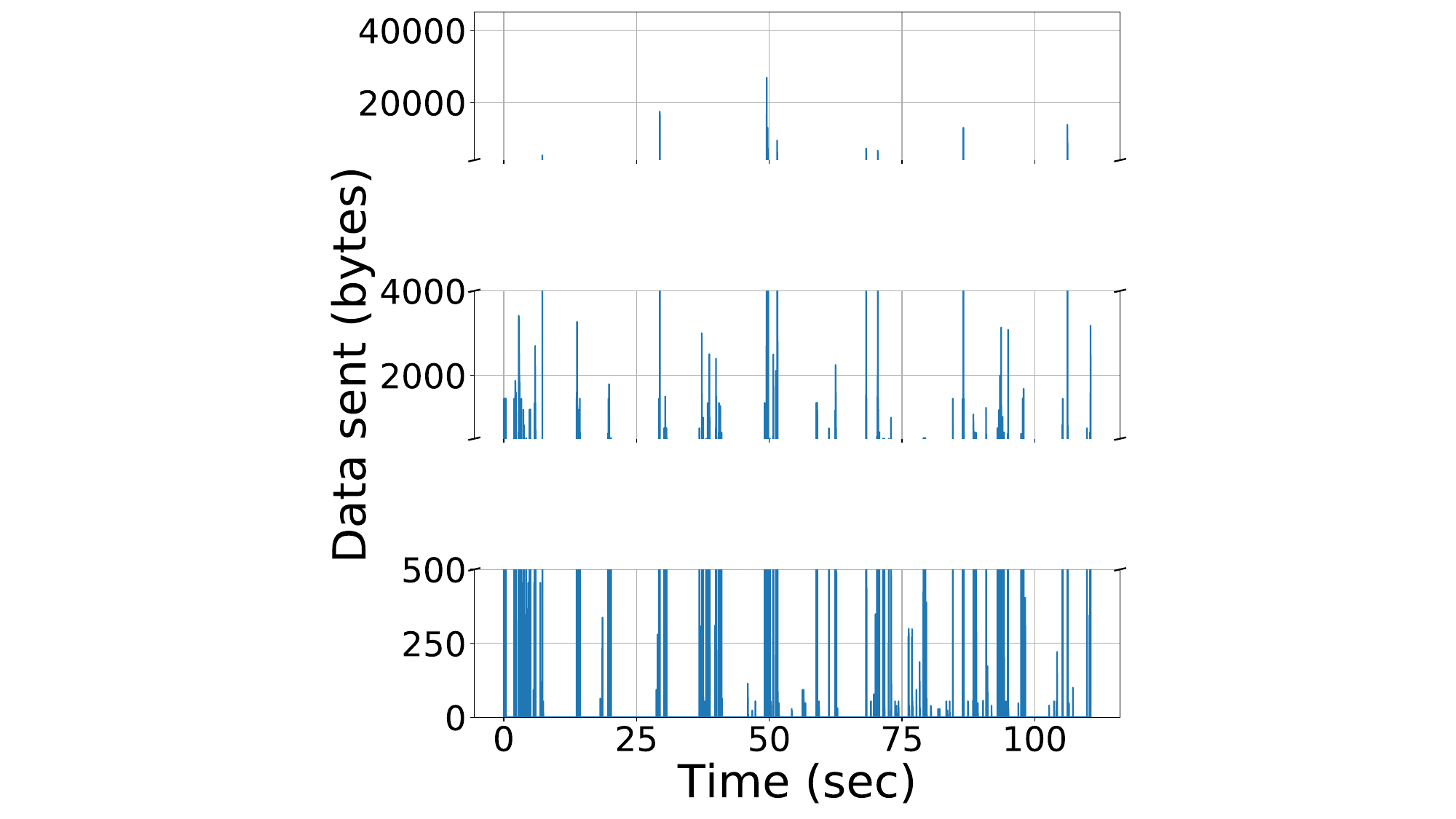}  
        \label{fig:tiktokin}
        \caption{Tiktok}
    \end{subfigure}
    %\vskip\baselineskip
    \begin{subfigure}[b]{0.19\textwidth}   
        \centering 
        \includegraphics[width=1\textwidth,trim={7.5cm 0cm 7.3cm 0cm},clip]{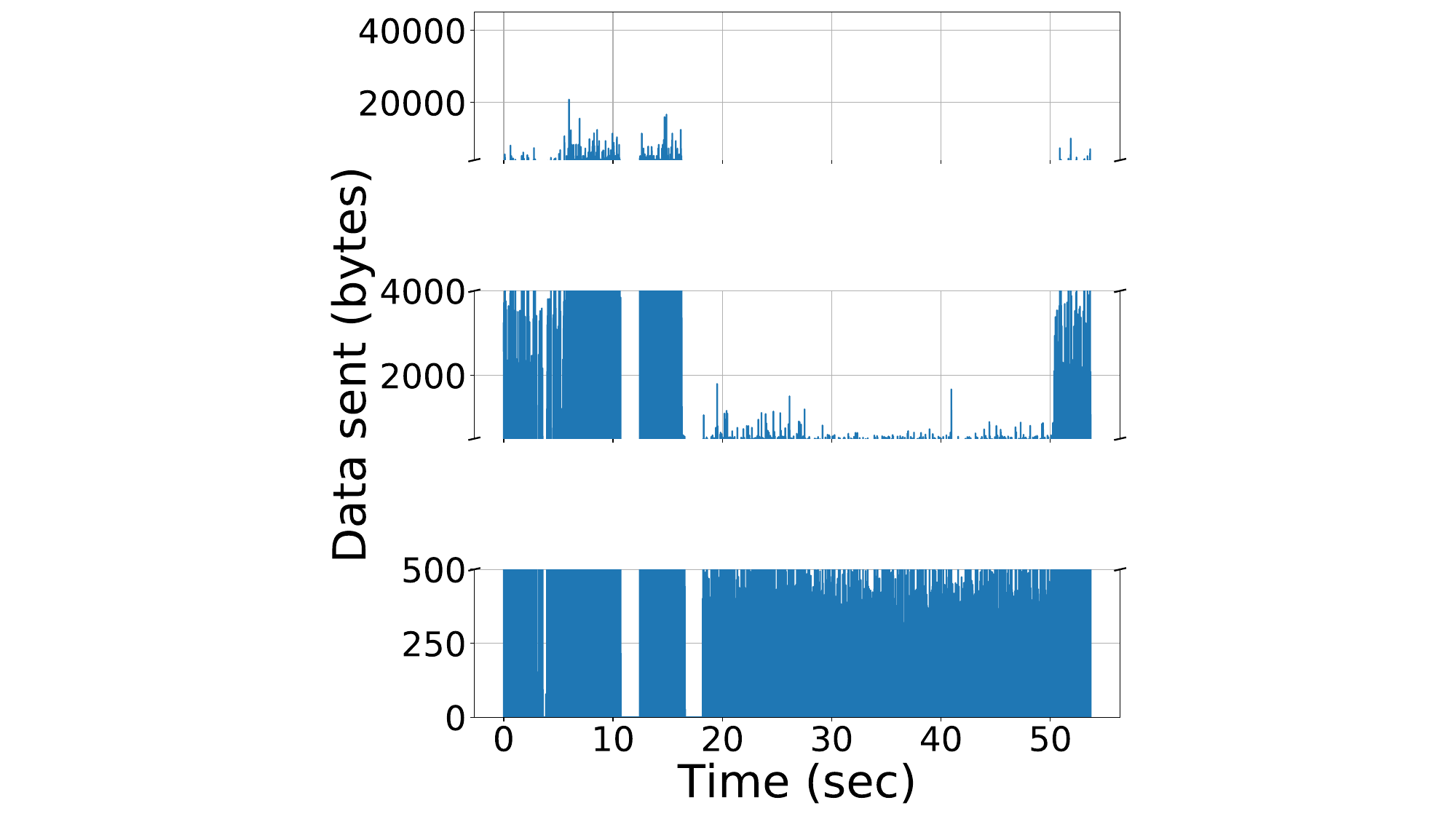}   
        \label{fig:zoomout}  
        \centering 
        \includegraphics[width=1\textwidth,trim={7.5cm 0cm 7.3cm 0cm},clip]{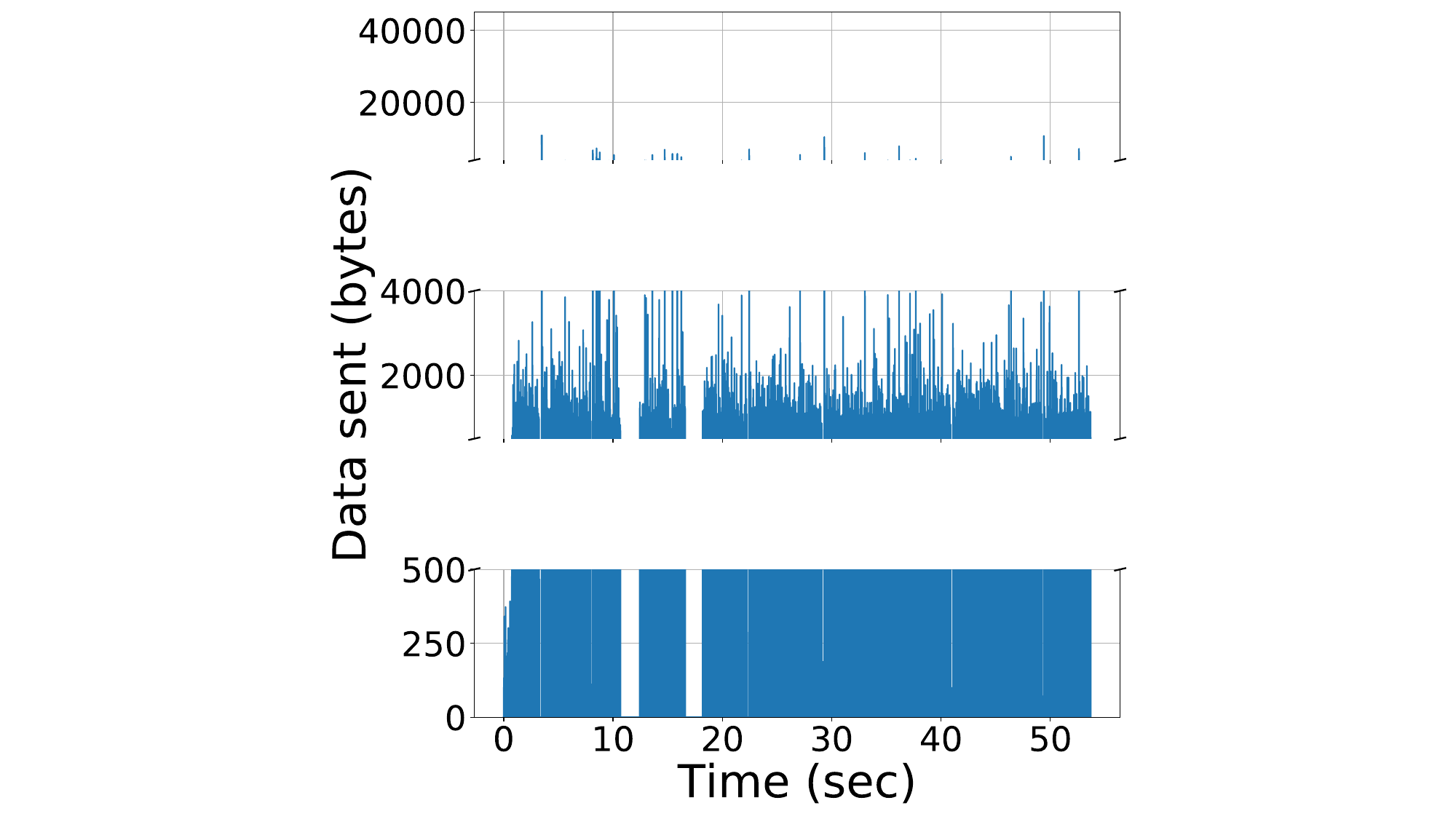}  
        \label{fig:zoomin}
        \caption{Zoom}
    \end{subfigure}
    %\vskip\baselineskip
    %\begin{subfigure}[b]{0.32\textwidth}   
    %    \centering 
    %    \includegraphics[width=0.49\textwidth]%{trafficpatterns/whatsapp_outmod.eps}   
   %     \label{fig:whatsappout}  
   %     \centering 
   %     \includegraphics[width=0.49\textwidth]%{trafficpatterns/whatsapp_inmod.eps}  
  %      \label{fig:whatsappin}
  %      \caption{Whatsapp -- Voice call}
   % \end{subfigure}
    %\hfill
    \begin{subfigure}[b]{0.19\textwidth}   
        \centering 
        \includegraphics[width=1\textwidth,trim={7.5cm 0cm 7.3cm 0cm},clip]{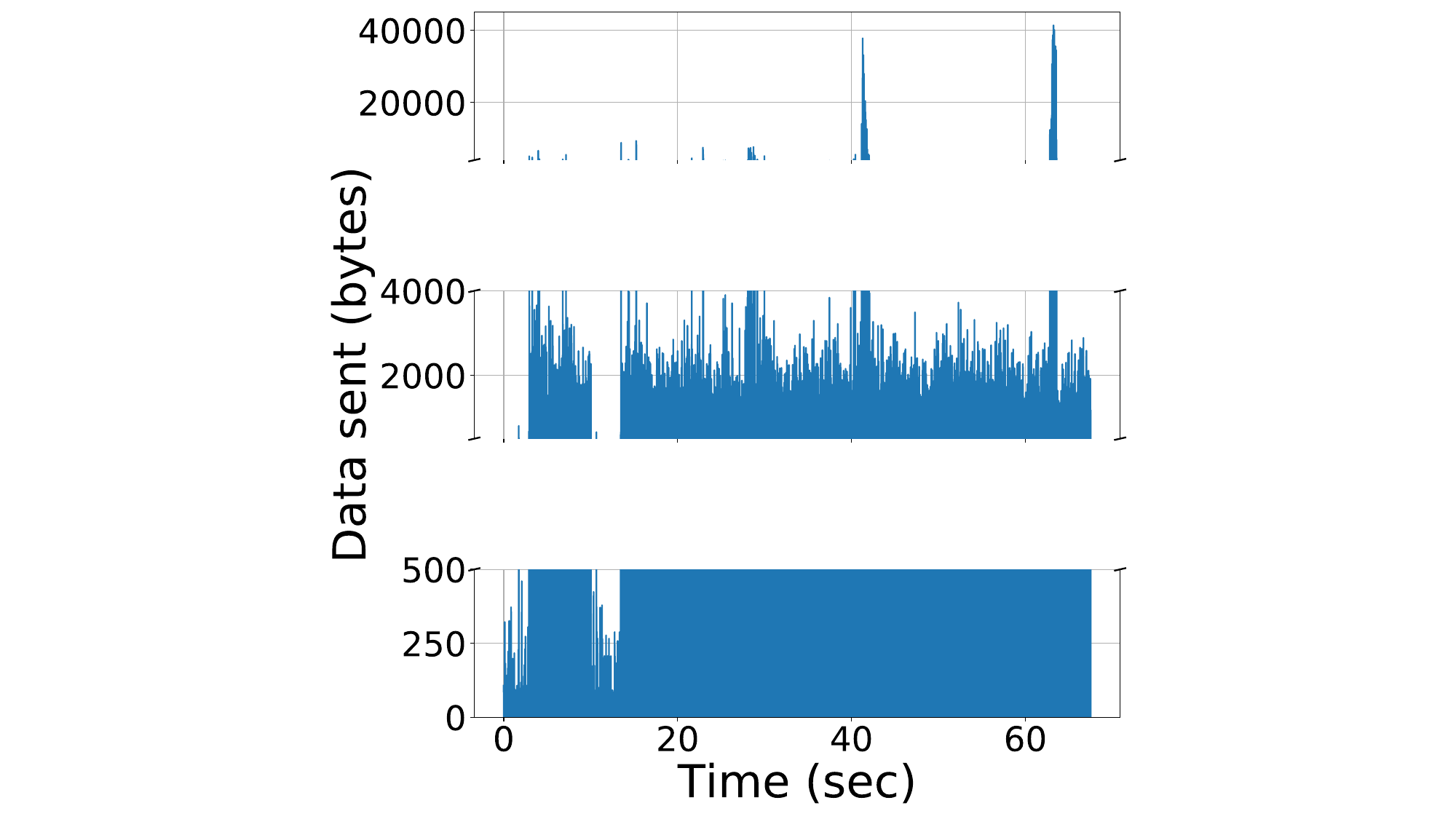}   
        \label{fig:oculusout}  
        \centering 
        \includegraphics[width=1\textwidth,trim={7.5cm 0cm 7.3cm 0cm},clip]{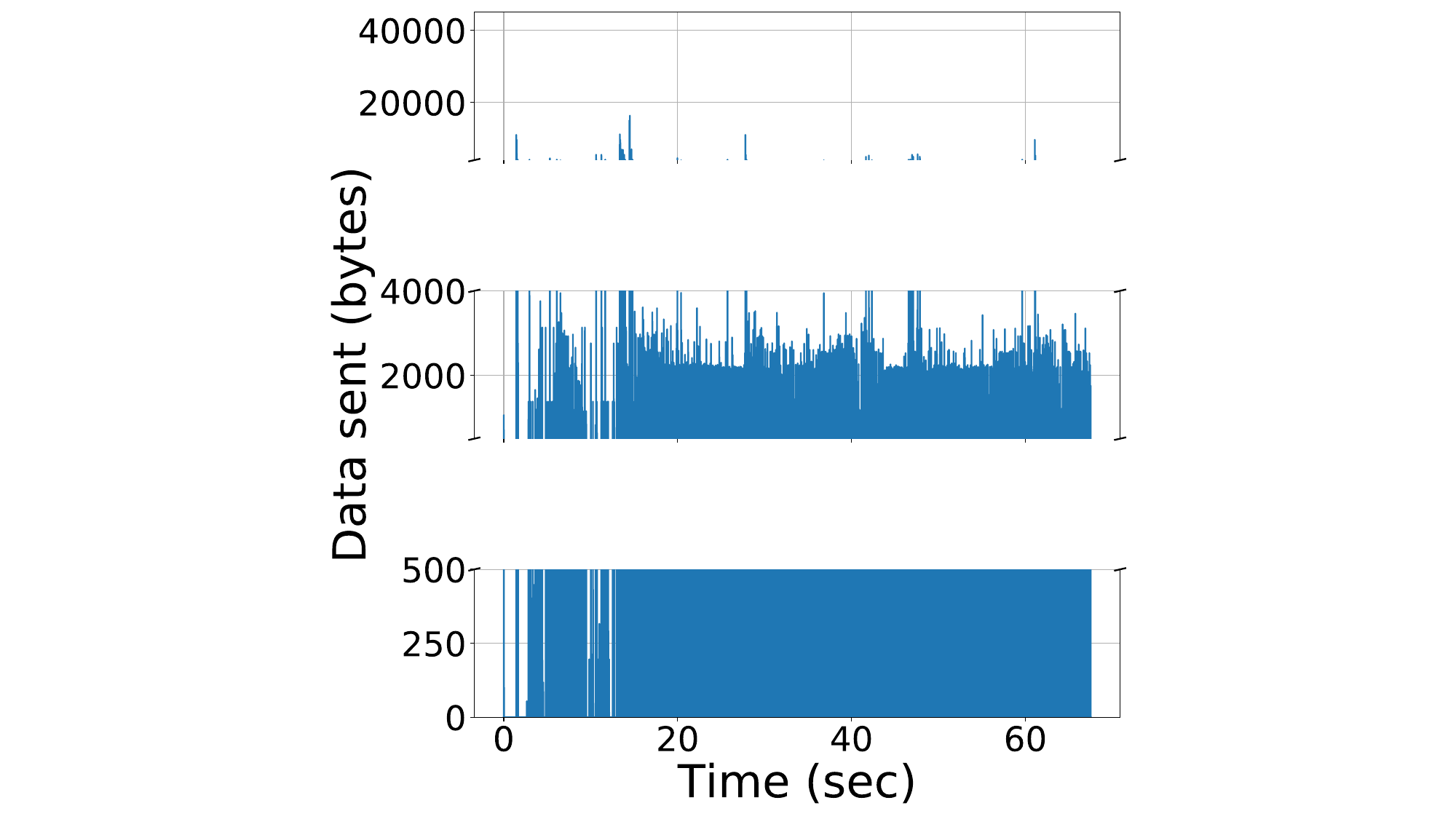} 
        \label{fig:oculusoin}
        \caption{Horizon Venues}
    \end{subfigure}
    %\vskip\baselineskip
    \caption{The downlink (top) and uplink (bottom) traffic patterns captured from different use cases for user plane experiments}
    \label{fig:5gtrafficpatterns}
\end{figure*}

\section{Raw Throughput Benchmarks} \label{app:rawbench}
\begin{figure}[h]
    \centering
    \begin{subfigure}[t]{0.445\columnwidth}
        \centering
        \includegraphics[width=\textwidth,trim={6.5cm 0cm 6.6cm 0cm},clip]{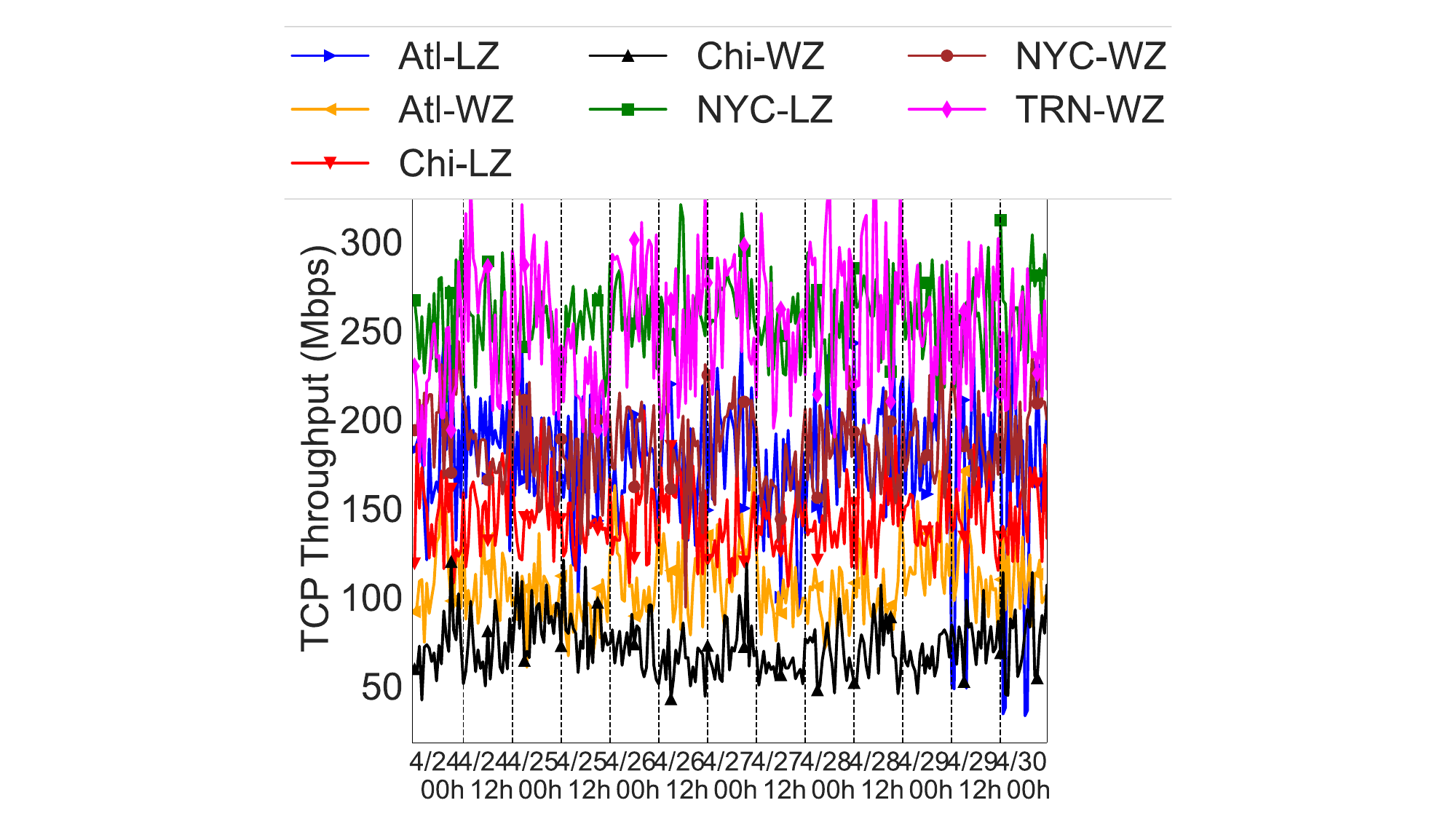} 
        \caption{North America - East}
        \label{fig:benchallthrnae}
    \end{subfigure}%
    \begin{subfigure}[t]{0.445\columnwidth}
        \centering
        \includegraphics[width=\textwidth,trim={6.5cm 0cm 6.6cm 0cm},clip]{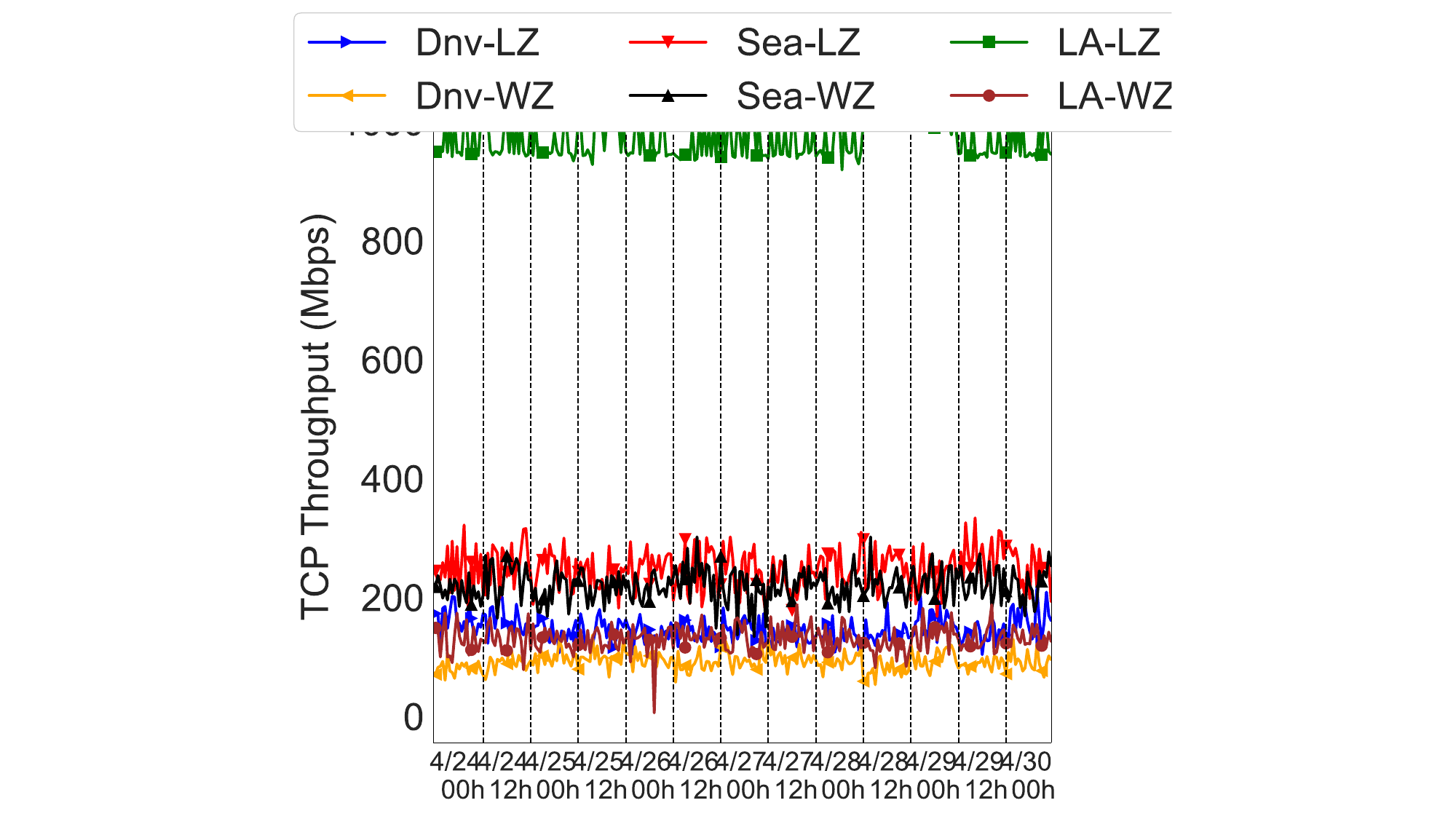}        
        \caption{North America - West}
        \label{fig:benchallthrnaw}
    \end{subfigure}
    \begin{subfigure}[t]{0.445\columnwidth}
        \centering
        \includegraphics[width=\textwidth,trim={6.5cm 0cm 6.6cm 0cm},clip]{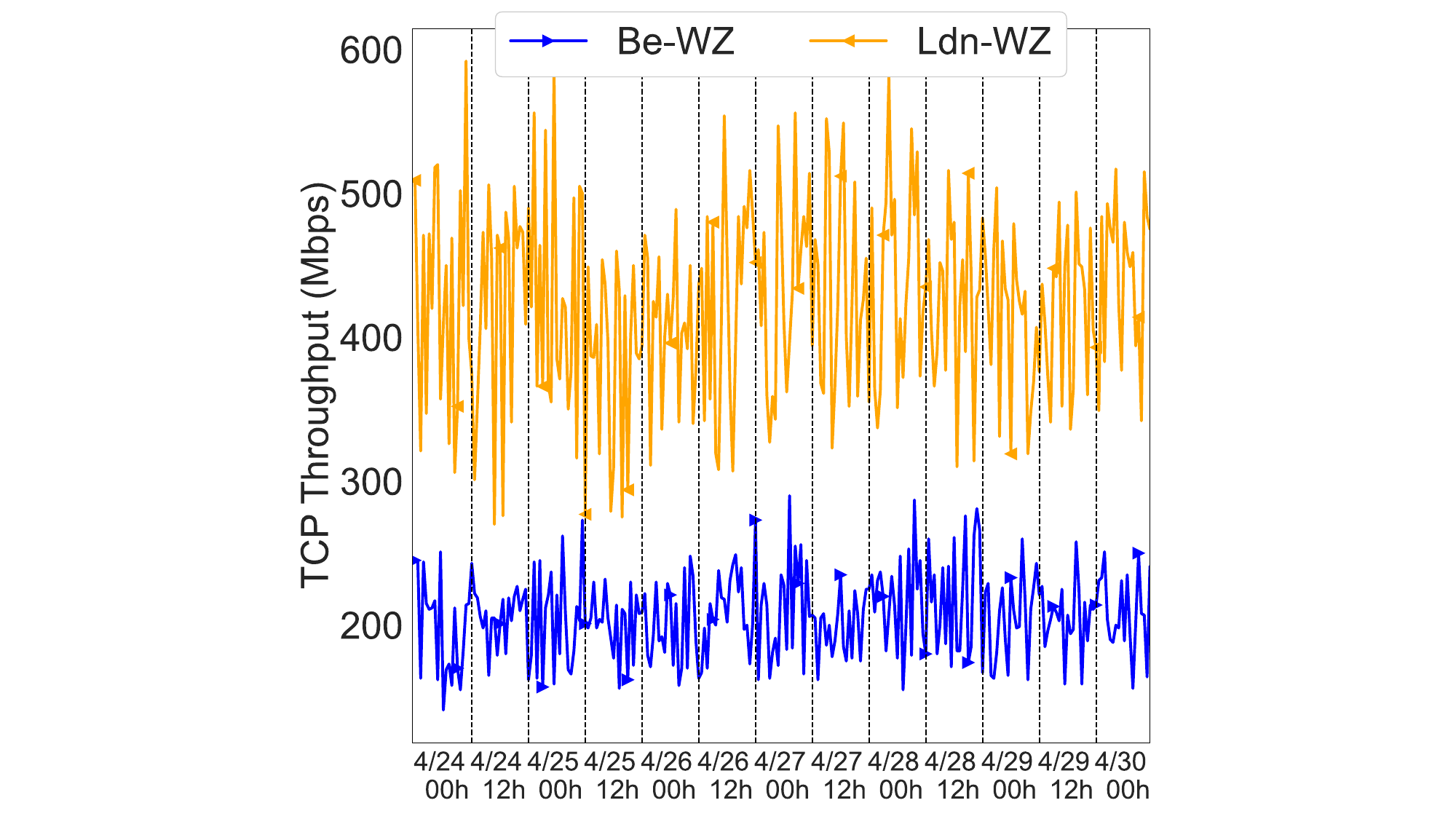}        
        \caption{Europe}
        \label{fig:benchallthreu}
    \end{subfigure}
    \begin{subfigure}[t]{0.445\columnwidth}
        \centering
        \includegraphics[width=\textwidth,trim={6.5cm 0cm 6.6cm 0cm},clip]{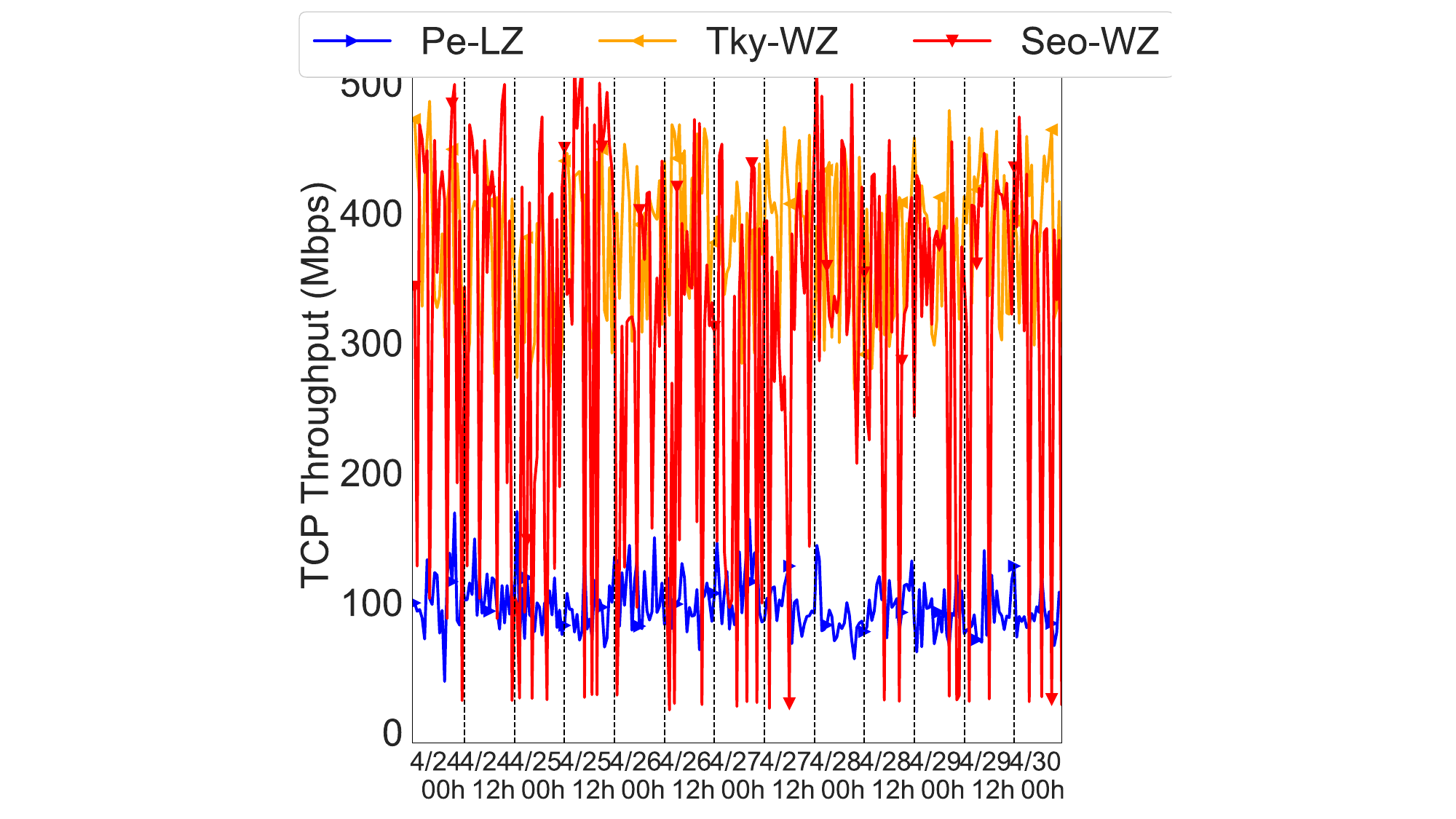}        
        \caption{Asia-Pacific}
        \label{fig:benchallthrap}
    \end{subfigure}
    \caption{Raw measurements of weekly TCP throughput between AZs and edge zones. (April 24-30th 2023)}
    \label{fig:benchthrall}
\end{figure}

\begin{figure}[h]
    \centering
    \begin{subfigure}[t]{0.445\columnwidth}
        \centering
        \includegraphics[width=\textwidth,trim={6.5cm 0cm 6.6cm 0cm},clip]{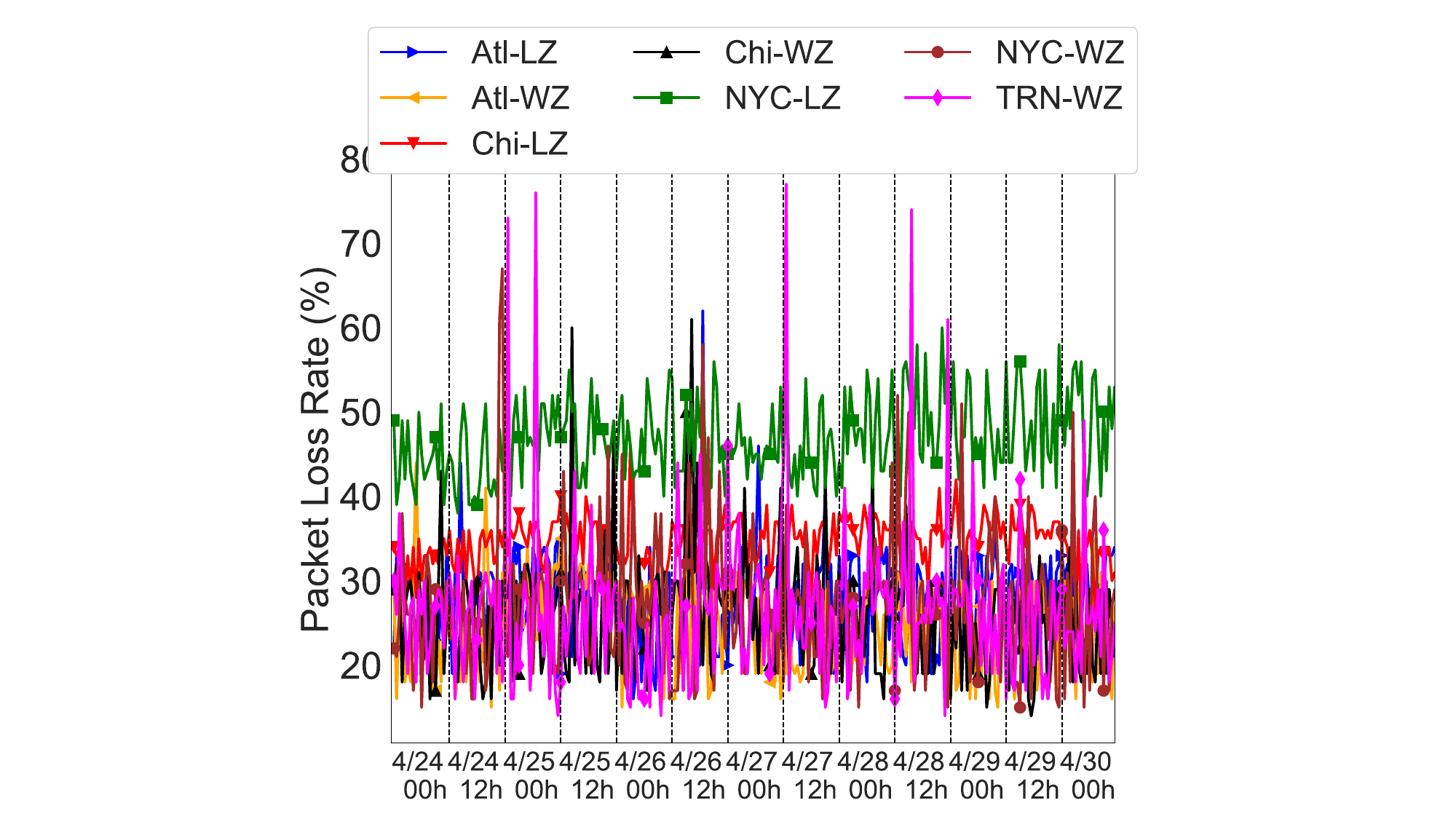} 
        \caption{North America - East}
        \label{fig:benchallplnae}
    \end{subfigure}%
    \begin{subfigure}[t]{0.445\columnwidth}
        \centering
        \includegraphics[width=\textwidth,trim={6.5cm 0cm 6.6cm 0cm},clip]{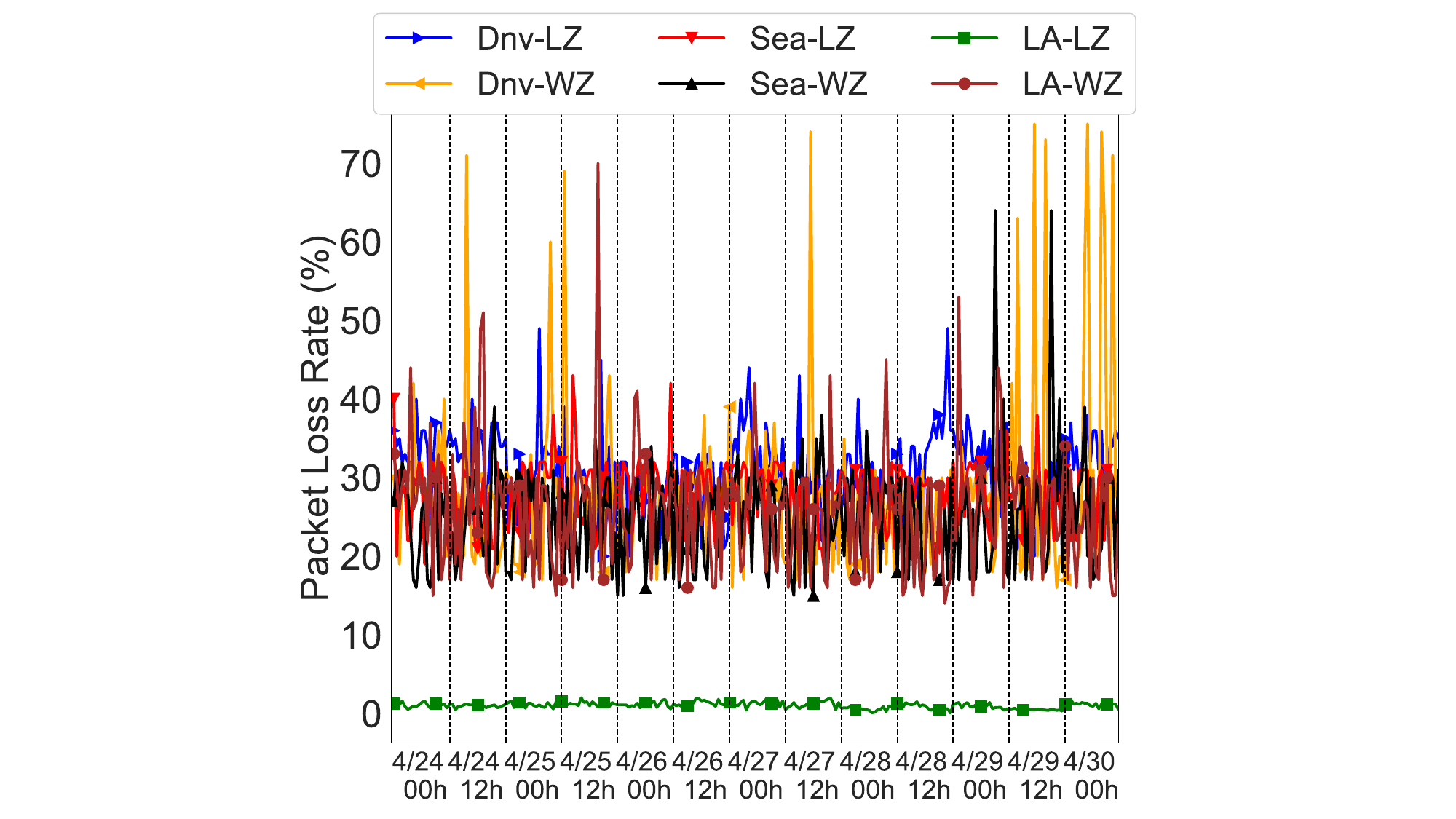}        
        \caption{North America - West}
        \label{fig:benchallplnaw}
    \end{subfigure}
    \begin{subfigure}[t]{0.445\columnwidth}
        \centering
        \includegraphics[width=\textwidth,trim={6.5cm 0cm 6.6cm 0cm},clip]{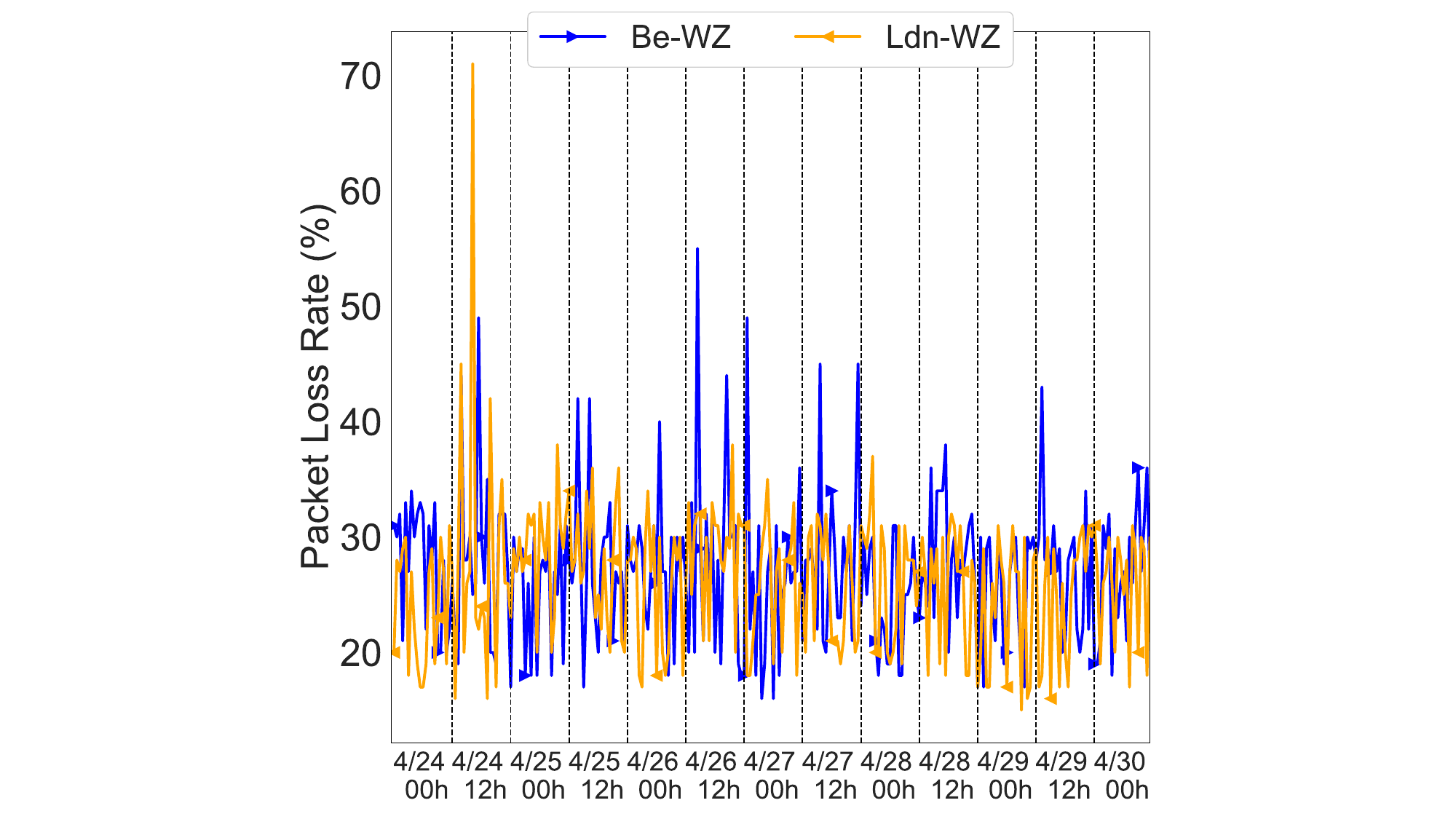}        
        \caption{Europe}
        \label{fig:benchallpleu}
    \end{subfigure}
    \begin{subfigure}[t]{0.445\columnwidth}
        \centering
        \includegraphics[width=\textwidth,trim={6.5cm 0cm 6.6cm 0cm},clip]{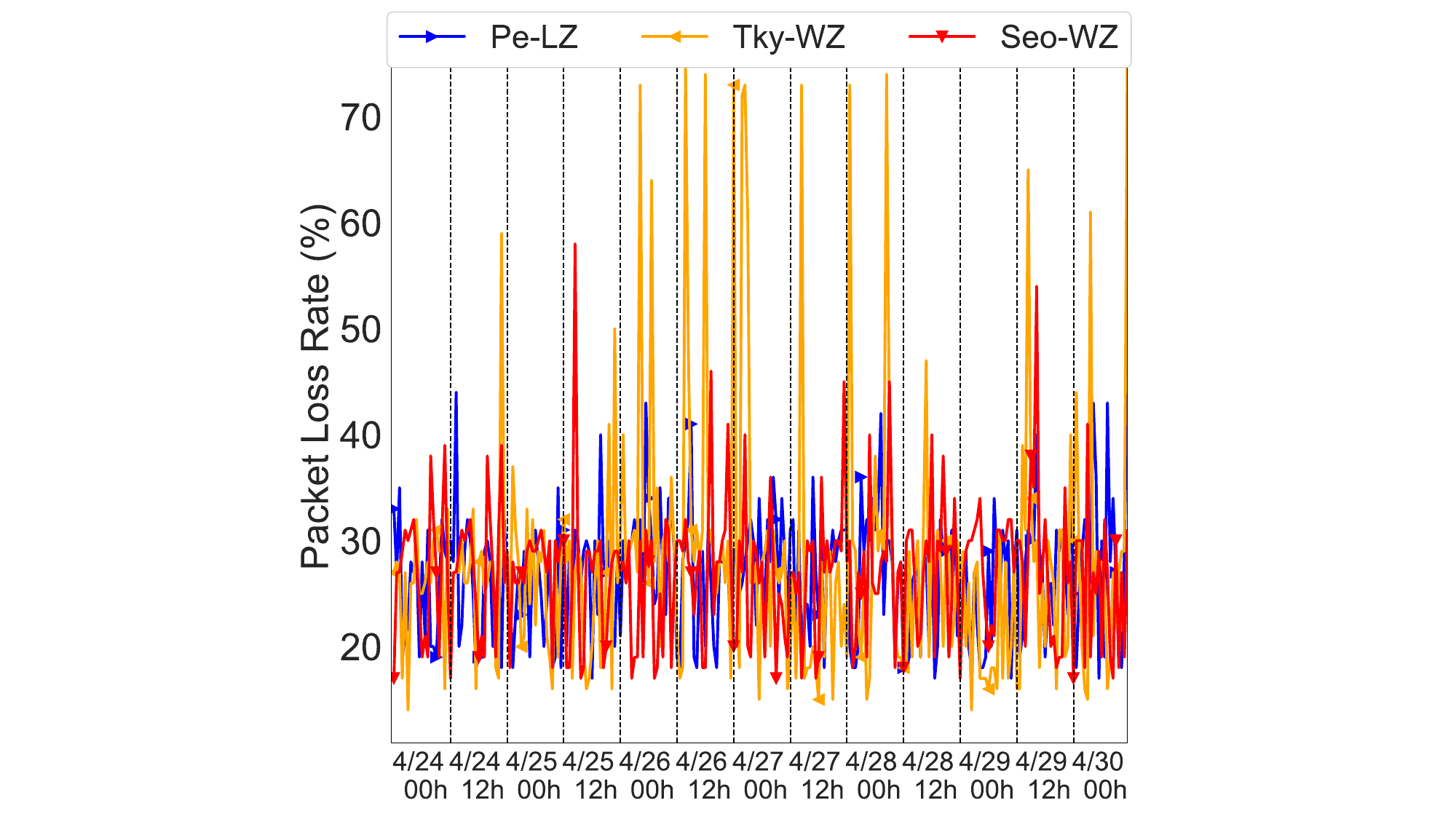}        
        \caption{Asia-Pacific}
        \label{fig:benchallplap}
    \end{subfigure}
    \caption{Raw measurements of weekly UDP packet loss between AZs and edge zones. (April 24-30th 2023)}
    \label{fig:benchplall}
\end{figure}

Our entire set of AWS edge zone benchmark measurements are given in Figures~\ref{fig:benchthrall} and ~\ref{fig:benchplall} for TCP throughput and UDP packet loss rate respectively. However, for the TCP and UDP benchmarks, each data point has been obtained using an iperf3 stream with 100 samples where the final average is reported. Due to the inherent volatile nature of the TCP and UDP connections between AZs and edge zones, this method yields a more reliable average.

While with sufficient measurements the average indeed reflects a consistent pattern both for TCP throughput and UDP packet loss, the variations are more apparently observed in Figures~\ref{fig:benchthrall},~\ref{fig:benchplall} from the weekly CDFs. This allows us to account for the time-related fluctuations in both TCP and UDP connections and obtain a reliable average. 

\section{AWS Experimentation Cost} \label{app:costtot}

\begin{figure}[t]
    \centering
    \includegraphics[width=\columnwidth,trim={4.5cm 5.5cm 8cm 2cm},clip]{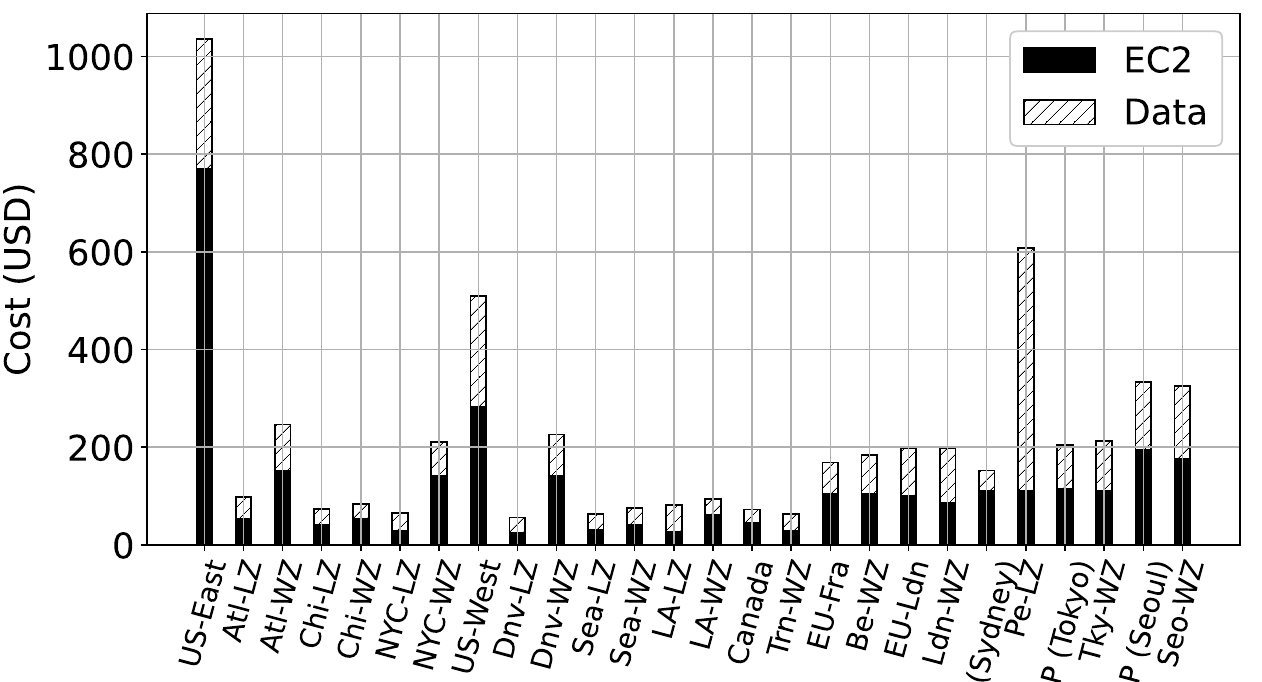}
     \caption{Total cost of the experiments in this paper broken down into EC2 and data transfer cost.}
     \label{fig:totcost}
\end{figure}

Conducting the experiments in this paper required instantiating over 200 EC2 instances distributed across multiple locations. Furthermore, to create 5G user plane traffic for our measurements in Section~\ref{sec:thrplresults}, large amounts of data was transferred between edge zones and the respective parent regions. The EC2 instance and data transfer costs for each edge zone are given in Table~\ref{tbl:cost}. We can see that the data transfer rates for all the North America and Europe regions are the same. For the Asia-Pacific, these rates increase significantly. The inbound data transfer is free for all zones in all regions.

\begin{table}[t]
\vspace{12pt}
\centering
\small\selectfont
\caption{EC2 instance and data transfer costs for all edge locations and AZs. The reported data transfer rate is for outbound traffic to the governing AZ of the edge location.} 
\label{tbl:cost}
  \begin{tabular}{c|c|c|c}
    \multirow{2}{*}{\textbf{City}} & \multirow{2}{*}{\textbf{Zone}} 
                                   & \multicolumn{2}{c}{\textbf{Cost (USD) }} \\
     & &
    \multicolumn{1}{|c}{\textbf{t3.xlarge (hourly) }} & 
    \multicolumn{1}{|c}{\textbf{Data (per GB) }} \\
    \hline \hline
    \multirow{3}{*}{Atlanta} & AZ & 0.1664 & 0.02 \\
                         & LZ & 0.208 & 0.02 \\
                         & WZ & 0.224 & 0.02 \\ \hline 
    \multirow{3}{*}{New York City} & AZ & 0.1664 & 0.02 \\
                         & LZ & 0.208 & 0.02 \\
                         & WZ & 0.224 & 0.02 \\ \hline
    \multirow{3}{*}{Chicago} & AZ & 0.1664 & 0.02 \\
                         & LZ & 0.208 & 0.02 \\
                         & WZ & 0.224 & 0.02 \\ \hline
    \multirow{3}{*}{Denver} & AZ & 0.1664 & 0.02 \\
                         & LZ & 0.208 & 0.02 \\
                         & WZ & 0.224 & 0.02 \\ \hline
    \multirow{3}{*}{Seattle} & AZ & 0.1664 & 0.02 \\
                         & LZ & 0.208 & 0.02 \\
                         & WZ & 0.224 & 0.02 \\ \hline
    \multirow{3}{*}{Los Angeles}  & AZ & 0.1664 & 0.02 \\
                         & LZ & 0.208 & 0.02 \\
                         & WZ & 0.224 & 0.02 \\ \hline
    \multirow{2}{*}{Toronto} & AZ & 0.1856 & 0.02   \\
                         & WZ & 0.2506 & 0.02  \\ \hline
    \multirow{2}{*}{London} & AZ & 0.1888 & 0.02  \\
                         & WZ & 0.236 & 0.02 \\ \hline
    \multirow{2}{*}{Berlin}  & AZ & 0.192 & 0.02   \\
                         & WZ & 0.24 & 0.02  \\ \hline
    \multirow{2}{*}{Tokyo} & AZ & 0.2176 & 0.09  \\
                         & WZ & 0.2938 & 0.09  \\ \hline
    \multirow{2}{*}{Seoul} & AZ & 0.208 & 0.08  \\
                         & WZ & 0.26 & 0.08  \\ \hline
    \multirow{2}{*}{Perth}  & AZ & 0.2112 & 0.10  \\
                         & LZ & 0.2851 & 0.10   \\ \hline
  \end{tabular}
\end{table}

We share the total cost of the experimentation process in Figure~\ref{fig:totcost}. It is important to note that for the chosen regions and locations, these numbers should not be interpreted as an indication of what actual deployment costs would look like. In other words, real life 5G deployments may have different costs. We simply report the total experimentation cost as it is shown in the AWS billing dashboard of the account that was used to conduct the experiments. Our goal is to provide any interested parties with an idea regarding the cost of building and operating such a testbed. 

For specific regions and locations, the experiments had to be repeated while addressing various bugs in the experimentation scripts and environment. The majority of the debugging process took place in the US-East (Northern Virginia) and the edge zones attached to it. Thus, we can see that the cost is much higher than other regions. In conclusion, we had to repeat the experiments in certain regions due to various bugs. Figure~\ref{fig:totcost} is presented here as a testament to the trial and error process we went through while gathering high fidelity results. %Furthermore, these costs are for data transmissions between regions. Transferring data to/from the Carrier RAN is more expensive. %The interested parties can check the relevant costs through the AWS Pricing Calculator~\cite{AWSPrici45online}.

\begin{figure*}[t]
  \centering
  \includegraphics[width=\textwidth,trim={0cm 7cm 12cm 0cm},clip]{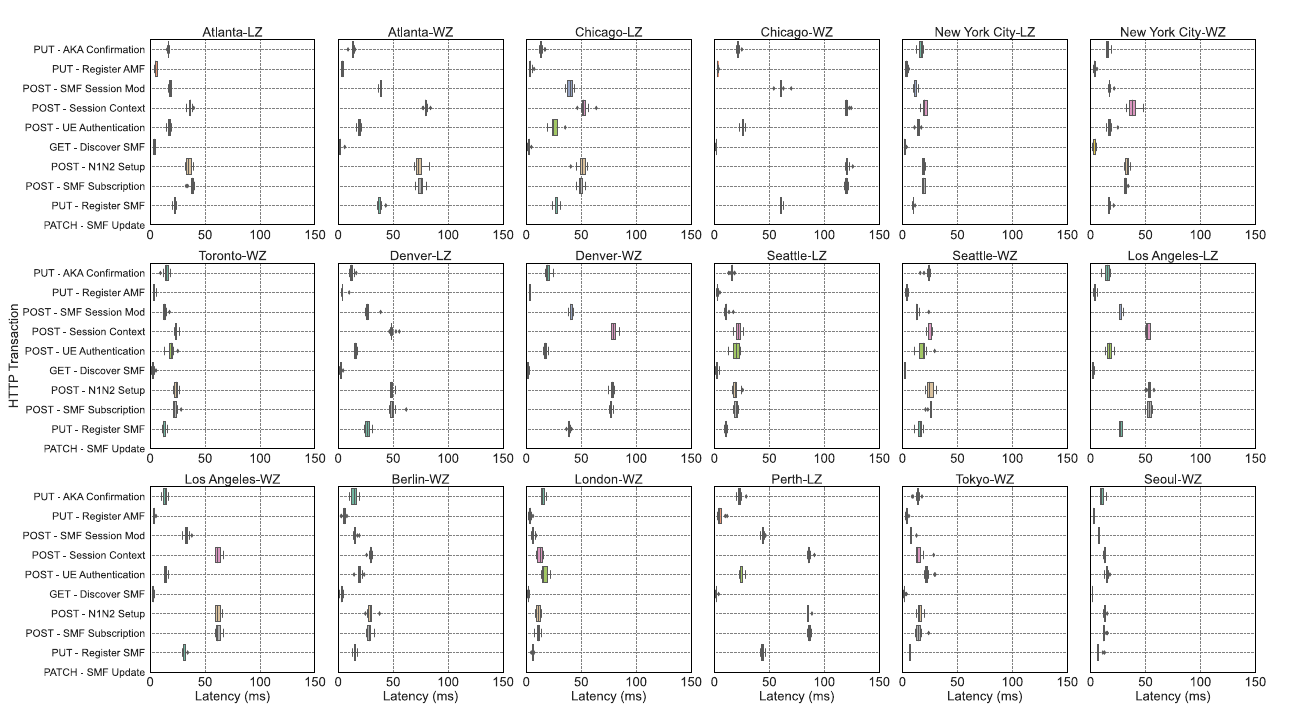}
  \caption{All HTTP transaction measurements for strategy 1 (URLLC User)}
  \label{fig:httpallstr1}
\end{figure*}

\section{AWS Cluster Deployment Details} \label{app:clusterdet}
%Throughout the process of experimenting on AWS, we found ourselves faced with some administrative and technical challenges.

\textbf{Setup Selection:} We use the Ranchers Kubernetes Engine (RKE)~\cite{RKE72online} to set up a self-managed production-grade Kubernetes cluster on top of 15 EC2 instances. This allows us to have greater visibility into the container management infrastructure. We are able to manipulate worker node labels during setup to manage the location-specific VNF deployment during experimentation.  In our flavor selection, we are limited to the t3.medium, t3.xlarge, and r5.2xlarge general-purpose compute nodes at WZs~\cite{5GEdgeCo55online}. With only 2 vCPUs, t3.medium is not ideal for worker nodes and can lead to instability in the Kubernetes control plane. Therefore, we choose to use the more conservative t3.xlarge with 4 vCPUs, instead of the r5.2xlarge with 8 vCPUs. This enables us to spawn more EC2 instances within our regional vCPU quotas. The same flavor is used in AZs and LZs as well to preserve equality across the experiments.
    
\textbf{EKS vs EC2:} When setting up a Kubernetes cluster in a public cloud, users have the choice to either rely on an automation pipeline offered by the provider (e.g., AWS) or set up their own clusters from scratch. To that end, AWS offers the Elastic Kubernetes Service (EKS)~\cite{ManagedK20online} as a popular option for directly granting users access to a Kubernetes cluster by abstracting the setup process. While EKS is a robust solution, we were apprehensive about this usage, due to the lack of flexibility compared to self-managed clusters on top of EC2 instances. Especially for conducting host-to-host performance benchmarks, without having access to the VMs directly, we felt that our measurements could be restricted. Furthermore, being able to label nodes during cluster creation provided us with flexibility in deploying the 5G core VNFs in our chosen edge locations. 

\textbf{Secondary Network Interface:} With our in-lab 5G testbed deployed on COTS hardware, the control plane communication among VNFs took pace over secondary interfaces through Multus~\cite{k8snetwo60online}. Multus is a Container Network Interface (CNI) plugin orchestrator, which allows for additional network interfaces to be attached to a pod. This is achieved through the creation of an additional Multus subnet, independent of the Kubernetes cluster Classless Inter-Domain Routing (CIDR) range. Moving to AWS, we noticed that none of the packets on this Multus subnet were being routed. This is because, routing in AWS, takes place through routing table rules in the relevant Virtual Private Cloud (VPC). Without any prior configuration, the VPC cannot route packets between the private IP addresses of Kubernetes pods assigned to them using Multus. To deal with this issue, AWS has prepared an elaborate guide to support Multus pods on top of EKS~\cite{awssampl36online}. While the guide is intended for an EKS cluster, we modified the automation scripts so that they functioned properly for the EC2 instances of our self-managed cluster. Essentially, the process involves three major steps: (1) creating a new secondary interface in the network stack of the guest-host; (2) running the Multus-pod and then attaching the Multus-IP of the pod to the new network interface created in step 1; and (3) updating the private IP address of the EC2 instance through AWS CLI to include the newly created interface. This way, the VPC can perform routing using the Multus-IPs. Unfortunately, for any operating system (OS) that is not Amazon Linux, the first step above cannot be automated through the provided guide~\cite{awssampl36online}. The OS of our EC2 instances are Ubuntu 22.04. Thus, effectively applying the Multus solution devised by AWS required the creation and application of a new Netplan for the required secondary interface of each pod. After doing this, Multus routing was successful across EC2 instances. 
However, we quickly realized that while experimenting with network slicing, there would be hundreds of Multus-pods being deployed and this solution could create scalability issues down the line. Especially, since there would be a secondary interface for each pod, this implies that a single EC2 instance can have 100s of private IPs as the deployment scaled. 
Therefore, we had to configure our 5G testbed to utilize DNS-based resolutions among the 5G core VNFs using Kubernetes cluster services.

%\textbf{Regional Quotas:} The default vCPU quota for each region is set to 5 for All Standard instances in AWS. For our experiments, we requested a quota increase to 128 vCPUs for each of the 8 regions used in our experiments. This required the submission of a report, detailing the specific task that would warrant such a request. In less than one and a half weeks, we were granted the quota increase.

\textbf{Wavelength Zone - No Internet :} Currently, WZs do not have incoming Internet access. This means that no Docker image can be pulled onto them. To bypass this issue, we created an Amazon Machine Image (AMI) that contained all our required Docker images and started our EC2 instances in WZs using that AMI. This included the instrumentation environment images, Kubernetes cluster images (i.e., specifically RKE variants), and our custom OAI images. Since this AMI is bound to the AWS dashboard, we can easily share the AMI with other researchers if they attempt to re-create this testbed.

\section{HTTP Transaction Measurements} \label{app:httptran}
\begin{figure*}[h]
  \centering
  \includegraphics[width=\textwidth,trim={0cm 7cm 12cm 0cm},clip]{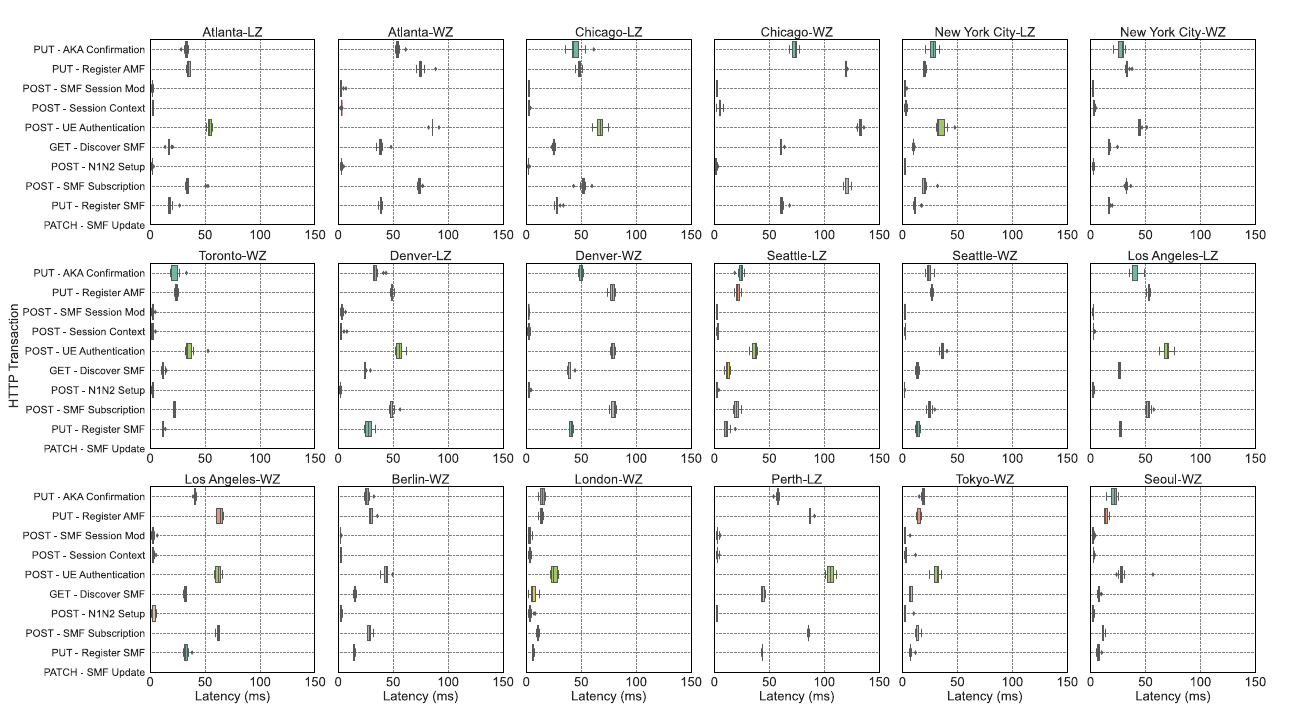}
  \caption{All HTTP transaction measurements for strategy 2 (MCS Static)}
  \label{fig:httpallstr2}
\end{figure*}

\begin{figure*}[h]
  \centering
  \includegraphics[width=\textwidth,trim={0cm 7cm 12cm 0cm},clip]{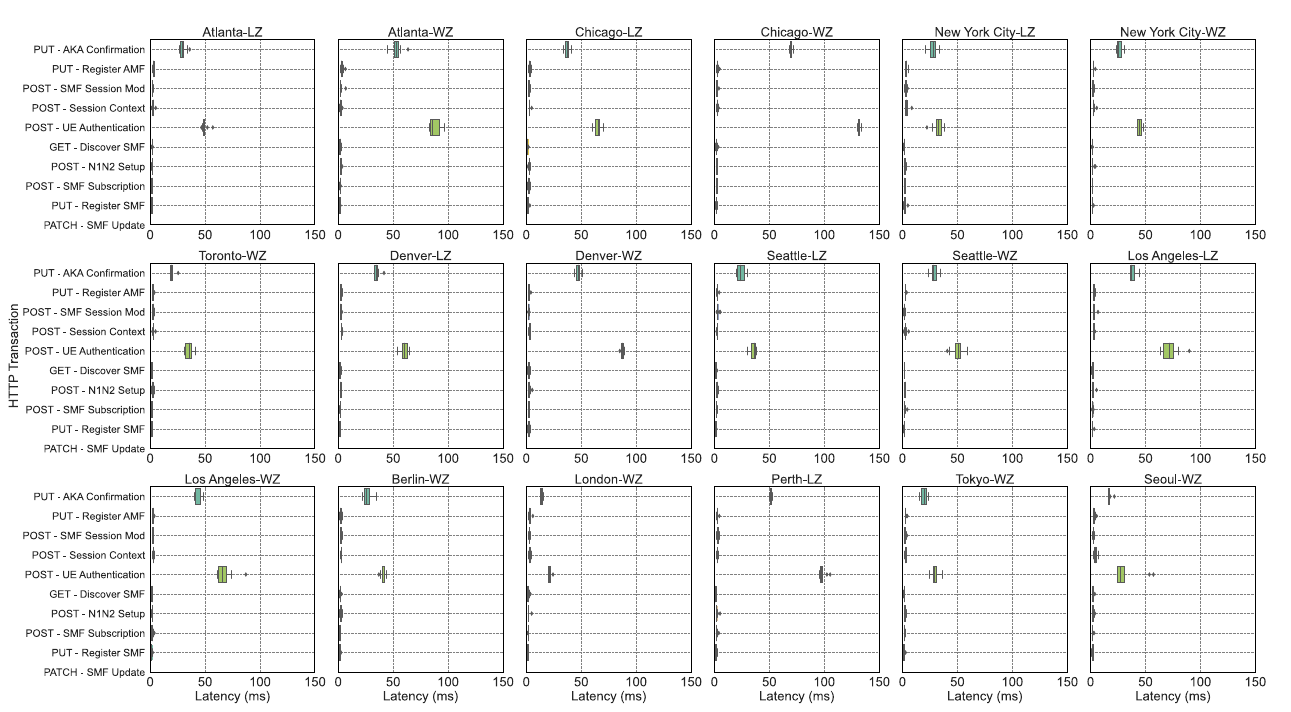}
  \caption{All HTTP transaction measurements for strategy 3 (MCS Mobile)}
  \label{fig:httpallstr3}
\end{figure*}

We provide the entire set of HTTP transaction measurements that lead to the averages presented in Figure~\ref{fig:opbreak}. All the HTTP measurements are given in Figures~\ref{fig:httpallstr1}, ~\ref{fig:httpallstr2}, ~\ref{fig:httpallstr3} for the URLLC User, MCS Static and MCS Mobile strategies respectively.
Our goal in providing these measurements is to show the variation in HTTP transaction latency values. Each transaction corresponds to ten traces. For the majority of the use cases, the latency fluctuations are less than 5ms. The highest relative deviation is observed in the London and Seattle-LZ measurements in Figures~\ref{fig:httpallstr1}, ~\ref{fig:httpallstr2} for the \texttt{POST - UE Authentication} HTTP message. For the London measurements, the error bars show higher variance but in a small interval. Since this specific HTTP message is associated with a compute-intensive process in the 5G-AKA service chain, this behavior is explained by the higher relative difference in software processing times within the VNFs. Looking at the 5G-AKA service chain in Figure~\ref{fig:oaimsg}, the \texttt{POST - UE Authentication} HTTP message triggers the sub-processes of the HN 5G-AKA VNFs (i.e., AUSF, UDM and UDR). After the HN VNFs process the request, the reply is sent to the AMF from the AUSF. Thus, the variation in the latency overhead of the \texttt{POST - UE Authentication} HTTP message stems from the HN 5G-AKA VNF processing.

\end{document}